\documentclass[11pt, notitlepage, tightenlines,reprint, aps, prd, superscriptaddress]{revtex4-2}
\usepackage[utf8]{inputenc}
\usepackage{placeins}
\usepackage{mathtools}
\usepackage{amsfonts}
\usepackage{amssymb,amsmath}
\usepackage{color}
\usepackage{xcolor}
\definecolor{lcolor}{rgb}{0.5,0,0}
\definecolor{citcolor}{rgb}{0,0.3,0.0}
\usepackage[breaklinks,colorlinks,urlcolor=blue,citecolor=citcolor,linkcolor=lcolor]{hyperref}

\usepackage{graphicx}
\usepackage{subfigure}
\graphicspath{ {image/} }
\usepackage{multirow}

\newcommand*\diff{\mathop{}\!\mathrm{d}}
\usepackage{physics}
\usepackage{bm}
\usepackage{txfonts}
\usepackage{mathrsfs} % in order to use mathscr
\usepackage{slashed}
\usepackage{cleveref}

\newcommand{\GeV}{{{\,}\textrm{GeV}}}

\begin{document}

\bibliographystyle{apsrev4-2}

\title{Scattering and gluon emission in a color field: A light-front Hamiltonian approach}
\author{Meijian Li}
\email{mliy@jyu.fi}
\affiliation{Department of Physics, P.O. Box 35, FI-40014 University of Jyv\"{a}skyl\"{a},
Finland}
\affiliation{
Helsinki Institute of Physics, P.O. Box 64, FI-00014 University of Helsinki,
Finland
}

\author{Tuomas Lappi}
\email{tuomas.v.v.lappi@jyu.fi}
\affiliation{Department of Physics, P.O. Box 35, FI-40014 University of Jyv\"{a}skyl\"{a},
Finland}
\affiliation{
Helsinki Institute of Physics, P.O. Box 64, FI-00014 University of Helsinki,
Finland
}

\author{Xingbo Zhao}
\email{xbzhao@impcas.ac.cn}
\affiliation{Institute of Modern Physics, Chinese Academy of Sciences, Lanzhou 730000, China}
\affiliation{University of Chinese Academy of Sciences, Beijing 100049, China}

\begin{abstract}
We develop a numerical method to nonperturbatively study scattering and gluon emission of a quark from a colored target using a light-front Hamiltonian approach. The target is described as a classical color field, as in the color glass condensate effective theory. The Fock space of the scattering system is restricted to the  $\ket{q}+\ket{qg}$ sectors, but the time evolution of this truncated system is solved exactly.  This method allows us to study the interplay between coherence and multiple scattering in gluon emission. It could be applied both to studying subeikonal effects in high-energy scattering and to understanding jet quenching in a hot plasma.
\end{abstract}
\maketitle

\section{Introduction}
The general picture of a high-energy dilute probe scattering off a color field is a commonly used approach for many different processes in QCD phenomenology. Scattering processes that probe the color glass condensate (CGC)~\cite{Gelis:2010nm} state of small-$x$ gluons inside a high-energy hadron or nucleus are described in terms of infinitely energetic partons passing through an infinitesimally thin color field sheet, using the eikonal approximation. In order to study the phenomenon of jet quenching and radiative energy loss, one studies the situation when a high-energy parton passes through an extended colored medium and loses energy by gluon emission~\cite{Casalderrey-Solana:2011ule,Mehtar-Tani:2011lic,Armesto:2012qa,Armesto:2013fca,Kajantie:2019nse,Kajantie:2019hft}. In both cases, one often performs analytical calculations in a kinematical approximation where the probe has an infinitely large energy. For realistic collider phenomenology in both physical situations, it is important, however, to be able to relax this approximation. 
For scattering off a CGC color field, subeikonal effects~\cite{Altinoluk:2020oyd,Chirilli:2021lif} can be important at realistic collider energies, such as at the upcoming Electron-Ion Collider~\cite{Accardi:2012qut}. 
This is the case, in particular, for the physics of spin at high energies~\cite{Kovchegov:2015pbl,Kovchegov:2017lsr,Kovchegov:2018znm,Jalilian-Marian:2019kaf,Adamiak:2021ppq}. Also for jet quenching, understanding the interplay between the coherence time of the emission and the timescales of the scattering centers of the medium is an area of active study~\cite{Blaizot:2012fh,CasalderreySolana:2012ef,Blaizot:2013vha,Mehtar-Tani:2019ygg,Barata:2021byj}. 
Here, we address this problem using a nonperturbative approach. We consider the scattering of a highly energetic quark off a strong classical background field, and we treat the quark in a Fock space consisting of $\ket{q}$ and $\ket{qg}$ sectors. 
We explicitly solve the time evolution of this system with the light-front Hamiltonian formalism, using the time-dependent basis light-front quantization approach (tBLFQ)~\cite{Zhao:2013cma}. 

The tBLFQ approach is a nonperturbative computational method to investigate time-evolution problems. 
It is based on light-front quantum field theory and the Hamiltonian formalism. 
The implementation of the basis function representation allows one to choose a basis with the same symmetries of the system under investigation, and is therefore advantageous for carrying out efficient numerical calculations.
This method has been previously  applied to  nonlinear Compton scattering~\cite{Zhao:2013cma, Hu:2019hjx}, to the interaction of an electron with intense electromagnetic fields~\cite{Chen:2017uuq}, and to quark-nucleus scattering~\cite{Li:2020uhl}.  

In the earlier treatment of quark-nucleus scattering with tBLFQ presented in Ref.~\cite{Li:2020uhl}, the Fock space of the quark was truncated to the leading sector $\ket{q}$. In this limit, the subeikonal effect was revealed in the transverse coordinate distribution of the quark.  
In this work, we extend the Fock space to $\ket{q}+\ket{qg}$, thus including gluon emission and absorption in the process.
We treat the target nucleus as a classical SU(3) color field given by the McLerran-Venugopalan (MV) model~\cite{McLerran:1993ni,McLerran:1993ka,McLerran:1994vd}. 
In the usual CGC treatment, the scattering only depends on the field integrated over the longitudinal coordinate. 
The method introduced here, however, can be applied to a more general situation where the process can be sensitive to the structure of the field in the longitudinal direction.
We explicitly solve for the time evolution of the quark as a quantum state inside the target color field. The time dependence is sensitive to all three parts of the Hamiltonian of our system: the interaction with the background field, gluon emission and absorption, and phase rotation with the light-front energy of the state. The phase rotation is neglected in the eikonal limit usually used in CGC calculations, and it encodes the physics of the formation time of the radiated gluon. In our full nonperturbative treatment, we can smoothly vary the magnitudes of these three effects separately. 

 We study the evolution of the quark by looking into its distribution in phase space, including the longitudinal momentum, the transverse momentum, light-front helicity, and color. Our focus in this paper is on presenting and testing the numerical method, and demonstrating it in different physical regimes.  
 For clarity, we use an initial condition of a pure $\ket{q}$ state with a specific color and helicity so that the $\ket{qg}$ components are generated only by the interactions. The only exception is when studying the sole effect from the interaction with the background field, where we also include a $\ket{qg}$ component in the initial state. 
 In the future, we aim to apply this numerical method to different physical situations, such as a high-energy scattering with subeikonal effects, which requires choosing initial conditions and measured observables corresponding to the physical process of interest. 
 The layout of this paper is as follows. We first introduce the formalism of tBLFQ for the case of a quark emitting/absorbing a gluon and scattering on a color field in Sec.~\ref{sec:method}. We then present and discuss numerical results in Sec.~\ref{sec:results}, highlighting the effects of the three different parts of the Hamiltonian separately and together. We conclude the work in Sec.~\ref{sec:summary}.

 \begin{figure}[t]
  \centering
  \includegraphics[width=0.35\textwidth]{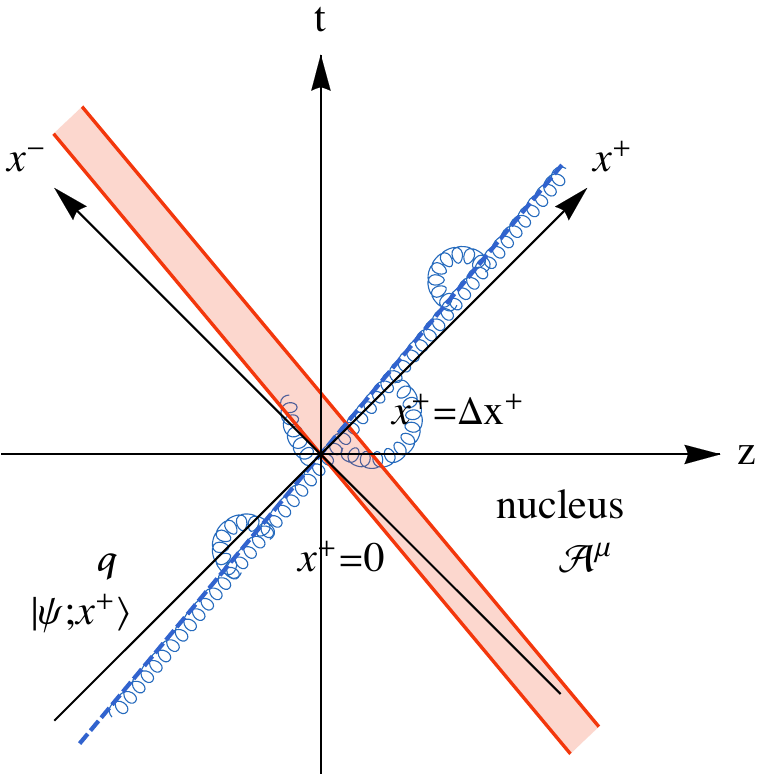}
  \caption{
  The quark is moving along the positive $z$ direction and it scatters on the nucleus which moves along the negative $z$ direction. 
  The dashed line is the worldline of the quark, $z=\beta_q t$ with $\beta_q$ the speed of the quark. 
  The quark state is a superposition of the $\ket{q}$ and the $\ket{qg}$ states.
  The quark line is dressed by helical lines representing the gluon in the $\ket{qg}$ state.
  The band represents the worldlines of the target nucleus, bounded by $z=-\beta_A t$ and $z=-\beta_A t + d'$.
  Here, $\beta_A$ is the speed of the nucleus and $d'=d\sqrt{1-\beta_A^2}$ with $d$ the width of the nucleus in its rest frame. 
  In the ultrarelativistic limit of $\beta_A\to 1$, the red band in the diagram shrinks to a single line aligned with $x^+=0$.
  }
 \label{fig:dis_zt}
\end{figure}

\section{Methodology: time-dependent basis light-front quantization (tBLFQ)}\label{sec:method}
The basic physical situation in our study is a high-energy quark moving in the positive $z$ direction, scattering on a high-energy nucleus moving in the negative $z$ direction, as shown in Fig.~\ref{fig:dis_zt}. 
The quark has momentum $P^\mu$ with $P^+\gg P^-, P_\perp$ whereas the nucleus has momentum $P_{\mathcal{A}}^\mu$ with $P_{\mathcal{A}}^- \gg P_{\mathcal{A}}^+, P_{\mathcal{A},\perp}$. 
We treat the quark state at the amplitude level and the nucleus as an external background field. 
The quark state is a superposition of the $\ket{q}$ and the $\ket{qg}$ states.
The quark interacts with the nuclear field over a finite distance in light-front time $0\le x^+\le L_\eta$. 
The light-front quantization formalism is manifestly boost invariant in the $z$ direction. 
Thus the same physical process can be described in different Lorentz frames with equivalent results. In practice, this means that the change of the $P^+$ momentum of the incoming quark can be compensated by a corresponding Lorentz contraction of the $x^+$ dependence of the target (both its size and internal structure). The physically genuinely different regimes correspond to different relative timescales of coherence and the background field interactions. For practical simulations, however, we choose specific numerical values, expressed here in $\GeV$ for concreteness.

\subsection{The light-front Hamiltonian}\label{sec:LFH}
The Lagrangian for the process we are considering is the QCD Lagrangian with an external field,
\begin{align}\label{eq:Lagrangian}
 \mathcal{L}=-\frac{1}{4}{F^{\mu\nu}}_a F^a_{\mu\nu}+\overline{\Psi}(i\gamma^\mu  D_\mu -  m_q)\Psi\;,
\end{align}
where $F^{\mu\nu}_a\equiv\partial^\mu C^\nu_a-\partial^\nu C^\mu_a-g f^{abc}C^\mu_b C^\nu_c$ is the field strength tensor, $D^\mu\equiv \partial_\mu +ig C^\mu$ the covariant derivative, and $ C^\mu= A^\mu + \mathcal{A}^\mu$ is the sum of the quantum gauge field $ A^\mu$ and the background gluon field $\mathcal{A}^\mu$.  

The light-front Hamiltonian is derived from the Lagrangian through the standard Legendre transformation~\cite{Brodsky:1997de} in the light-cone gauge of the quark, i.e.,$A^+=\mathcal A^+ = 0$, and we show the detailed derivation in Appendix~\ref{app:Hamiltonian}. 
Here, we focus on the Hamiltonian in the truncated Fock space that we are actually working with. 

The interacting quark state admits an infinite Fock space expansion in terms of the bare states. The dimensionality of this Fock space grows with the number of basis states (color, helicity, and momentum states) to the power of the number of particles. 
This growth makes it intractable when numerically going beyond higher orders in the Fock state expansion. 
Here, we truncate this expansion to the leading two sectors, $\ket{q}$ and $\ket{qg}$,
\begin{align}
 \ket{q}_{\text{dressed}}=\psi_q\ket{q}+\psi_{qg}\ket{qg}+\cdots
   \;,
\end{align}
where $\psi_q$ ($\psi_{qg}$) is the probability amplitude of the $\ket{q}$ ($\ket{qg}$) sector, and ``$\cdots$" includes all the other Fock sectors with gluons and sea quarks, such as $\ket{qgg}$ and $\ket{qq\bar q}$, which are not considered in this work.
In the truncated Fock space, the light-front Hamiltonian consists of two parts, $P^-(x^+)=P_{KE}^- + V(x^+)$, where $P_{KE}^-$ is the kinetic energy and $V(x^+)$ the interaction. 
Note that we do not consider the kinetic energy of the background field. 
The kinetic energy part of the Hamiltonian is a sum of single particle energies,
\begin{align}
 \begin{split}
   P_{KE}^-=&\int\diff x^-\diff^2 x_\perp
   \bigg\{
   -\frac{1}{2}A^j_a{(i\nabla)}^2_\perp A_j^a\\
   &+\frac{1}{2}\overline{\Psi}\gamma^+\frac{m_q^2-\nabla_\perp^2}{2i\partial_-}\Psi
   \bigg\}
   \;.
 \end{split}
\end{align}
The interaction part of the Hamiltonian consists of two terms, $V(x^+)=V_{qg}+V_{\mathcal{A}}(x^+)$, and its diagrammatic representation is illustrated in Table~\ref{tab:H}. 
The first term $V_{qg}$ is the interaction between the quark and the dynamical gluon: 
\begin{align}
 \begin{split}
   V_{q g}=&\int \diff x^- \diff^2 x_\perp
    g\overline{\Psi}\gamma^\mu T^a\Psi A^a_\mu
   \;.
 \end{split}
\end{align}
It accounts for gluon emission and absorption inside the dressed quark state.
The second term $V_{\mathcal{A}}(x^+)$ includes the interaction of the background field with the quark and that with the dynamical gluon, $ V_{\mathcal{A}}(x^+)= V_{\mathcal{A},q}(x^+)+ V_{\mathcal{A},g}(x^+)$, with
\begin{align}
 \begin{split}
     V_{\mathcal{A},q}(x^+)=&
     \int \diff x^- \diff^2 x_\perp
      g\overline{\Psi}\gamma^+ T^a\Psi\mathcal{A}^a_+(x^+)\;,\\
      V_{\mathcal{A},g}(x^+)= &
      \int \diff x^- \diff^2 x_\perp
      g f^{abc} \partial^+ A^i_b A^c_i \mathcal{A}^a_+(x^+)
      \;.    
 \end{split}
\end{align}
It admits an explicit time dependence introduced by the background field. 
Note that for an infinitesimal time step, the quark and the gluon interacting with the background field are separate interaction terms. The usual CGC picture of both the quark and gluon being rotated by the shockwave field of the target arises after iterating these interactions over several time steps. 
\begin{table}[t] 
 \centering
 \caption{ The interaction matrix $V(x^+)$ for a dressed quark in the Fock space $\ket{q} + \ket{q g}$. The matrix elements are represented by diagrams. The straight line represents the quark, the black curled line the gluon, and the red curled line the background field. }
 \label{tab:H}
 % Use: \begin{tabular{|lcc|} to put table in a box
 \begin{tabular}{|c |c| c|} %m{3cm} m{3cm}
 \hline
 % 1st row
 \begin{tabular}{c}Fock \\ sector\end{tabular}
 & \multicolumn{1}{c|}{$\ket{q}$}
 & \multicolumn{1}{c|}{$\ket{q g}$}\\ 
 \hline
 % 2nd row
\begin{tabular}{c} $\bra{q}$\end{tabular}
 & 
 \begin{tabular}{c}
   \includegraphics[width=.1\textwidth]{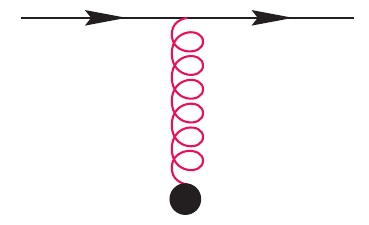}
 \end{tabular}
 & 
 \begin{tabular}{c}
   \includegraphics[width=.1\textwidth]{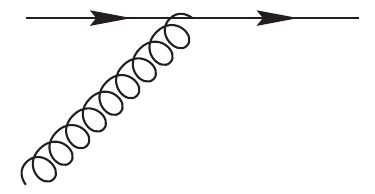}
 \end{tabular}
 \\
 \hline
 % 3rd row
 $\bra{q g}$
 & 
 \begin{tabular}{c}
   \includegraphics[width=.1\textwidth]{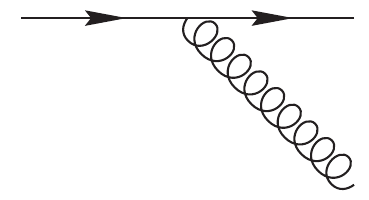}
 \end{tabular}
 & 
 \begin{tabular}{c c}
   \includegraphics[width=.1\textwidth]{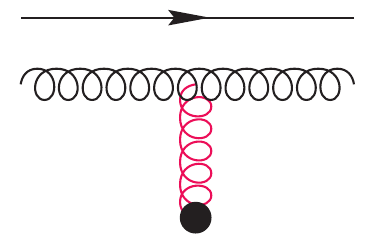}
   &\includegraphics[width=.1\textwidth]{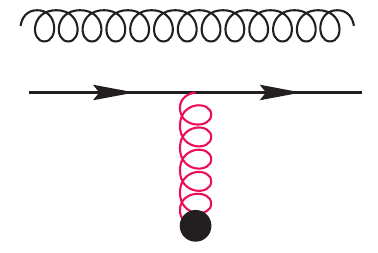}
 \end{tabular}
\\
   \hline
 \end{tabular}
\end{table}

The background field $\mathcal{A}^\mu$ accounts for the target, and we describe it using the MV model~\cite{McLerran:1993ni,McLerran:1993ka,McLerran:1998nk}. This is a classical field satisfying the reduced Yang-Mills equation,
\begin{align}\label{eq:poisson}
 (m_g^2-\nabla^2_\perp )  \mathcal{A}^-_a(\vec{x}_\perp,x^+)=\rho_a(\vec{x}_\perp,x^+)\;,
\end{align}
and it has only one nonzero component $\mathcal{A}^-$. In the MV model, one assumes that the target field is independent of $x^-$ (the light-front time for a left-moving target). This is justified by the probe's large momentum $P^+$, which means that the $x^-$ dependence of the probe is larger than the $x^-$ scales of the target. The consequence of this approximation is that the longitudinal momentum $P^+$ of the probe is preserved in the interaction.
The gluon mass $m_g$ is introduced to regularize the infrared (IR) divergence in the field, which simulates color neutrality on the source distribution~\cite{krasnitz2003gluon}. 
We take $m_g = 0.1\GeV$ in the numerical simulations.
The background field can be expressed in terms
of Green’s function as 
\begin{align}
 \begin{split}
   \mathcal{A}_a^-(\vec x_\perp, x^+)
   =\int
   \diff^2 z_{\perp}G_0(\vec x_\perp-\vec z_{\perp})        
   \rho_a(\vec z_\perp, x^+)
   \;,
 \end{split}
\end{align}
where
\begin{align}
 G_0(\vec x_\perp-\vec y_\perp)=\int \frac{\diff^2 k_\perp}{{(2\pi)}^2}\frac{e^{-i \vec k_\perp \cdot (\vec x_\perp-\vec y_\perp)}}{m_g^2+\vec k_\perp^2}\;.
\end{align}

The color charges are treated as Gaussian stochastic variables that are uncorrelated between different points in the transverse plane and between different points in light-front time. They satisfy the correlation relation
\begin{equation}\label{eq:chgcor}
 \expval{\rho_a(\vec{x}_\perp,x^+)\rho_b(\vec{y}_\perp,y^+)}=g^2\tilde\mu^2\delta_{ab}\delta^2(\vec{x}_\perp-\vec{y}_\perp)\delta(x^+-y^+)\;.
\end{equation}
Note that the parameter $\tilde\mu^2$ has dimensions of $\GeV^3$, consisting of $\GeV^2$ for the transverse dimension $\vec x_\perp$ and $\GeV$ for the target's longitudinal dimension $x^+$. 
This corresponds to the transport coefficient $\hat{q}$ in jet quenching~\cite{Blaizot:2012fh}. 
For a high-energy scattering process, what matters is the charge density $(g^2\tilde\mu)^2$ integrated over the extension of the field along $x^+$~\cite{Dumitru:2002qt,Fukushima:2007dy}. 
This integrated quantity, corresponding to the typical transverse momentum transferred by the target color field, is known as the saturation scale $Q_s^2$.
For a field with constant charge density, it can be obtained, up to logarithmic corrections, from the product of $(g^2\tilde\mu)^2$ and the duration of the field $L_\eta$ \cite{Lappi:2007ku}.
The conventions regarding factors of $\pi$ and $2$ differ between different sources in the literature. 
Here,  we use the fundamental representation saturation scale, which we take to be given by the relation
\begin{align}\label{eq:Qs}
 Q_s^2=C_F \frac{(g^2\tilde\mu)^2L_\eta}{2\pi}\;,
\end{align}
neglecting the logarithmic corrections. 
Here, $C_F=(N_c^2-1)/(2N_c)=4/3$ is the second-order Casimir invariant in the fundamental representation.

\subsection{Time evolution of the state}
The evolution of quantum states is governed by the time-evolution equation on the light front. 
Since we are interested in how the quark evolves under the interaction, it is natural to use the interaction picture (denoted by the subscript $I$),
\begin{align}\label{eq:ShrodingerEq}
 i\frac{\partial}{\partial x^+}\ket{\psi;x^+}_I=\frac{1}{2}V_I(x^+)\ket{\psi;x^+}_I\;.
\end{align}
In the interaction picture, the interaction Hamiltonian is $V_I(x^+)=e^{i\frac{1}{2}P^-_{KE}x^+}V(x^+)e^{-i\frac{1}{2}P^-_{KE}x^+}$, and the interaction picture state is related to the Schrödinger picture state by $\ket{\psi;x^+}_I=e^{i\frac{1}{2}P^-_{KE}x^+}\ket{\psi;x^+}$.

The solution of Eq.~\eqref{eq:ShrodingerEq} describes the state of the investigated system at any given light-front time $x^+$,
\begin{align}\label{eq:ShrodingerEqSol}
 \ket{\psi;x^+}_I=\mathcal{T}_+\exp\left[-\frac{i}{2}\int_0^{x^+}\diff z^+V_I(z^+)\right]\ket{\psi;0}_I\;,
\end{align}
where $\mathcal{T}_+$ denotes light-front time ordering. 
In perturbative calculations, the time-ordered exponential is written as an expansion in powers of $V_I(z^+)$, and is approximated by retaining the leading terms in the series. 
However, in cases where the external fields are strong, a perturbative treatment may not be sufficient. 

One possible nonperturbative treatment is decomposing the time-evolution operator into many small steps of the light-front time $x^+$, then solving each time step in the sequence numerically,
\begin{align}\label{eq:time_evolution_exp}
 \begin{split}
   \mathcal{T}_+ &\exp[-\frac{i}{2}\int_0^{x^+}\diff z^+V_I(z^+)]\\
   =&\lim_{n\to\infty}\prod^n_{k=1}\mathcal{T}_+ \exp\left[-\frac{i}{2}\int_{x_{k-1}^+}^{x_k^+}\diff z^+V_I(z^+)\right]
 \;.
 \end{split}
\end{align}
The step size is $\delta x^+ \equiv x^+/n$, and the intermediate time is $x_k^+=k\delta x^+ (k=0,1,2,\ldots,n)$ with $x_0^+=0$ and $x_n^+=x^+$. 
This product sequence is equivalent to the time-ordered exponential in the continuum limit  $n\to\infty$. 

In practice, the calculation is carried out in a finite-dimensional basis space, where the state becomes a column vector, and the interaction operator is in matrix form.
The choice of the numerical method, to some extent, depends on the basis representation of the system. 
Here,  we consider two typical treatments for general purposes, and we will discuss the numerical method in solving this problem after introducing the basis in the next section. 

Knowing that Eq.~\eqref{eq:ShrodingerEq} is an ordinary differential equation, one primary group of numerical methods is the finite-difference method (FDM).
FDM approximates the derivatives with finite differences in each small time step. 
Typical methods of the group include the Euler method, the second-order difference scheme MSD2~\cite{MSD2}, and Runge-Kutta methods~\cite{press2007numerical}. 
For example, with the most straightforward method, the forward Euler, one would treat the evolution in each time step as
\begin{multline}\label{eq:time_evolution_exp_FDM}
   \lim_{\delta x^+\to 0}\mathcal{T}_+ \exp\left[-\frac{i}{2}\int_{x_{k}^+}^{x_k^+ + \delta x^+}\diff z^+V_I(z^+)\right]\\
   \to 
   \left[1-\frac{i}{2}V_I(x_k^+)\delta x^+\right]
 \;.
\end{multline}
This method is, however, not numerically stable since the formula is not invariant under time reversal. 
For practical use, stable methods such as MSD2 and the fourth-order Runge-Kutta methods are recommended. 
The Runge-Kutta methods propagate a solution over each step by combining the information from several smaller Euler-style steps and eliminating lower-order errors.
Thus it has the advantage of simulating the time dependence even inside each time step. 
One could adjust the step size $\delta x^+$ to achieve a desired accuracy in the calculation.
There are also implicit methods, such as the Crank–Nicholson method, which uses the backward difference in time and is always stable. 
However, in these cases, one might need to pay the price of inverting the interaction matrix in a large basis space, which is not always an easy task, especially when the interaction matrix is complicated. 

Another treatment is to compute the exponential directly, which is automatically unitary.
When the time step is sufficiently small, the interaction during every single step can be considered as constant in time, and the evolution operator reduces to an ordinary exponential,
\begin{multline}\label{eq:time_evolution_exp_ord}
   \lim_{\delta x^+\to 0}\mathcal{T}_+ \exp\left[-\frac{i}{2}\int_{x_{k}^+}^{x_k^+ + \delta x^+}\diff z^+V_I(z^+)\right]\\
   \to 
   \exp\left[-\frac{i}{2}V_I(x_k^+)\delta x^+\right]
 \;.
\end{multline}
However, this way, one loses the time dependence of $V_I(z^+)$ within each time step.
This method would be favorable if the matrix exponential is straightforward to evaluate, which is the case especially when the interaction matrix is diagonal.

These introduced methods all simulate the time evolution by computing the interaction in a sequence of time steps, and they provide a nonperturbative solution. Our algorithm is a combination of the Runge-Kutta method for the gluon emission and absorption and the matrix exponentiation for interaction with the background field, as is explained in detail in Sec.~\ref{sec:basistimeevol}.

\subsection{Basis representation}
\subsubsection{Constructing the basis}
We are interested in how the momentum states, i.e., eigenstates of the kinetic energy part of the Hamiltonian $ P^-_{KE}$, evolve due to gluon emission/absorption and interactions with a background field.
Therefore we choose the basis state $\ket{\beta}$ as the eigenstates of the free Hamiltonian $P^-_{KE}$
\begin{align} 
   P^-_{KE}\ket{\beta}=P^-_\beta\ket{\beta}\;,
\end{align}
i.e., the ``bare'' 1- and 2-particle Fock states. 
The quark state is a sum over the basis states
\begin{align}
   \ket{\psi;x^+}_I=\sum_\beta c_\beta (x^+)\ket{\beta}\;,
\end{align}
where $c_\beta(x^+)\equiv\expval{\beta|\psi;x^+}_I$ are the basis coefficients. 
The initial state at $x^+=0$ can be specified by assigning values of $c_\beta(0)$, and the information of a state at $x^+$ is encoded in the column vector $\bm{c}(x^+)$.

In each Fock sector, the many-particle basis states are direct products of single particle states. The basis state in the $\ket{qg}$ sector is in the format of $\ket{\beta_{qg}}=\ket{\beta_q}\otimes\ket{\beta_g}$. 
Each single particle state carries five quantum numbers,
\begin{align}\label{eq:basis_beta}
 \beta_l = \{k^+_l, k^x_l, k^y_l, \lambda_l, c_l \}, \qquad l=q \text{ or } g\;. 
\end{align}
The first quantum number, $k^+_l$, labels the longitudinal momentum of the particle. 
For this degree of freedom, we employ the usual plane-wave basis states, i.e., eigenstates of the longitudinal momentum operator $P^+$, with corresponding eigenvalues $p_l^+$. In this paper, we compactify $x^-$ to a circle of length $2 L$ (i.e., $x_+$ to a circle of length $L$ ).
We impose (anti-)periodic boundary conditions on (fermions) bosons. As a result, the longitudinal momentum $p_l^+$ in the basis states takes discrete values as
\begin{align}\label{eq:basis_ppl}
 p_l^+ = \frac{2\pi}{ L} k_l^+ \;,
\end{align}
with the dimensionless quantity $k_g^+ = 1, 2, 3,\ldots$ for bosons  (neglecting the zero mode) and $k_q^+ =1/2, 3/2, 5/2,\ldots$ for fermions. 

For each Fock state, let $K = \sum_l k_l^+$ be the total $k^+$ of all the $l$ particles in that state.
Since the background field that we are considering does not provide extra longitudinal momentum to the state, the total $p^+$ of the system and $K$ are conserved.
In the $\ket{q}$ sector, the quarks in all basis states have $k_q^+ = K$. 
In the $\ket{qg}$ sector, there are a number of $(K - 0.5)$ K-segments, where in each K-segment, the quark and the gluon have definite values of $k_q^+$ and $k_g^+$. 
For example, with $K =8.5$, the quark in the $\ket{q}$ sector has $k^+ = K =8.5$, and the $\ket{qg}$ sector comprises eight K-segments, each with $\{k_q^+ =0.5, k_g^+ = 8\}$, $\{k_q^+ = 1.5, k_g^+ = 7\}$, $\ldots$, $\{k_q^+ = 7.5, k_g^+ = 1\}$, respectively. 

The next two quantum numbers, $k^x_l$ and $k^y_l$, label the momentum components in the transverse directions. 
The two-dimensional transverse space is a lattice extending from $-L_\perp$ to $L_\perp$ in each direction  with periodic boundary conditions. 
The number of transverse lattice sites in each dimension is $2 N_\perp$, so the lattice spacing is $a_\perp=L_\perp/N_\perp$. 
Thus the transverse coordinate vector $\vec r_\perp=(r_x,r_y)$ is discretized as
\begin{align}\label{eq:basis_rperp}
 r_i= n^i a_\perp ~(i = x,y) , \quad n^i=-N_\perp,-N_\perp+1,\ldots,N_\perp-1 \;.
\end{align}
The corresponding momentum space is also discrete with periodic boundary conditions. 
The transverse momentum vector $\vec p_\perp=(p^x,p^y)$ on the momentum grid reads
\begin{align}\label{eq:basis_pperp}
 p^i=k^i d_p~ (i = x,y), \quad k^i=-N_\perp,-N_\perp+1,\ldots,N_\perp-1\;,
\end{align}
where $d_p\equiv \pi/L_\perp$ is the resolution in the transverse momentum space, which effectively acts as an IR cutoff $\lambda_{IR}=d_p$.
The ultraviolet~(UV) cutoff from the transverse momentum grid is $\lambda_{UV}=N_\perp d_p=\pi/a_\perp$.
The transverse coordinate and the transverse momentum spaces are related through the Fourier and the inverse Fourier transformations.
For the interaction with the background field, we go to transverse coordinate space, where the basis states are characterized by the quantum numbers
\begin{align}\label{eq:basis_betabar}
 \bar\beta_l = \{k^+_l, n^x_l, n^y_l, \lambda_l, c_l\}, \qquad l=q \text{ or } g\;. 
\end{align}

The fourth quantum number, $\lambda_l$, labels the light-front helicity~\cite{Soper:1972xc}.
The quark helicity takes the values $\lambda_q=\pm 1/2$ (also represented as $\lambda_q=\uparrow,\downarrow$) and the gluon helicity takes the values $\lambda_g=\pm 1$ (also represented as $\lambda_g=\uparrow,\downarrow$).
The last quantum number, $c_l$, labels the particle’s color index. 
For the quark, $c_q=1,2,3$, and for the gluon, $c_g=1,2,\ldots,8$.

The basis states are eigenstates of the kinetic energy operator $P^-_{KE}$. 
For each Fock state, the total kinetic energy sums over all the constituent particles $l$ in that state, $P_{\beta}^- = \sum_l p_l^-$.
The kinetic energy of the quark is $p^-_q=(\vec p_{\perp,q}^2+m_q^2)/p^+_q$ and that of the gluon is $p^-_g=\vec p_{\perp,g}^2/p^+_g$.

The number of basis states $N_{tot}$ for the Fock space $\ket{q}+\ket{qg}$ is therefore
\[N_{tot} = (2N_\perp)^2 \times 2 \times 3 + (K-0.5) \times (2N_\perp)^4 \times 4\times 24 \;.\]
This is the number that controls the overall numerical complexity of the calculation.

In the numerical simulation, we take $L_\perp=50~\GeV^{-1}(=9.87~\text{fm})$ and $N_\perp=16$.
Exceptions are separately noted.
This translates into a rather large lattice spacing $a_\perp$ in physical units. 
In order to stay safe from lattice effects, we must use rather small values of $g^2\tilde{\mu}$ and $m_g$ in physical units to stay close enough to the continuum, i.e., with $Q_s a_\perp \lesssim \pi$. 
However, since the actual physical behavior of the system only depends on dimensionless combinations of the parameters, one can directly reinterpret our results as valid for larger values (in physical units) of $g^2\tilde{\mu}$ on a correspondingly smaller (in physical units) lattice size $L_\perp$. The main purpose of this paper is the development of the numerical method, and while we quote values for the parameters in physical units for convenience, the exact values of the parameters should not be interpreted as precisely matching a specific collision system.

\subsubsection{Gluon emission and absorption matrix elements}\label{sec:int_basis_qg}
In the basis space, the quark state is represented as a column vector $\bm{c}(x^+)$.
The interaction operator $V_I(x^+)$ is represented as a matrix, which we denote as $\mathcal{V}(x^+)$.
Each matrix element encodes the transition amplitude between two basis states,
\begin{align}\label{eq:V_beta_betap}
   \begin{split}
       \mathcal V_{\beta\beta'}(x^+)\equiv & \expval{\beta|V_I(x^+)|\beta'}\\
       =&\expval{\beta|V(x^+)|\beta'}\exp\left[\frac{i}{2}(P^-_\beta-P^-_{\beta'})x^+
       \right]. 
   \end{split}
\end{align}
Recall that the interaction operator contains two terms, $V(x^+)=V_{qg}+V_{\mathcal{A}}(x^+)$ (see discussions in Sec.~\ref{sec:LFH}).
In constructing the basis representation, we have discretized the transverse and the longitudinal spaces, and the light-front Hamiltonian is quantized on the same discrete space (see further details in Appendix~\ref{app:modes}).
We write out the matrix element of $V_{qg}$ in the basis representation in this section, and we discuss that of $V_{\mathcal{A}}(x^+)$ in the next section.

The $V_{qg}$ operator acts between the $\ket{q}$ and the $\ket{qg}$ sectors. In the following  expressions, $p_l=(p^+_l,p^x_l,p^y_l)$ is the three momentum of particle $l$.
The symbol $\beta_l$ denotes the collective quantum numbers defined in Eq.~\eqref{eq:basis_beta}, and the relation between the integer (half-integer) momentum quantum numbers $k_l$ and their associated momenta $p_l$ are given by Eqs.~\eqref{eq:basis_ppl} and \eqref{eq:basis_pperp}.
The interaction operator is
\begin{align}
 \begin{split}
     V_{q g}
     =& \sum_{\beta_1,\beta_2,\beta_3}
     \frac{g}{\sqrt{ p_1^+ p_2^+ p_3^+ 2L(2L_\perp)^2}}\\
     & b^\dagger_{\beta_2} \bar{u}(p_2,\lambda_2)
     \gamma^\mu T_{c_2,c_1}^{c_3}
     b_{\beta_1} u(p_1,\lambda_1)\\
     &\times
     [ a_{\beta_3}
     \epsilon_\mu(p_3,\lambda_3)
     \delta_{k_2^+, k_1^+ + k_3^+}
     \delta_{k_2^x, k_1^x + k_3^x}
     \delta_{k_2^y, k_1^y + k_3^y}\\
     &+ a^\dagger_{\beta_3}\epsilon_\mu^*(p_3,\lambda_3)
     \delta_{k_1^+, k_2^+ + k_3^+}
     \delta_{k_1^x, k_2^x + k_3^x}
     \delta_{k_1^y, k_2^y + k_3^y}
     ]
     \;.
 \end{split}
\end{align}

The matrix element for a transition from a $\ket{q}$ state to a $\ket{qg}$ state reads
\begin{align}\label{eq:M_q_qg}
 \begin{split}
   &\bra{\beta_{qg}(k_q^+,k_q^x,k_q^y,\lambda_q,c_q,k_g^+,k_g^x,k_g^y,\lambda_g,c_g)} V_{q g}\\ &
   \quad \quad \times \ket{\beta_q(k_Q^+,k_Q^x,k_Q^y,\lambda_Q,c_Q)}\\
   &\quad = 
   \frac{g}{\sqrt{ p_Q^+ p_q^+ p_g^+ 2L(2L_\perp)^2}}
  \bar{u}(p_q,\lambda_q)
   \gamma^\mu T_{c_q,c_Q}^{c_g}
    u(p_Q,\lambda_Q)\\
    & \quad \quad \times \epsilon_\mu^*(p_g,\lambda_g)
    \delta_{k_Q^+,k_q^+ + k_g^+}
    \delta_{k_Q^x,k_q^x + k_g^x}
    \delta_{k_Q^y,k_q^y + k_g^y}
 \;,
 \end{split}
 \end{align}
and that for a gluon absorption process is the Hermitian conjugate
\begin{align}\label{eq:M_qg_q}
 \begin{split}
   &\bra{\beta_q(k_Q^+,k_Q^x,k_Q^y,\lambda_Q,c_Q)} V_{q g} \\
   &\quad \quad \times\ket{\beta_{qg}(k_q^+,k_q^x,k_q^y,\lambda_q,c_q,k_g^+,k_g^x,k_g^y,\lambda_g,c_g)}\\
   &\quad =
 \frac{g}{\sqrt{ p_q^+ p_Q^+ p_g^+ 2L(2L_\perp)^2}}
\bar{u}(p_Q,\lambda_Q)
 \gamma^\mu T_{c_Q,c_q}^{c_g}
  u(p_q,\lambda_q)\\
  &\quad \quad \times \epsilon_\mu(p_g,\lambda_g)
  \delta_{k_Q^+,k_q^+ + k_g^+}
  \delta_{k_Q^x,k_q^x + k_g^x}
  \delta_{k_Q^y,k_q^y + k_g^y}
 \;.
 \end{split}
\end{align}
Here, $u(p,\lambda)$ is the spinor of the fermion, and $\epsilon_\mu(p,\lambda)$ is the polarization vector of the vector boson. Their expressions can be found in Appendix.~\ref{app:spinor}.
We use the subscripts ``$Q$'' and ``$q$'' to distinguish between the quark in the $\ket{q}$ state and that in the $\ket{qg}$ state.
For convenience, let us define the longitudinal momentum fraction of the gluon as $z\equiv p_g^+/p_Q^+$, so that $p^+_g = z p_Q^+$ and $p^+_q = (1-z) p_Q^+$. 
Let us also define the momentum difference between the quark (gluon) in the $\ket{qg}$ state and the quark in the $\ket{q}$ state as
\begin{align}\label{eq:Dq_Dg}
 \vec\Delta_q \equiv \vec p_{\perp,q} -  \vec p_{\perp,Q}
 \;,
 \qquad
   \vec\Delta_g \equiv \vec p_{\perp,g} -  \vec p_{\perp,Q}
   \;.
\end{align}

The spinor-polarization vector contraction parts of the matrix elements in Eqs.~\eqref{eq:M_q_qg} and ~\eqref{eq:M_qg_q} are summarized in Table~\ref{tab:uue} in Appendix~\ref{app:spinor}. 
They depend on the relative center-of-mass momentum, 
\begin{equation}\label{eq:Dm}
   \vec \Delta_m
 \equiv -(1-z)\vec\Delta_q
 +z \vec\Delta_g \;,
\end{equation}
instead of separately on the single particle transverse momenta $\vec p_{\perp,l}$. 
The energy difference from the phase factor in Eq.~\eqref{eq:V_beta_betap} also depends on $\Delta_m = |\vec \Delta_m|$,
\begin{align}\label{eq:Delta_P_min}
   p^-_{qg} - p^-_Q 
   = \frac{\Delta_m^2 + z^2 m_q^2}{z(1-z)p^+_Q}
   \;.
\end{align}
Thus the matrix element of $V_{qg}$ does not depend separately on the individual momenta of the particles but on the transferred momentum.

The periodic boundary condition implemented on the transverse momentum grid should also apply to the determination of the momentum conservation $\delta^2(\vec p_{\perp,Q} - \vec p_{\perp,q} - \vec p_{\perp,g}) $ on the lattice and the calculation of the transferred momenta $\vec\Delta_q$ and $\vec\Delta_g$.
Due to the periodicity, $p^i_Q$ and $ p^i_q + p^i_g$ ($i=x,y$) are equal if either they have the same value or they are different by a period in the transverse momentum space, $2\lambda_{UV}$.
Consequently, a transition process on the lattice could correspond to more than one different physical process, so one must decide which copy of the periodical momentum space lattice should be used to evaluate the momentum differences $\vec\Delta_q$ and $\vec\Delta_g$ that determine the matrix element. 
For example, a $\ket{qg}$ state with the quark and the gluon each carrying a transverse momentum close to the boundary $\lambda_{UV}$ can merge to a $\ket{q}$ state with a large total transverse momentum close to $ 2\lambda_{UV}$, which is outside of the fundamental Brillouin zone. 
However, on a periodic lattice, we could interpret the same gluon as having a transverse momentum just beyond the opposite boundary $-\lambda_{UV}$, merging with a quark close to $\lambda_{UV}$ into a quark with a momentum close to zero. 
With the first interpretation, the momentum difference vectors $\vec\Delta_q$ and $\vec\Delta_g$ point in the same direction, whereas for the second one, they are opposite. 
Thus, the relative center-of-mass momentum $\vec \Delta_m  = -(1-z)\vec\Delta_q +z \vec\Delta_g$ that determines the matrix element and light-front energy difference will be very different with the two interpretations.  

To get rid of ambiguities due to the periodicity, we choose the following prescription. 
We always use the value of $ \vec p_{\perp,Q}$ within the fundamental Brillouin zone as $ \vec p_{\perp,Q}$ in calculating the quark momentum transfer $\vec \Delta_q$. 
We then use the value of the momentum sum $ \vec p_{\perp,q} + \vec p_{\perp,g}$ (which might lie outside of the fundamental Brillouin zone) as $ \vec p_{\perp,Q}$ in calculating the gluon momentum transfer $\vec \Delta_g$. 
For the configuration discussed above, this corresponds to the second interpretation of a back-to-back $\ket{qg}$ state merging into a small momentum $\ket{q}$ state. 
The reason for this choice is precisely to maintain this interpretation of back-to-back splitting and merging, which is the physically most relevant process for the physical situations we are interested in.
We discuss the periodic boundary condition and explain our prescription in detail in Appendix~\ref{app:periodic}. 

\subsubsection{Background field interaction matrix elements}\label{sec:int_basis_A}

The $V_{\mathcal{A}}(x^+)$ term is introduced by the chosen background field, and it contains two parts, one acting on the quark and the other on the gluon:
\begin{align}
 \begin{split}
     V_{\mathcal{A}}(x^+)
     =&
     \sum_{\beta_1,\beta_2} 
     \int
     \frac{\diff^2 x_\perp }{{(2L_\perp)}^2}
     \delta_{k_2^+,k_1^+}
     \delta_{\lambda_1,\lambda_2}
     e^{i(p_2^\perp-p_1^\perp) \cdot x_\perp}
     \\
     & g\mathcal{A}^a_+ (\vec x_\perp , x^+)
   \left(
       2 T_{c_2,c_1}^a  
       b^\dagger_{\beta_2} 
       b_{\beta_1}
       -i2 f^{a c_1 c_2}
   a^\dagger_{\beta_2}
   a_{\beta_1}
   \right)
      \;.
 \end{split}
\end{align}
Here,  the symbol $\beta_l$ denotes the collective quantum numbers defined in Eq.~\eqref{eq:basis_beta}, and the relation between the integer (half-integer) momentum quantum numbers $k_l$ and their associated momenta $p_l$ are given by Eqs.~\eqref{eq:basis_ppl} and \eqref{eq:basis_pperp}.
The $V_{\mathcal{A}}(x^+)$ term does not contain the quantum gauge field and therefore does not directly connect different Fock sectors, so matrix elements of the type $\bra{qg}V_{\mathcal{A}}\ket{q}$ and $\bra{q}V_{\mathcal{A}}\ket{qg}$ are zero.
The background field does not change the particle's longitudinal momentum $p_l^+$ either, so the matrix elements between two $\ket{qg}$ states from different K-segments are also zero.

The background field is local in coordinate space, so it is convenient to evaluate the matrix element in the coordinate basis.
The matrix element for a transition from a $\ket{q}$ basis state to another $\ket{q}$ basis state reads
\begin{multline}\label{eq:M_qA}
 \bra{\bar \beta_{q'}(k_{q'}^+, n_{q'}^x, n_{q'}^y, \lambda_{q'},c_{q'})} V_{\mathcal{A}}(x^+) \ket{\bar \beta_{q}(k_q^+, n_q^x, n_q^y, \lambda_q,c_q)}\\
   =
  2g \mathcal{A}^a_+ (\vec r_{\perp,q}, x^+)
    T_{c_{q'},c_q}^{a}
    \delta_{\lambda_q,\lambda_{q'}}
    \delta_{k_q^+,k_{q'}^+}
    \delta_{n_q^x, n_{q'}^x}
    \delta_{n_q^y, n_{q'}^y}
 \;.
 \end{multline}
 The collective basis number $\bar \beta_l$ is defined in Eq.~\eqref{eq:basis_betabar}, and the relation between the basis numbers ($k^+_l$, $n^x_l$, and $n^y_l$) and their associated momenta/locations ($p^+_l$, $r^x_l$, and $r^y_l$) are given by Eqs.~\eqref{eq:basis_ppl} and \eqref{eq:basis_rperp}.
 The matrix element for a transition from a $\ket{g}$ basis state to another $\ket{g}$ basis state reads
\begin{multline}\label{eq:M_gA}
 \bra{\bar \beta_{g'}(k_{g'}^+, n_{g'}^x, n_{g'}^y, \lambda_{g'},c_{g'})}
   V_{\mathcal{A}}(x^+) 
   \ket{\bar \beta_{g}(p_g^+, n_g^x, n_g^y, \lambda_g,c_g)}\\
= - i 2g f^{a c_g c_{g'}}
\mathcal{A}^a_+ (\vec r_{\perp,g}, x^+)
\delta_{\lambda_g,\lambda_{g'}}
\delta_{k_g^+,k_{g'}^+}
\delta_{n_g^x, n_{g'}^x}
\delta_{n_g^y, n_{g'}^y}
  \;.
\end{multline}

The background field  $\mathcal{A}_+^a(=\mathcal{A}^{-,a}/2)$ is generated from the sampled color charges on the same discretized transverse lattice of the Fock state.
The longitudinal dimension of the color charge in $x^+$ (note that this is the light-front time of the incident quark) is taken to consist of $N_\eta$ independent layers~\cite{Lappi:2007ku}. 
The color charge, as well as the generated background field, extend from $0$ to $L_\eta$ along $x^+$. Thus each layer has a thickness of $\tau=L_\eta/N_\eta$, with
the $n_\tau$-th ($n_\tau=1,2,\ldots,N_\eta$) layer spanning $x^+ = [(n_\tau-1)\tau,n_\tau \tau]$. 
The correlation relation of the color charge as defined in Eq.~\eqref{eq:chgcor} now takes a discrete form 
\begin{multline}\label{eq:chgcor_dis}
   \expval{\rho_a(n^x,n^y,n_\tau)\rho_b({n'}^x,{n'}^y,n_\tau')}\\
    =g^2\tilde{\mu}^2\delta_{ab}\frac{\delta_{n^x,{n'}^x}\delta_{n^y,{n'}^y}}{a_\perp^2}\frac{\delta_{n_\tau,n_\tau'}}{\tau}\;.
\end{multline}
The Kronecker delta dividing the discrete resolution would become the Dirac delta in Eq.~\eqref{eq:chgcor} in the continuous limits of $a_\perp\to 0$ and $\tau\to 0$.
For generality, we allow the time step $\delta x^+$ to be smaller than the layer thickness $\tau$; this allows one to continuously go from scattering off a large coherent (independent of $x^+$) background field to scattering off independent scattering centers represented by separate layers in $x^+$.

\subsubsection{Time evolution in the basis}
\label{sec:basistimeevol}
We now look at the time evolution in this basis representation.
The solution of the time-evolution equation, Eq.~\eqref{eq:ShrodingerEqSol}, acquires a matrix form.
In each time step, the evolution reads
\begin{align}\label{eq:ShrodingerEqSolBasis}
 \bm{c}(x^+ + \delta x^+)=\mathcal{T}_+\exp\left[
   -i\frac{1}{2}\int_{x^+}^{x^+ + \delta x^+}\diff z^+ \mathcal{V}(z^+)
   \right]\bm{c}(x^+)\;,
\end{align}
where $\mathcal{V}(x^+)$ is the interaction matrix in the basis representation, and we have already discussed its matrix element $\mathcal V_{\beta\beta'}(x^+)$ in  Secs.~\ref{sec:int_basis_qg} and~\ref{sec:int_basis_A}. 
We could now select a suitable numerical method that takes advantage of the interaction matrix's structure. 

First, we notice that since our interaction matrix is a sum of two terms, $V_I(x^+)=V_{qg, I}(x^+)+V_{\mathcal{A}, I}(x^+)$, we can decompose the evolution over an infinitesimally short interval into two successive operations
\begin{align}\label{eq:V_split}
 \begin{split}
   \mathcal{T}_+ & \exp\left[-\frac{i}{2}\int_{x^+}^{x^++\delta x^+}\diff z^+\mathcal{V}_I(z^+)\right]\\
   \approx &
   \exp\left[-\frac{i}{2}\int_{x^+}^{x^++\delta x^+}\diff z^+ \mathcal{V}_{qg, I}(x^+)\right]\\
 &\times \exp\left[-\frac{i}{2}\int_{x^+}^{x^++\delta x^+}\diff z^+ \mathcal{V}_{\mathcal{A}, I}(x^+)\right]\;.
   \end{split}
\end{align}
Then we use different numerical methods for the two different kinds of interactions.  

The gluon emission/absorption operator $V_{qg}$ is off-diagonal in the Fock space and is thus challenging to exponentiate. Therefore we use the fourth-order Runge-Kutta (RK4) method to calculate its contribution in the time evolution,
\begin{align}\label{eq:M_qg}
   \mathcal{M}_{qg} (x^+;\delta x^+)
   \equiv &
   U_{RK4}
   \left(
   -\frac{i}{2} 
   \mathcal{V}_{qg,I}; x^+, x^+ + \delta x^+
   \right)\;.
\end{align}
The explicit form of the RK4 operator $U_{RK4}$ can be found in Appendix.~\ref{app:RK4}.
In terms of computational complexity, the RK4 method on the basis space is, in principle, $O(N_{tot}^2)$, but it is more like $O(N_{tot})$ for $\mathcal{V}_{qg}$. 
That is because the gluon emission/absorption interaction is nonzero only when the momentum is conserved, so the matrix $\mathcal{V}_{qg}$ is very sparse. In practice, we organize the numerical computation to iterate over only the matrix elements allowed by momentum conservation, which achieves this $O(N_{tot})$ complexity. 

On the contrary, the interaction with the background field $V_{\mathcal{A}}(x^+)$ is diagonal in the Fock space: it does not cause transitions between $\ket{q}$ and $\ket{qg}$ sectors. 
Moreover, the background field in our simulation is eikonal, meaning that the interaction is diagonal in coordinate space and in helicity space. 
One only needs to exponentiate a $N_c\times N_c$ color matrix to achieve a unitary evolution over a time step, which can be calculated analytically with the Cayley-Hamilton theorem~\cite{Curtright:2015iba}.
Therefore, it is feasible to do the calculation in the exponential form by Fourier transforming the wave function into coordinate space and then back again as
\begin{multline}
       \mathcal{M}_{\mathcal{A}} (x^+;\delta x^+)\\
 \equiv e^{i\frac{1}{2}P^-_{KE}x^+} 
 \left[
 \mathcal{F}^{-1}
 e^{- i \frac{1}{2}\mathcal{V}_{\mathcal{A}}(x^+) \delta x^+}
\mathcal{F}
\right]
e^{-i\frac{1}{2}P^-_{KE}x^+} 
 \;.
\end{multline}
Here, $\mathcal{F}=\mathcal{F}(\vec p_\perp\to \vec r_\perp)$ and $\mathcal{F}^{-1}=\mathcal{F}^{-1}(\vec r_\perp\to \vec p_\perp)$ are the Fourier and the inverse Fourier transformation operators, respectively (see further details in Appendix.~\ref{app:modes}). 
Note that the kinetic energy operator is diagonal in momentum, not coordinate space.  
Thus the kinetic energy phase part of the interaction picture interaction needs to be evaluated in momentum, not coordinate space. 
The computational complexity of the kinetic energy part is $O(N_{tot})$.
The (inverse) Fourier transform is carried out through the fast Fourier transform algorithm, which has a complexity of $O(N_{tot}\log{N_{tot}})$, and the interaction in coordinate space is $O(N_{tot})$. Thus the overall complexity of the
background field interaction is $O(N_{tot}\log{N_{tot}})$.

The full evolution for each time step combines the two contributions as
\begin{align}\label{eq:evolution_basis}
 \bm{c}(x^+ + \delta x^+)=
 \mathcal{M}_{\mathcal{A}} (x^+;\delta x^+)  \mathcal{M}_{qg} (x^+;\delta x^+)
 \bm{c}(x^+)\;,
\end{align}
The total computational complexity of each time step is, therefore, $O(N_{tot}\log{N_{tot}})$, much more efficient than the $O(N_{tot}^2)$ operations that a momentum space interaction with the background field would be. 
Thus splitting the interaction into two successive steps by Eq.~\eqref{eq:V_split} and using a Fourier transform for the background field allow for a very efficient time-evolution algorithm.

\section{Results}\label{sec:results}

By carrying out the explicit time evolution of the state, we are able to access the information about its time development as a function of $x^+$.
In this section, we study the time evolution of the quark state by looking into its longitudinal momentum, transverse momentum, helicity, and color.

We simulate three different cases. 
In the first case, the interaction contains just the gluon emission/absorption term $V=V_{qg}$; in the second case, the interaction contains just the background field term $V=V_{\mathcal{A}}$. Finally, we consider the full interaction $V=V_{qg}+V_{\mathcal{A}}$.

In the cases with nonzero transitions between the $\ket{q}$ and the $\ket{qg}$ sectors, we start with an initial condition as a single quark state with a definite color, helicity and momentum.
When studying the effect just from the background field, i.e., no transitions between the $\ket{q}$ and the $\ket{qg}$ sectors, we choose a superposition of a single $\ket{q}$ and a single $\ket{qg}$ state as the initial state to study their respective evolutions under the interaction. 
These initial states do not correspond exactly to those in a physical high-energy scattering process, where the quark would have already developed a gluon cloud before the interaction. 
However, it enables us to test the physical effects of the different parts of the Hamiltonian, and our numerical method, in a cleaner and more tractable setup. 

\FloatBarrier

\begin{figure}[tbp!]
  \centering
  \subfigure[\ $V_{I}(x^+)=V(x^+)$]
  {\label{fig:ppl_Vfree_nophase_8p5_N16}
    \includegraphics[width=0.4\textwidth]{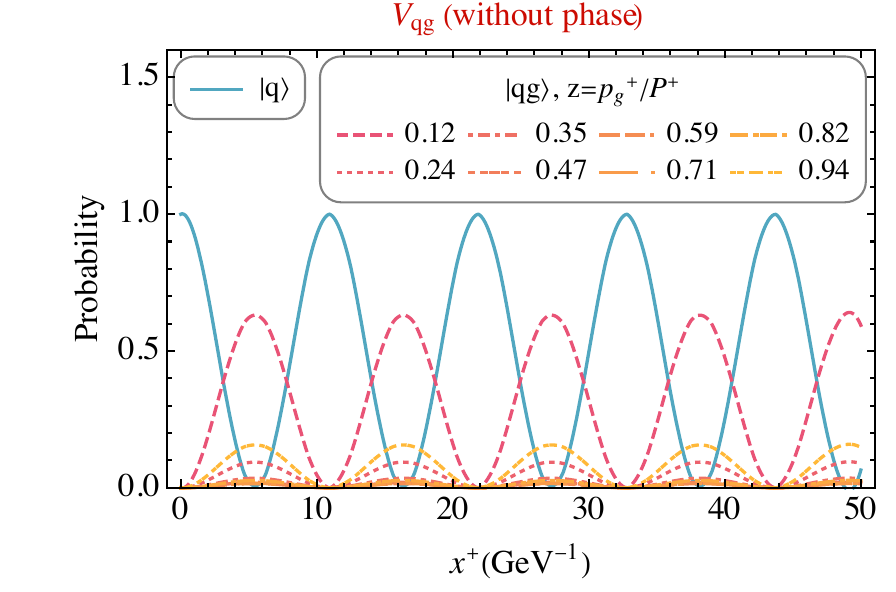}
  } 
  \qquad
  \subfigure[\ $V_{I}(x^+)=e^{i\frac{1}{2}P^-_{KE}x^+} V(x^+) e^{-i\frac{1}{2}P^-_{KE}x^+}$]
  {\label{fig:ppl_Vfree_phase_8p5_N16}
    \includegraphics[width=0.4\textwidth]{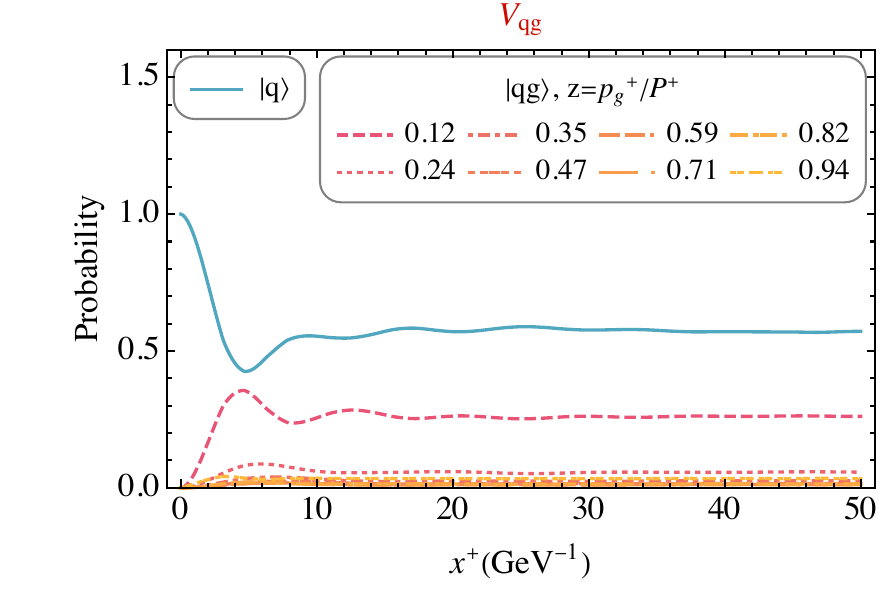}
  }
  \caption{
  The evolution of the probabilities of 
  different $p^+$ states, including the $\ket{q}$ sector and the K-segments of the $\ket{qg}$ sector characterized by the gluon longitudinal momentum fraction $z$.
  The interaction contains just the gluon emission/absorption term $V(x^+)=V_{qg}$, (a) without the phase factor and (b) with the phase factor.
  The initial state of the quark is a single quark state with $\vec p_{\perp,Q} = \vec 0_\perp$, $p_Q^+=P^+=8.5~\GeV$, light-front helicity $\lambda_Q=1/2$, and color $c_Q=1$. 
  Parameters in the two simulations: $N_\perp=16$, $L_\perp=50~\GeV^{-1}$, $L_\eta=50~\GeV^{-1}$, $m_g=0.1~\GeV$, $m_q=0.02~\GeV$, $K=8.5$.
   The duration of each time step in the simulation is $\delta x^+=0.39~\GeV^{-1}$.
  }
  \label{fig:ppl_Vqg_evolve_8p5_16}
\end{figure}

\subsection{Gluon emission and absorption}\label{sec:Vfree}
The interaction $V_{qg}$ excites transitions between the $\ket{q}$ and the $\ket{qg}$ sectors. 
This effect is intertwined with the phase rotation generated by the free part of the Hamiltonian $P^-_{KE}$, which in the interaction picture is manifested by the time evolution of the interaction matrix  $V_{qg,I}(x^+)=e^{i\frac{1}{2}P^-_{KE}x^+} V_{qg} e^{-i\frac{1}{2}P^-_{KE}x^+}$. 
This phase factor leads to a decoherence between emissions separated by a long enough light-front time.
To understand the effects from the gluon emissions/absorptions and the phase factor separately, we run the simulations in two cases: with the phase factor, in which we take $V_{I}(x^+)$ as $V_{qg,I}(x^+)$; and without the phase factor, in which we take $V_{I}(x^+)$ as $V_{qg}$.

We first study the evolution of the quark state in the longitudinal momentum $p^+$ phase space.
Figure~\ref{fig:ppl_Vqg_evolve_8p5_16} shows the evolution of the probabilities of different $p^+$ states, including the $\ket{q}$ sector and the K-segments of the $\ket{qg}$ sector characterized by the gluon longitudinal momentum fraction $z$.
The probability of each $p^+$ state sums over all states in the transverse momentum space, helicity space, and color space.
The initial state of the quark is a single quark state with $\vec p_{\perp,Q} = \vec 0_\perp$, $p_Q^+=P^+=8.5~\GeV$, light-front helicity $\lambda_Q=1/2$, and color $c_Q=1$. 
In Fig.~\ref{fig:ppl_Vfree_nophase_8p5_N16}, in the absence of the phase factor, the system oscillates between the initial $\ket{q}$ state and all the $p^+$ states in the $\ket{qg}$ sector.
In addition, those different $p^+$ states oscillate with the same frequency but with different amplitudes.
In Fig.~\ref{fig:ppl_Vfree_phase_8p5_N16}, with the phase factor restored, the probability for each $p^+$ state behaves as a damped oscillation.

To understand the oscillational patterns observed in the simulation via Fig.~\ref{fig:ppl_Vqg_evolve_8p5_16},  we study a simplified two-mode problem analytically.
Let us consider the state in a two-dimensional vector space, corresponding to the two Fock sectors. 
The state vector reads
\begin{align}
    \ket{\psi;x^+}_I= 
        \begin{bmatrix}
        \psi_q(x^+)\\
        \psi_{qg}(x^+)
        \end{bmatrix}
    \;.
\end{align}
The interaction operator $V_{qg,I}(x^+)=e^{i\frac{1}{2}P^-_{KE}x^+} V_{qg} e^{-i\frac{1}{2}P^-_{KE}x^+}$ in the matrix form reads
\begin{align}
  V_{qg}=
  \begin{bmatrix}
    0 & u \\
    u^* & 0
    \end{bmatrix},
    \qquad
  P_{KE}^-=
  \begin{bmatrix}
    p^-_q & 0 \\
    0 & p^-_{qg}
    \end{bmatrix}
    \;.
\end{align}
By solving the time evolution equation as Eq.~\eqref{eq:ShrodingerEq}, we obtain the probabilities of the states as sinusoidal functions of the evolution time:
\begin{align}\label{eq:Vqg_phase_WF}
  \begin{split}
  |\psi_q(x^+)|^2
  &=\left[1-
  \frac{4w^2}{\eta^2}
  \sin^2\left(
    \frac{\eta x^+}{4} 
  \right)\right]
  |\psi_q(0)|^2\\
  &+ \frac{4w^2}{\eta^2}
  \sin^2\left(
    \frac{\eta x^+}{4} 
  \right)|\psi_{qg}(0)|^2
  \;,\\
  |\psi_{qg}(x^+)|^2
  &=\frac{4w^2}{\eta^2}
  \sin^2\left(
    \frac{\eta x^+}{4} 
  \right)
  |\psi_q(0)|^2\\
  &+ \left[1-
  \frac{4w^2}{\eta^2}
  \sin^2\left(
    \frac{\eta x^+}{4} 
  \right)\right]
  |\psi_{qg}(0)|^2
  \;.
  \end{split}
\end{align}
For convenience, we have defined $w\equiv |u|$, $\Delta\equiv p^-_q - p^-_{qg}$, and $\eta\equiv \sqrt{\Delta^2+4w^2}$. The parameter  $w$ corresponds to  the magnitude of the gluon emission and absorption term, and $\Delta$ is the energy difference arising from the phase factor. 
The oscillation frequency depends on both the matrix element and the energy difference, as seen in the expression of $\eta$.
The oscillation amplitude depends on the ratio of the two terms $w^2/\eta^2$.
This two-mode process is essentially the Rabi oscillation, with a Rabi frequency of $2w$ and a detuning of $\Delta$~\cite{Rabi:1937dgo,fox2006quantum}.

The solution of the two-mode problem in Eq.~\eqref{eq:Vqg_phase_WF} helps understand the evolution of the extended $\ket{q}+\ket{qg}$ state in the basis space, which is essentially an $N_{tot}$-mode problem.
Let us consider the transition between one $\ket{q}$ state and $n$ different $\ket{qg}$ states, and the interaction operator is given by
\begin{subequations}
    \begin{align}\label{eq:n_modes_Vqg}
    V_{qg}=&
    \begin{bmatrix}
    0 & u_1 & u_2 & \cdots & u_n\\
    u_1^* & 0 & 0 & \cdots & 0\\
    u_2^* & 0 & 0 & \cdots & 0\\
    \vdots &  \vdots &  \vdots & \ddots &  \vdots\\
    u_n^* & 0 & 0 & \cdots & 0\\
    \end{bmatrix}\;,\\
    \label{eq:n_modes_PKE}
    P_{KE}^-=&
    \begin{bmatrix}
    p^-_q & 0 & 0 & \cdots & 0\\
    0 & p^-_{qg,1} & 0 & \cdots & 0\\
    0 & 0  & p^-_{qg,2} & \cdots & 0\\
    \vdots & \vdots & \vdots & \ddots& \vdots \\    
    0 & 0 & \cdots & 0 & p^-_{qg,n}\\
    \end{bmatrix}
    \;.
    \end{align}
\end{subequations}

\begin{figure}[tbp!]
  \centering
  \subfigure[\ evolution of $\ket{q}$ at different $N_\perp$]
  {
    \includegraphics[width=0.4\textwidth]{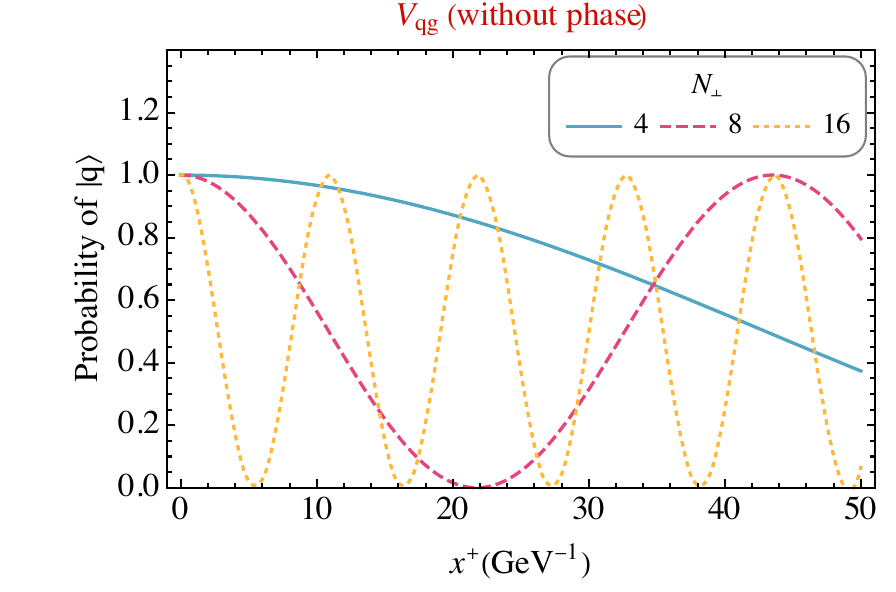}
  } 
  \subfigure[\ evolution of $\ket{q}$ at different $P^+$]
  {
    \includegraphics[width=0.4\textwidth]{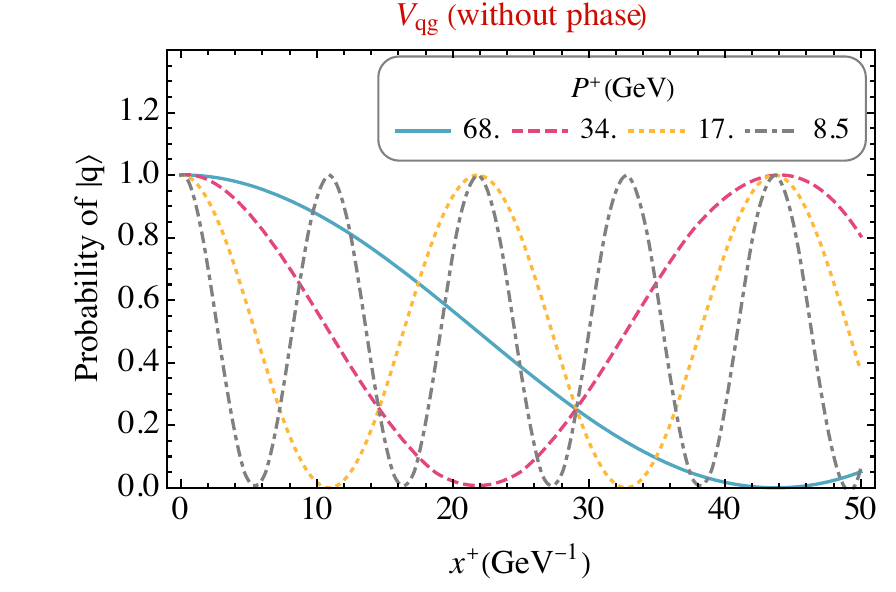}
  }
  \caption{
  The probability of the quark staying in the $\ket{q}$ sector at (a) various $N_\perp$ ($P^+=8.5~\GeV$) and (b) various $P^+$ ($N_\perp=16$).
  The interaction contains just the gluon emission/absorption term $V(x^+)=V_{qg}$, and phase factor is not included.
  The initial state of the quark is a single quark state with $\vec p_{\perp,Q} = \vec 0_\perp$, $p_Q^+=P^+$, light-front helicity $\lambda_Q=1/2$, and color $c_Q=1$. 
  Parameters in these simulations: $L_\perp=50~\GeV^{-1}$, $L_\eta=50~\GeV^{-1}$, $m_g=0.1~\GeV$, $m_q=0.02~\GeV$, $K=8.5$.   
  The duration of each time step in the simulation is $\delta x^+=0.39~\GeV^{-1}$.
}
  \label{fig:ppl_evolve_Nperp_ppl}
\end{figure}

The simulation without the phase factor corresponds to only considering the part $V_{qg}$ of the Hamiltonian. 
In this case, there are only two nonzero eigenvalues, $w_{\pm}=\pm \sqrt{|u_1|^2+|u_2|^2+\ldots+|u_n|^2}$, which are opposite to each other. 
Thus the situation is very similar to the two-mode problem.
As a result, each basis state oscillates with the same frequency $w=|w_{\pm}|$, although the amplitudes of those oscillations could be different, depending on the values of the interaction matrix elements $u_i$.
The probability for each $p^+$ state, summing over different transverse momentum modes, therefore also oscillates with the same frequency. 

In the full calculation, $u_i$s are the matrix elements of $V_{qg}$ in Eqs.~\eqref{eq:M_q_qg} and~\eqref{eq:M_qg_q}, and they depend on the transferred momentum. 
The frequency $w$ is dominated by the most significant transition mode, so it is approximately $ w\propto \lambda_{UV}^2/P^+$, where $P^+$ is the total longitudinal momenta of the state, and $\lambda_{UV}$ is the largest allowed transverse momentum on the lattice.
Figure~\ref{fig:ppl_evolve_Nperp_ppl} shows the evolution of probability of the $\ket{q}$ sector at different $\lambda_{UV}(=N_\perp \pi/L_\perp)$ by taking different $N_\perp$ at a fixed $L_\perp$, and at different $P^+$.  
The dependence of the oscillation frequency on $P^+$ and $\lambda_{UV}$ indeed agrees with the expectation $ w\propto \lambda_{UV}^2/P^+$. 

When the phase factor is restored, this corresponds to including both the $V_{qg}$ and the $P^-_{KE}$ terms in the Schr\"{o}dinger picture light-front Hamiltonian $P^-$.
Unlike in the case without the phase factor, there are now $n+1$ different eigenvalues. 
Each basis state is, in essence, a superposition of different eigenstates.  
The summation over these states leads to decoherence, which appears in Fig.~\ref{fig:ppl_Vfree_phase_8p5_N16} as a damped oscillation.
The probability of each $p^+$ state approaches an asymptotic value, which is related to the matrix elements of the Hamiltonian.

\begin{figure}[tbp!]
  \centering
  \includegraphics[width=0.43\textwidth]{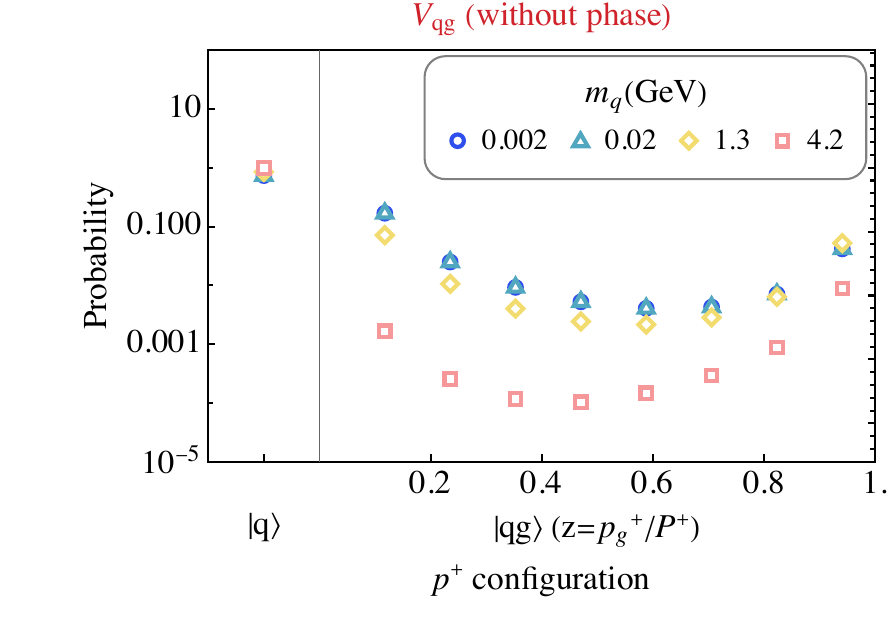}
  \caption{
  The probability of the quark state at different $p^+$ configurations after the evolution, with various quark masses. 
  The interaction is $V(x^+)=V_{qg}$, and the phase factor is not included, i.e., $V_{I}(x^+)=V(x^+)$.
   The initial state of the quark is a single quark state with $\vec p_{\perp,Q} = \vec 0_\perp$, $p_Q^+=P^+$, light-front helicity $\lambda_Q=1/2$, and color $c_Q=1$. 
  Parameters in these simulations: $L_\perp=50~\GeV^{-1}$, $L_\eta=50~\GeV^{-1}$, $m_g=0.1~\GeV$, $K=8.5$.
 The duration of each time step in the simulation is $\delta x^+=0.39~\GeV^{-1}$.
  }
  \label{fig:ppl_final}
\end{figure}

We present the probability distribution of the 
quark state in the $p^+$ space after the evolution in Fig.~\ref{fig:ppl_final}. 
The initial state is a single quark state, and those $\ket{qg}$ states with different $p^+$ configurations emerge during the evolution.
As we see from the result, the gluon emission/absorption process favors  $\ket{qg}$ states with either small or large $z$, the gluon longitudinal momentum fraction. 
The dependencies on $z$ and the quark mass can be understood by examining the spinor-polarization vector contractions in the matrix elements of $V_{qg}$, as in Table.~\ref{tab:uue} in Appendix~\ref{app:modes}. 
Quark light-front-helicity-conserving transitions to both gluon polarization states, $[\lambda_Q\to\lambda_q \lambda_g]=[\uparrow\to\uparrow \lambda_g], [\downarrow\to\downarrow\lambda_g]$ ($\lambda_g= \uparrow,\downarrow$), are enhanced at small gluon momentum fraction $z$. 
Overall emissions in this soft gluon limit, where the emission matrix element is independent of the gluon polarization, are the most likely ones. 
For large gluon momentum fraction $z$, on the other hand, the only surviving quark light-front-helicity-conserving emissions are the ones where also the gluon has the same helicity as the quark, $[\lambda_Q\to\lambda_q \lambda_g]=[\uparrow\to\uparrow\uparrow], [\downarrow\to\downarrow\downarrow]$. 
The quark helicity flip transitions $[\lambda_Q\to\lambda_q \lambda_g]=[\uparrow\to\downarrow\uparrow], [\downarrow\to\uparrow\downarrow]$, are proportional to the quark mass, and heavily weight large values of $z$, which can be seen in a comparison of the different mass results in Fig.~\ref{fig:ppl_final}. 

\begin{figure*}[tbp!]
  \centering
  \subfigure[\ the quark in $\ket{q}$  \label{fig:pT_Vfree_nophase_8p5a}]
  {\includegraphics[width=.9\textwidth]
  {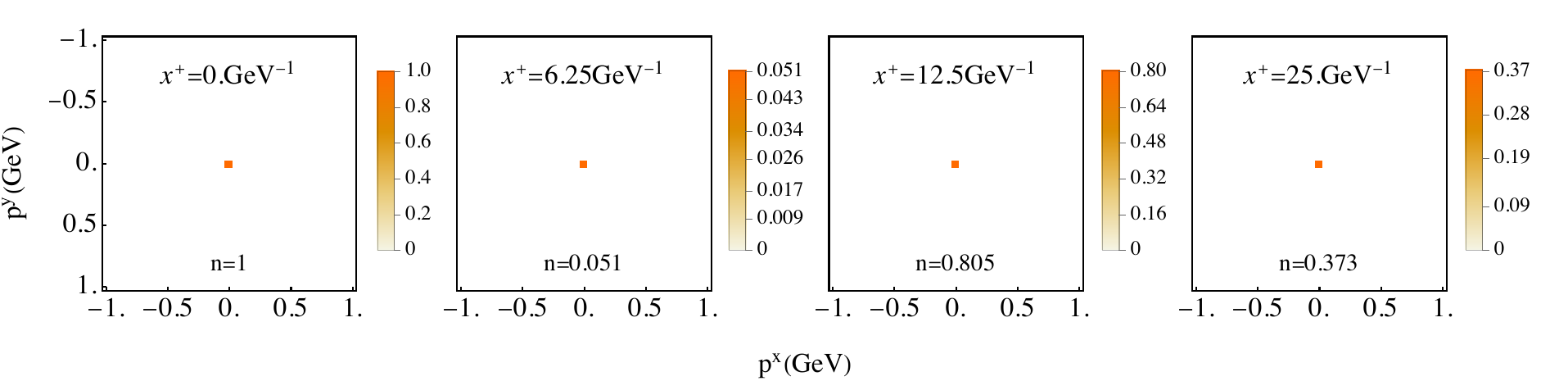}
  }
  \subfigure[\ the quark in $\ket{qg}$  \label{fig:pT_Vfree_nophase_8p5b}]
  {\includegraphics[width=.9\textwidth]{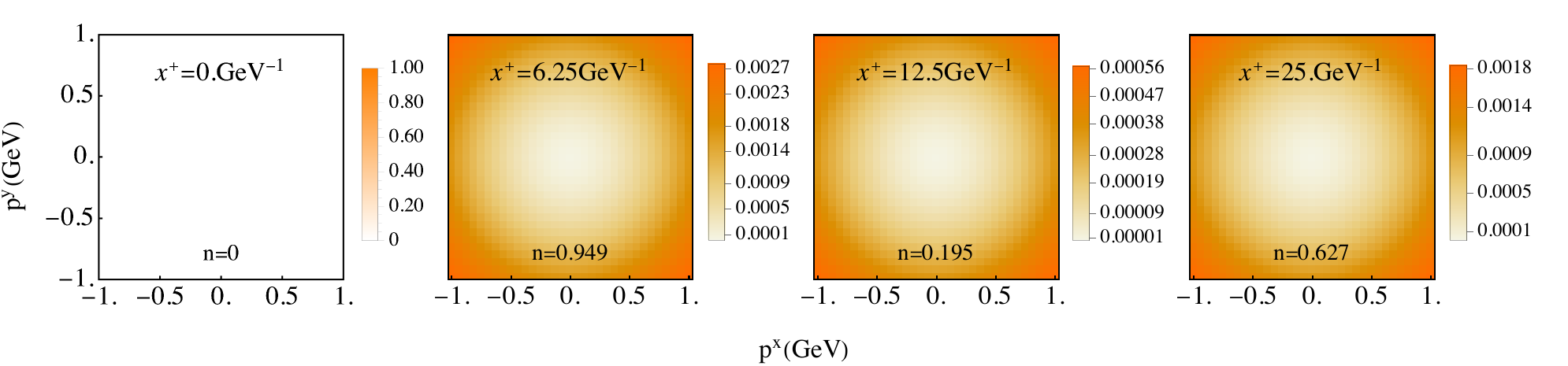}
  }
  \subfigure[\ the gluon in $\ket{qg}$  \label{fig:pT_Vfree_nophase_8p5c}]
  {\includegraphics[width=.9\textwidth]{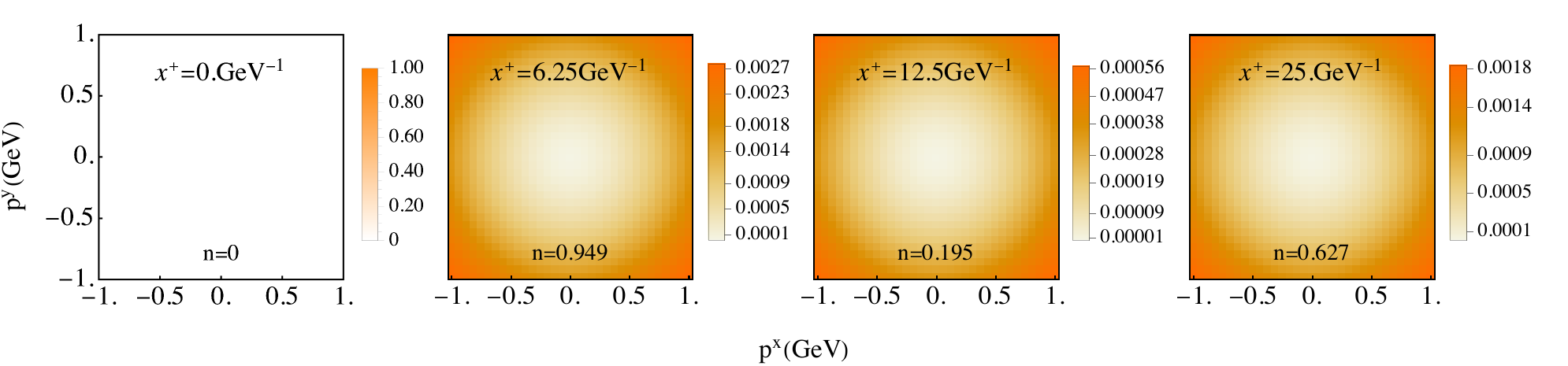}
  }
  \caption{
 The evolution of the transverse momentum distributions of (a) the quark in the $\ket{q}$ sector, (b) the quark in $\ket{qg}$ sector, and (c) the gluon in $\ket{qg}$ sector. 
  The interaction contains just the gluon emission/absorption term $V(x^+)=V_{qg}$, and the phase factor is not included.
  The initial state is a single quark state with $\vec p_{\perp,Q}=\vec 0_\perp$, $p_Q^+=P^+=8.5~\GeV$, light-front helicity $\lambda_Q=1/2$, and color $c_Q=1$.  
    From left to right, the transverse momentum distributions of the particle are shown at increasing light-front time instances. 
  The number at the bottom of each panel is the total probability of the plotted states.
  Parameters in the simulation: 
 $N_\perp=16$, $L_\perp=50~\GeV^{-1}$, $m_g=0.1~\GeV$, and $m_q=0.02~\GeV$. 
  The duration of each time step in the simulation is $\delta x^+=0.39~\GeV^{-1}$.
  }
  \label{fig:pT_Vfree_nophase_8p5}
\end{figure*}
\begin{figure*}[tbp!]
  \centering
  \subfigure[\ the quark in $\ket{q}$  \label{fig:pT_Vfree_phase_8p5a}]
  {\includegraphics[width=.9\textwidth]
  {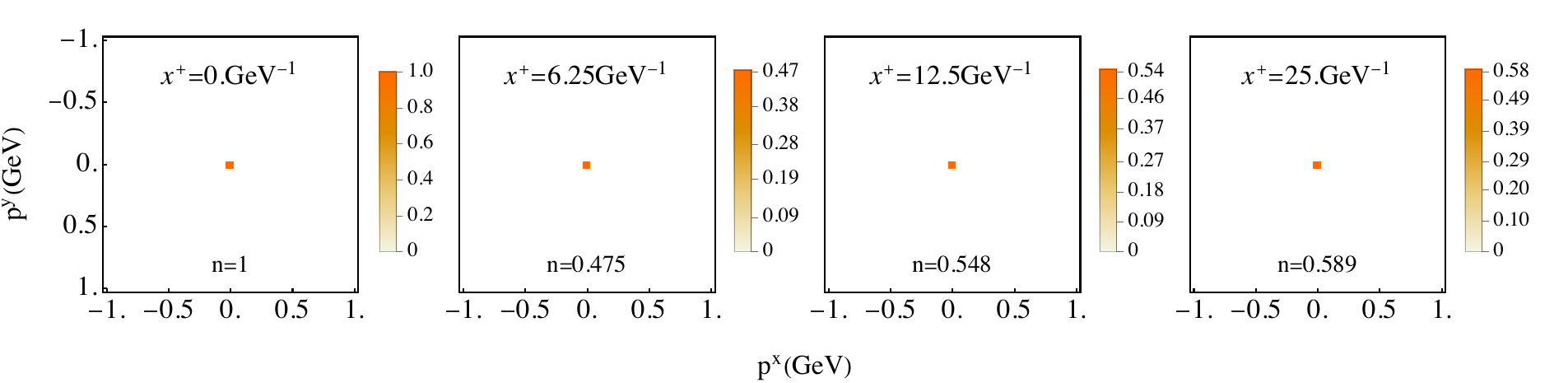}
  }
  \subfigure[\ the quark in $\ket{qg}$  \label{fig:pT_Vfree_phase_8p5b}]
  {\includegraphics[width=.9\textwidth]{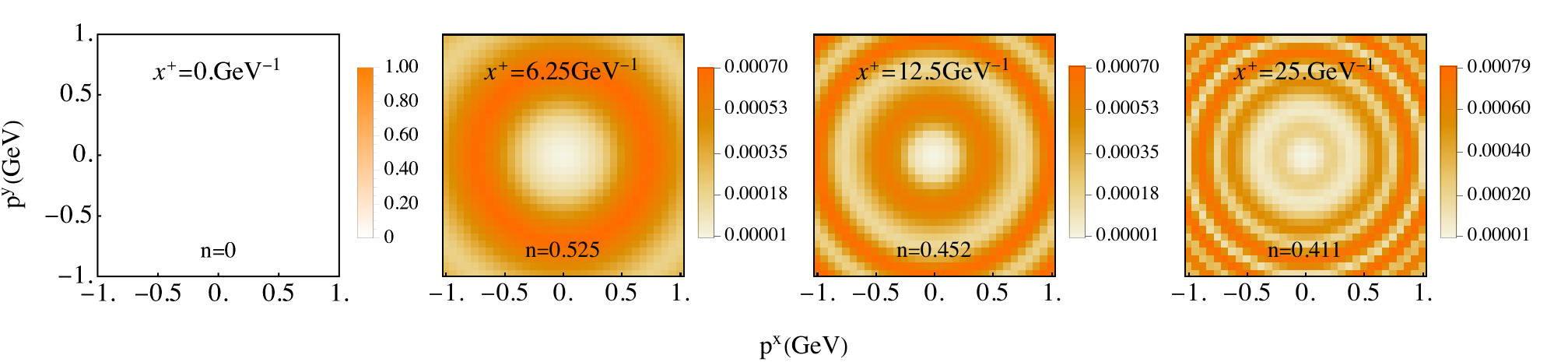}
  }
  \subfigure[\ the gluon in $\ket{qg}$  \label{fig:pT_Vfree_phase_8p5c}]
  {\includegraphics[width=.9\textwidth]{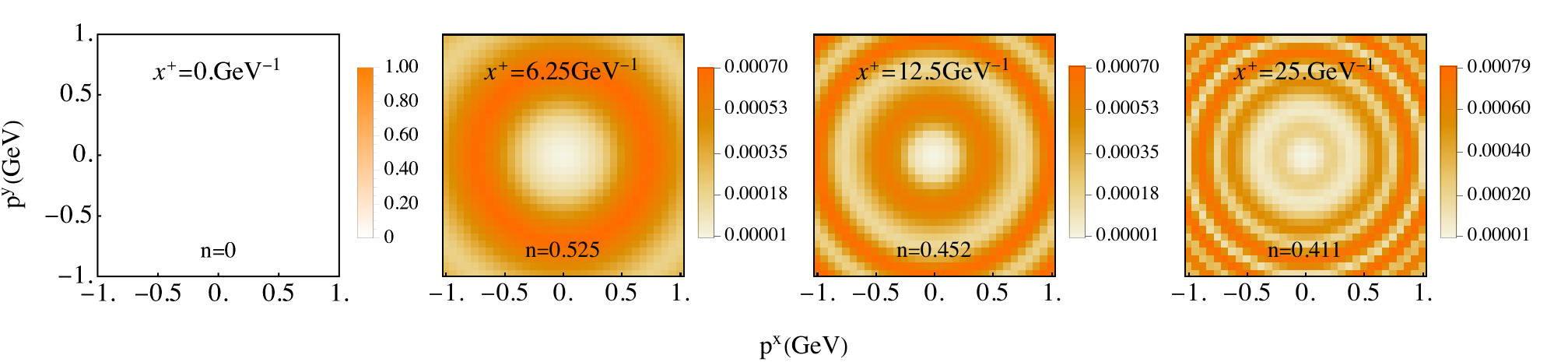}
  }
  \caption{ 
  The evolution of the transverse momentum distributions of (a) the quark in the $\ket{q}$ sector, (b) the quark in $\ket{qg}$ sector, and (c) the gluon in $\ket{qg}$ sector. 
  The interaction contains just the gluon emission/absorption term $V(x^+)=V_{qg}$, and the phase factor is included.
  The initial state is a single quark state with $\vec p_{\perp,Q}=\vec 0_\perp$, $p_Q^+=P^+=8.5~\GeV$, light-front helicity $\lambda_Q=1/2$, and color $c_Q=1$.  
From left to right, the transverse momentum distributions of the particle are shown at increasing light-front time instances. 
  The number at the bottom of each panel is the total probability of the plotted states.
  Parameters in the simulation: 
 $N_\perp=16$, $L_\perp=50~\GeV^{-1}$, $m_g=0.1~\GeV$, and $m_q=0.02~\GeV$. 
  The duration of each time step in the simulation is $\delta x^+=0.39~\GeV^{-1}$.
}
  \label{fig:pT_Vfree_phase_8p5}
\end{figure*}

Next, we study the evolution of the quark state in the transverse momentum space. Figures~\ref{fig:pT_Vfree_nophase_8p5} and~\ref{fig:pT_Vfree_phase_8p5} demonstrate the probability distributions in the transverse momentum plane for successive times. 
The transverse momentum distributions are shown separately for the quark in the $\ket{q}$ sector and the quark and the gluon in the $\ket{qg}$ sector. 
Now that we do not have a background field, the total transverse momentum is conserved. 
Thus, the quark in the $\ket{q}$ sector stays in its initial momentum state $\vec p_{\perp,Q}=\vec 0_\perp$. 
The quark and the gluon in the $\ket{qg}$ sector are back-to-back in momentum and have distributions that are symmetric around the origin, apart from the edges of the discrete transverse momentum lattice, where rotational invariance is lost.
The distributions without the phase factor are shown in Fig.~\ref{fig:pT_Vfree_nophase_8p5}. 
Here, the emitted quark and gluon both favor large transverse momentum modes, as we see in the sequential distributions in Figs.~\ref{fig:pT_Vfree_nophase_8p5b} and ~\ref{fig:pT_Vfree_nophase_8p5c}. 
The probabilities of different transverse momentum modes in the $\ket{qg}$ sector oscillate coherently, so they maintain their relative magnitudes while rising and falling as functions of $x^+$ through the evolution, as seen in Fig.~\ref{fig:ppl_Vfree_nophase_8p5_N16}. 
When the phase factor is included, the emitted quark and gluon show a changing concentric circular pattern in transverse momentum space, as we see in the sequential distributions in Figs.~\ref{fig:pT_Vfree_phase_8p5b} and ~\ref{fig:pT_Vfree_phase_8p5c}.
As we have discussed earlier in the context of the evolution of different $p^+$ states, here different transverse momentum states in the $\ket{qg}$ sector are also different superpositions of the eigenstates. 
Thus, the probabilities of different transverse momentum modes in the $\ket{qg}$ sector do not oscillate coherently, and their relative magnitudes change through the evolution. 
Additionally, the oscillation frequency of each eigenstate depends on the change of the light-front energy and the value of $V_{qg}$,   both depending on the transferred momentum squared. 
This explains why the pattern reflecting the relative magnitudes among different transverse momentum states is azimuthally symmetric and centered at the initial momentum mode of the quark.
The states at later times exhibit artificial effects from the periodic boundaries, and they are not presented here.

\begin{figure*}[tbp!]
  \centering
  \subfigure[\ $p_\perp$ distribution with and without the phase]
  { \label{fig:prob_pperp_phase_a}
    \includegraphics[width=0.4\textwidth]{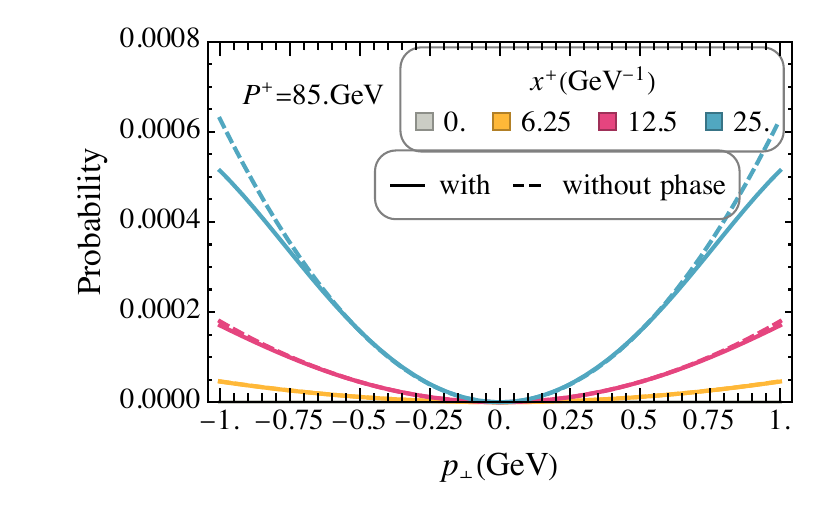}
  } 
  \subfigure[\ the ratio of the two cases in (a)]
  {
    \label{fig:prob_pperp_phase_b}
    \includegraphics[width=0.4\textwidth]{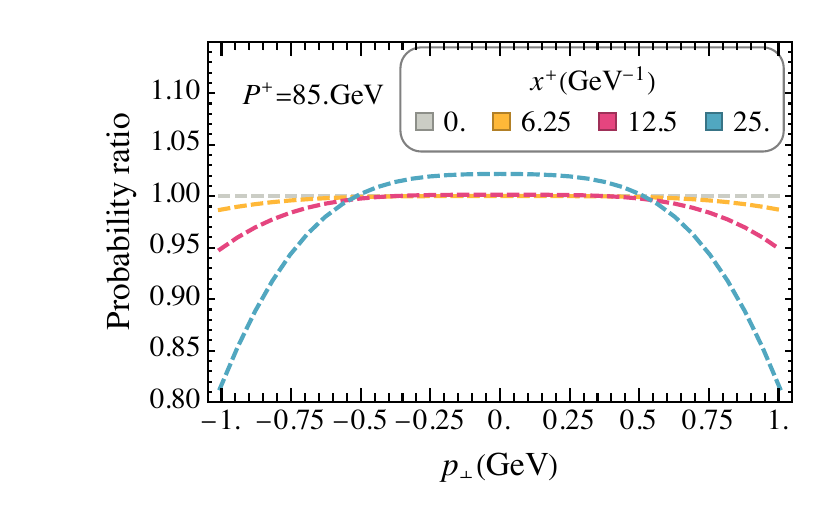}
  }
  \subfigure[\ $p_\perp$ distribution with and without the phase]
  { \label{fig:prob_pperp_phase_c}
    \includegraphics[width=0.4\textwidth]{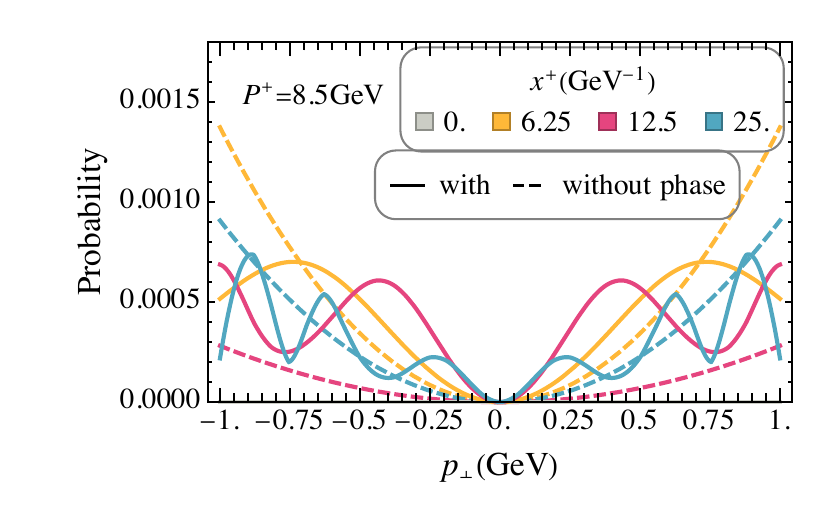}
  } 
  \subfigure[\ the ratio of the two cases in (c)]
  {
    \label{fig:prob_pperp_phase_d}
    \includegraphics[width=0.4\textwidth]{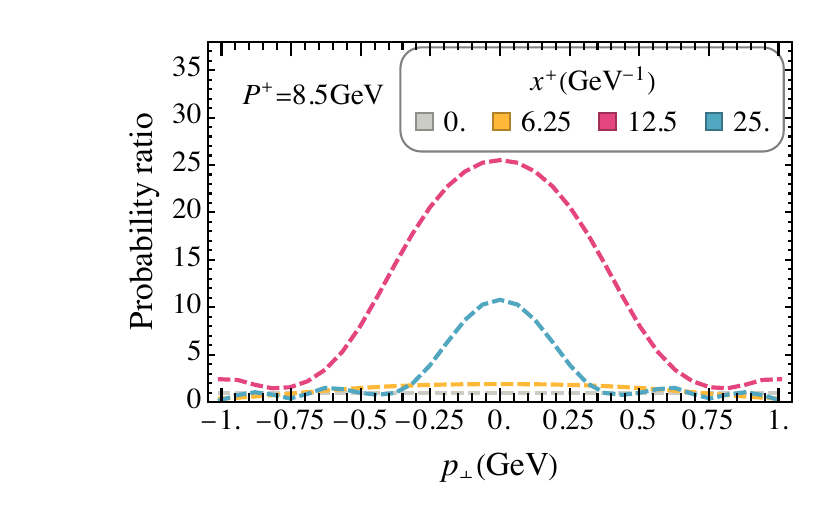}
  }
  \subfigure[\ $p_\perp$ distribution with and without the phase]
  { \label{fig:prob_pperp_phase_re}
    \includegraphics[width=0.4\textwidth]{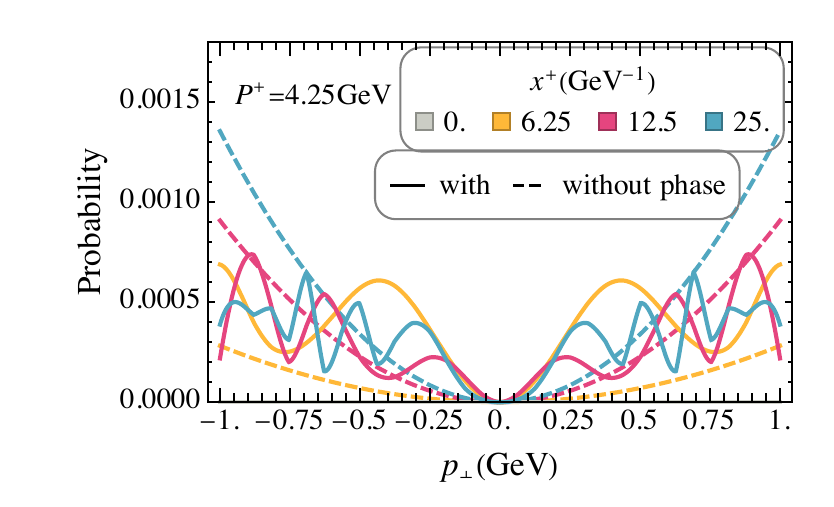}
  } 
  \subfigure[\ the ratio of the two cases in (e)]
  {
    \label{fig:prob_pperp_phase_e}
    \includegraphics[width=0.4\textwidth]{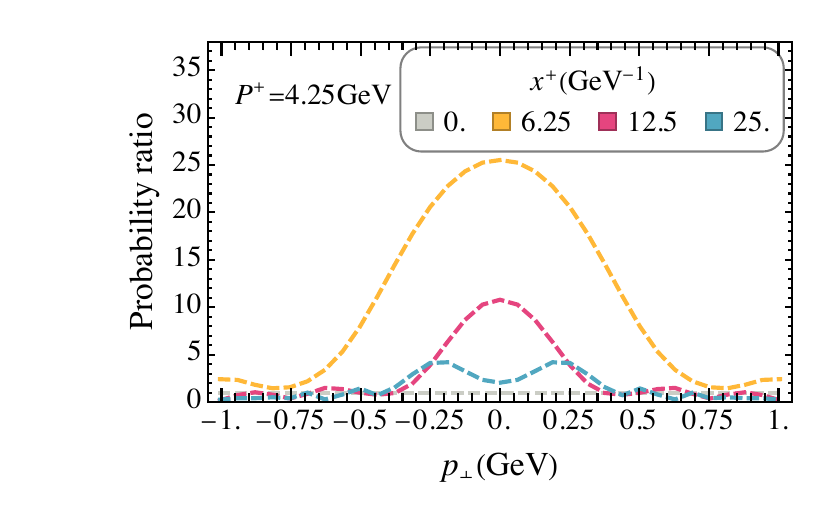}
  }
  \caption{The evolution of the quark transverse momentum distribution in the $\ket{qg}$ sector.
  The simulations run with the interaction containing just the gluon emission/absorption term $V=V_{qg}$, with and without the phase factors, at various $P^+$.
  Left panels (a,c,e): The probabilities as a function of
  $p_\perp =|\vec p_\perp|$ ($\arg \vec p_\perp = 0,\pi$) at a sequence of $x^+$s. The evolutions with (without) phase factors are in the solid (dashed) lines.
  Right panels (b,d,f): The ratio of the probability with the phase factor over that without the phase factor.
  From the top row to the bottom, $P^+=85~\GeV, 8.5~\GeV, 4.25~\GeV$ for (a,b), (c,d), and (e,f), respectively. 
  The initial state is a single quark state with $\vec p_{\perp,Q}=\vec 0_\perp$, $p_Q^+=P^+$, light-front helicity $\lambda_Q=1/2$, and color $c_Q=1$. 
  Parameters in these simulations: 
 $N_\perp=16$, $L_\perp=50~\GeV^{-1}$, $m_g=0.1~\GeV$, and $m_q=0.02~\GeV$. 
  The duration of each time step in the simulation is $\delta x^+=0.39~\GeV^{-1}$.
 }
  \label{fig:prob_pperp_phase}
\end{figure*}

To see the effect of the phase factor more clearly, we take the ratio of the probability distribution with the phase factor over that without the phase factor.
Since both the $V_{qg}$ interaction and the phase factors are azimuthally symmetric in the transferred $\vec p_\perp$ plane, we analyze the evolution of the $\vec p_\perp$ distribution at $p_\perp =|\vec p_\perp|$, $\arg \vec p_\perp = 0,\pi$.
We set the initial state as a single quark state with $\vec p_{\perp,Q}=\vec 0_\perp$, and run the simulations with and without the phase factors at various $P^+$. 
The results are shown in Fig.~\ref{fig:prob_pperp_phase}.

In the left panels, Figs.~\ref{fig:prob_pperp_phase_a},~\ref{fig:prob_pperp_phase_c}, and~\ref{fig:prob_pperp_phase_re}, the probability distributions of the quark in the $\ket{qg}$ sector are shown as a function of $p_\perp =|\vec p_\perp|$ ($\arg \vec p_\perp = 0,\pi$) at a sequence of $x^+$s, with $P^+=85~\GeV, 8.5~\GeV, 4.25~\GeV$, respectively. 
The distributions with the phase factor, as in the solid lines, show oscillational patterns, compared to those without the phase factor, as in the dashed lines.
In the plots of the ratio of the probability with the phase factor over that without the phase factor, as in Figs.~\ref{fig:prob_pperp_phase_b},~\ref{fig:prob_pperp_phase_d}, and~\ref{fig:prob_pperp_phase_e}, there is a peak around zero momentum transfer, and it gets narrower over time. 
(One exception is the $x^+=25~\GeV$ curve at $P^+=4.25~\GeV$ in Fig.~\ref{fig:prob_pperp_phase_e}: there is a dip instead of a peak in the center. 
But this is caused by artificial reflections from the periodic boundary, so we should neglect it for the purpose of this discussion.)
By comparing the three different $P^+$ cases, one can see that the peak narrows faster at a smaller $P^+$.
Moreover, the peak develops at a rate inversely proportional to $P^+$, which can be seen by comparing the $x^+=12.5~\GeV$ ($x^+=25~\GeV$) curve in Fig.~\ref{fig:prob_pperp_phase_d} to the $x^+=6.25~\GeV$ ($x^+=12.5~\GeV$) curve in Fig.~\ref{fig:prob_pperp_phase_e}. 
This is because a smaller $P^+$ leads to a larger kinetic energy $P^-_{KE}\propto 1/P^+$, making the decoherence faster. 
This behavior is a demonstration of the familiar effect leading to Fermi's golden rule. 
At late times $x^+\to \infty$, the only allowed transitions are the ones that conserve the light-front energy $P^-$.  
This energy conservation is enforced by the phase factor, canceling the energy nonconserving transitions, even when they are favored by large transition matrix elements.

\begin{figure*}[tbp!]
  \centering
  \subfigure [\ $V_{I}(x^+)= V(x^+)$]
  {\label{fig:color_evolution_Vfree_nophase_4p25_N16}
    \includegraphics[width=0.8\textwidth]{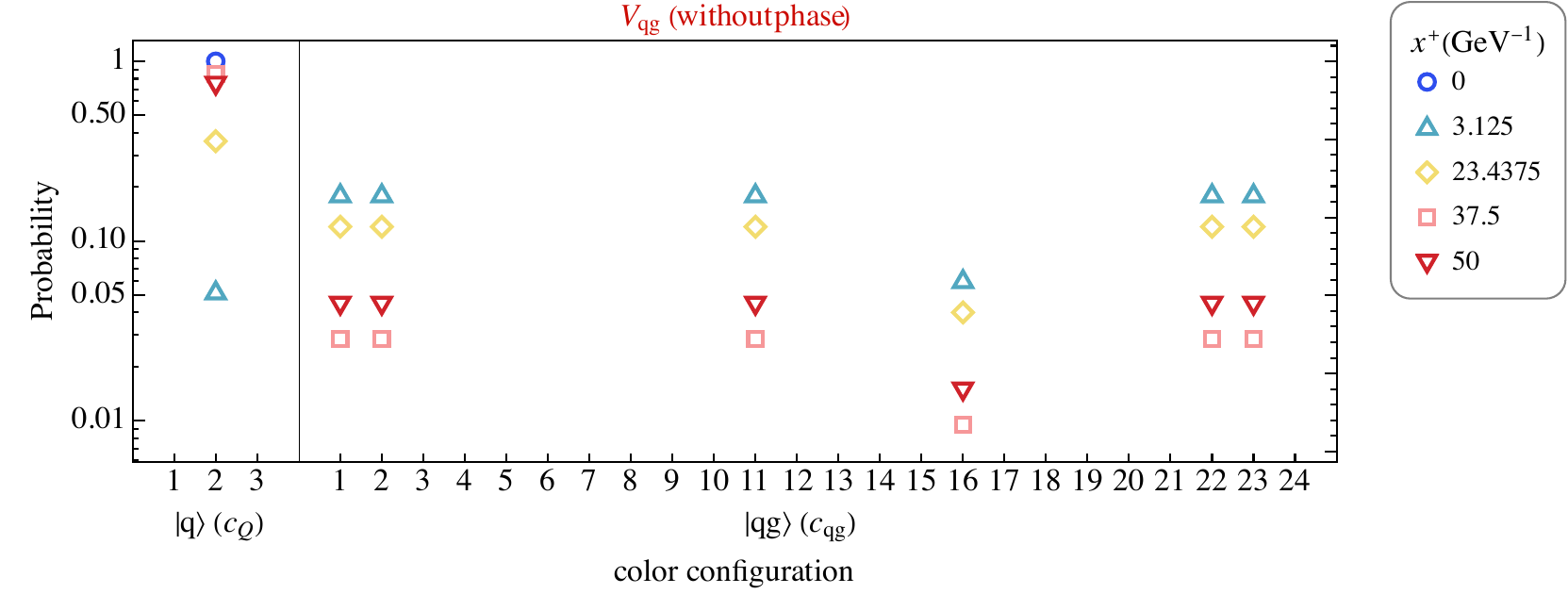}
  } 
  \subfigure [\ $V_{I}(x^+)=e^{i\frac{1}{2}P^-_{KE}x^+} V(x^+) e^{-i\frac{1}{2}P^-_{KE}x^+}$]
  {\label{fig:color_evolution_Vfree_phase_4p25_N16}
    \includegraphics[width=0.8\textwidth]{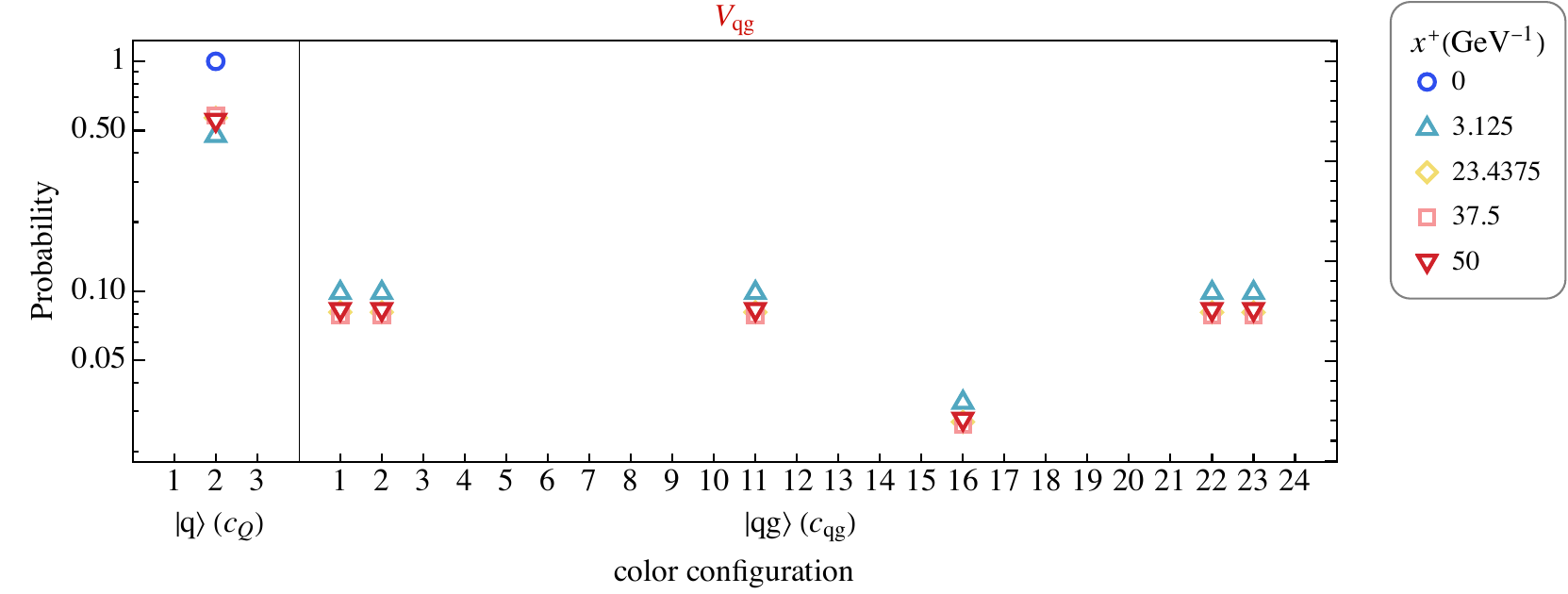}
  } 
  \caption{
  Evolution of the quark state in color space. 
  The interaction contains just the gluon emission/absorption term $V(x^+)=V_{qg}$, (a) without the phase factor and (b) with the phase factor.
  The color configuration in the $\ket{q}$ sector is labeled by the quark color index $c_Q=1,2,3$.
  The color configuration in the $\ket{qg}$ sector is labeled by the color index $c_{qg}=(c_q-1)8+c_g$, where the quark color index $c_q=1,2,3$ is the outer iterator and the gluon color index $c_g=1,\ldots,8$ is the inner iterator.
  The initial state of the quark is a single quark state with $\vec p_{\perp,Q} = \vec 0_\perp$, $p_Q^+=P^+=4.25~\GeV$, light-front helicity $\lambda_Q=1/2$, and color $c_Q=2$. 
  Parameters in these panels: $N_\perp=16$, $L_\perp=50~\GeV^{-1}$, $m_g=0.1~\GeV$, $m_q=0.02~\GeV$, and  $K=8.5$.   
  The duration of each time step in the simulation is $\delta x^+=0.39~\GeV^{-1}$.
  }
  \label{fig:color_evolution_Vfree}
\end{figure*}

We then look at the evolution of the quark state in color space, as in Fig.~\ref{fig:color_evolution_Vfree}. 
The initial state here is a single quark with color index $c_Q=2$. 
Only six of the $\ket{qg}$ color states are allowed in the transitions due to color conservation. 
Without the phase factor, the probabilities of those states oscillate over time, as in Fig.~\ref{fig:color_evolution_Vfree_nophase_4p25_N16}.
The oscillation is suppressed when the phase factor is restored, as in Fig.~\ref{fig:color_evolution_Vfree_phase_4p25_N16}.
This oscillation and its suppression have the same reason as in the $p^+$ distribution shown in Fig.~\ref{fig:ppl_Vqg_evolve_8p5_16}.
Without the phase factor, the probability of each momentum mode oscillates coherently, so the probability of each color state, which is a summation over all the momentum modes, also oscillates coherently.
However, with the phase factor, different momentum modes oscillate with different frequencies, eventually going out of phase. 
Thus, the probability of a color state, which sums over all the momentum modes, even acquiring an oscillation initially, could not maintain it.

\begin{figure}[tbp!]
  \centering
  \subfigure [\ $V_{I}(x^+)= V(x^+)$]
  {\label{fig:spin_evolution_Vfree_nophase_4p25_N16_v2}
    \includegraphics[width=0.4\textwidth]{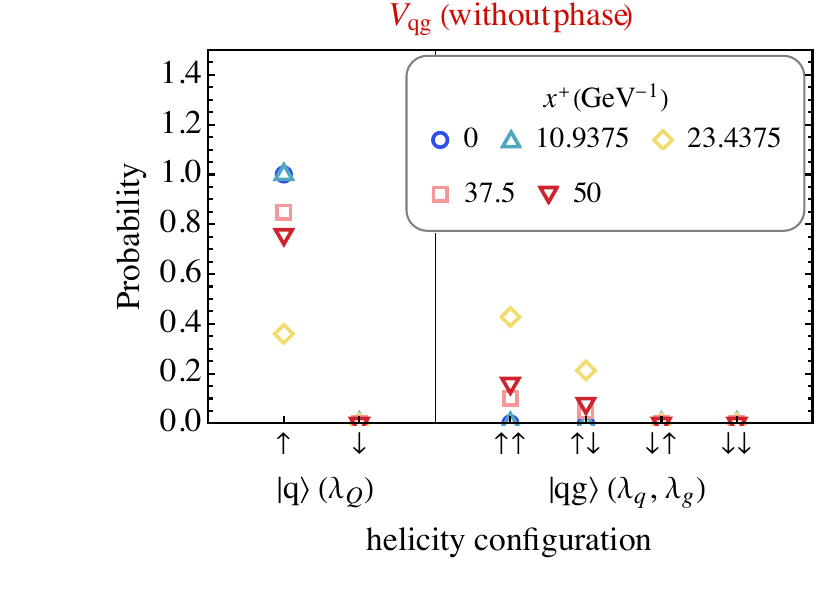}
  } 
  \subfigure [\ $V_{I}(x^+)=e^{i\frac{1}{2}P^-_{KE}x^+} V(x^+) e^{-i\frac{1}{2}P^-_{KE}x^+}$]
  {\label{fig:spin_evolution_Vfree_phase_4p25_N16_v2}
    \includegraphics[width=0.4\textwidth]{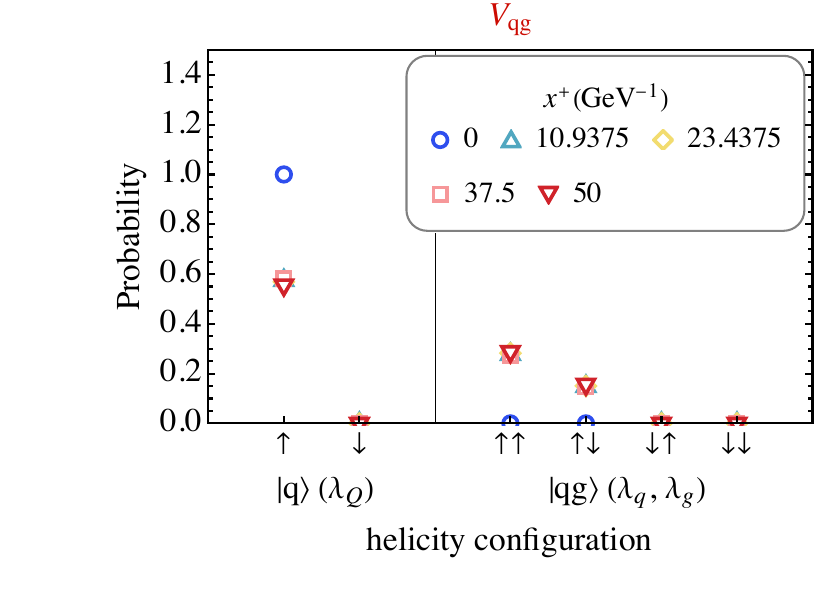}
  } 
  \caption{
  Evolution of the quark state in light-front helicity phase space.
  The interaction contains just the gluon emission/absorption term $V(x^+)=V_{qg}$, (a) without the phase factor and (b) with the phase factor.
   The helicity configuration in the $\ket{q}$ sector is labeled by the quark helicity $\lambda_Q=\uparrow,\downarrow$.
  The helicity configuration in the $\ket{qg}$ sector is labeled by the quark helicity $\lambda_q=\uparrow,\downarrow$ and the gluon helicity $\lambda_g=\uparrow,\downarrow$.
  The initial state of the quark is a single quark state with $\vec p_{\perp,Q} = \vec 0_\perp$, $p_Q^+=P^+=4.25~\GeV$, light-front helicity $\lambda_Q=1/2$, and color $c_Q=2$. 
  Parameters in the two simulations: $N_\perp=16$, $L_\perp=50~\GeV^{-1}$, $m_g=0.1~\GeV$, $m_q=0.02~\GeV$, and  $K=8.5$.
  The duration of each time step in the simulation is $\delta x^+=0.39~\GeV^{-1}$.
 }
  \label{fig:spin_evolution_Vfree}
\end{figure}

Lastly, we examine the evolution of the quark state in helicity phase space. The results are presented in Fig.~\ref{fig:spin_evolution_Vfree}.  
As the evolution time increases, states in the $\ket{qg}$ sector appear.
Since the initial state is a single quark state with $\lambda_Q=1/2$, the produced $\ket{qg}$ states favor the  $\{ \lambda_q=1/2, \lambda_g=\pm 1 \}$ configurations in which the quark helicity is preserved. 
The transition to the $\{\lambda_q=-1/2, \lambda_g=1\}$ state is weighted by the quark mass, which is relatively small in this case. 
The $\{\lambda_q=-1/2, \lambda_g=-1\}$ state is not allowed.
Very much like the evolution of probability distribution in the color space, the probabilities of those helicity states oscillate over time when the phase factor is not included, as in Fig.~\ref{fig:spin_evolution_Vfree_nophase_4p25_N16_v2}, and the oscillations are suppressed when the phase factor is restored, as in Fig.~\ref{fig:spin_evolution_Vfree_phase_4p25_N16_v2}.

From the above results and discussions, we see that the evolution with the gluon emission/absorption interaction contains two contributions: the transition between the $\ket{q}$ and the $\ket{qg}$ sectors by $V_{qg}$, and a phase rotation by $P^-_{KE}$. This interaction preserves the system's total momentum, and it changes the distribution of the state in both the $p^+$ and the $\vec p_\perp$ spaces, as well as in color and helicity spaces. Without the phase factor, the transitions happen as coherent oscillations between different states, but the phase factor causes the transitions to decohere. 

\FloatBarrier

\subsection{Interaction with background field}\label{sec:VA}
In this section, we study the effect from the background field without gluon emission or absorption.
The background field interacts with the $\ket{q}$ and the $\ket{qg}$ sectors separately, and does not in itself cause transitions between them.
The interaction with just the $\ket{q}$ sector was previously studied with the tBLFQ approach in Ref.~\cite{Li:2020uhl}.
The background field in the simulation has $P_{\mathcal{A}}^+=0$, so it does not change the $p^+$ configuration of the system.
The nonzero component of the background field is $\mathcal{A}^-$, which couples to the $J^+$ current of the fermion field, so the light-front helicity of the quark state is not affected either.
The background field only affects the distributions in transverse momentum space and in color space. 

\begin{figure*}[tbp!]
  \centering
  \subfigure[\ the quark in $\ket{q}$  \label{fig:pT_VA_phase_8p5a_g2dx3}]
  {\includegraphics[width=.9\textwidth]
  {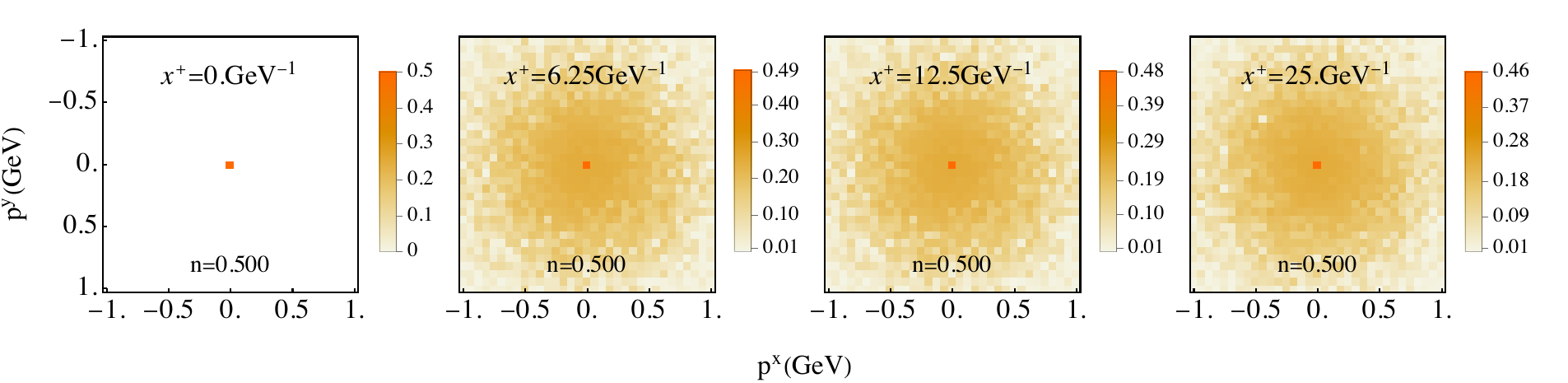}
  }
  \subfigure[\ the quark in $\ket{qg}$  \label{fig:pT_VA_phase_8p5b_g2dx3}]
  {\includegraphics[width=.9\textwidth]{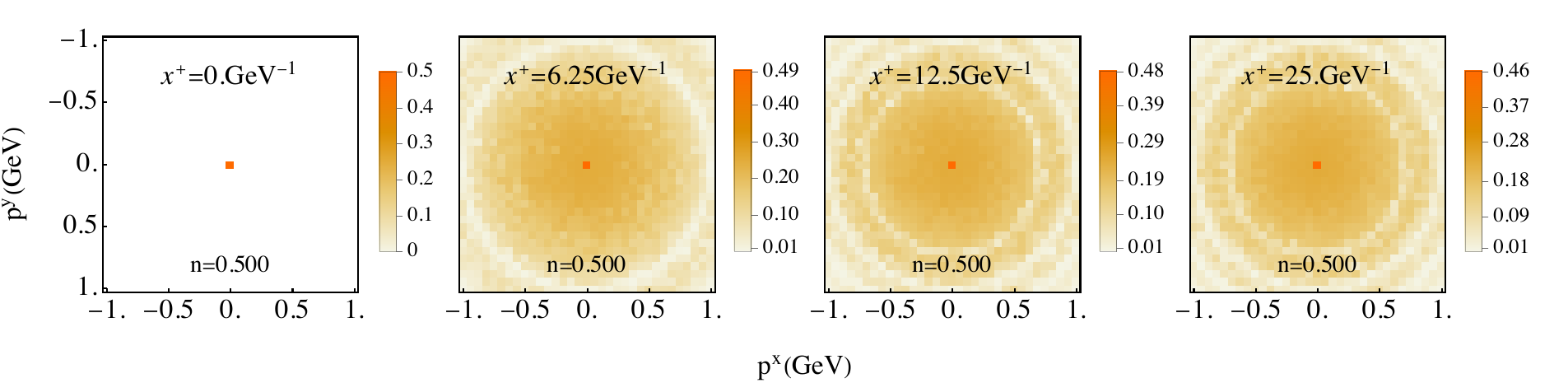}
  }
  \subfigure[\ the gluon in $\ket{qg}$  \label{fig:pT_VA_phase_8p5c_g2dx3}]
  {\includegraphics[width=.9\textwidth]{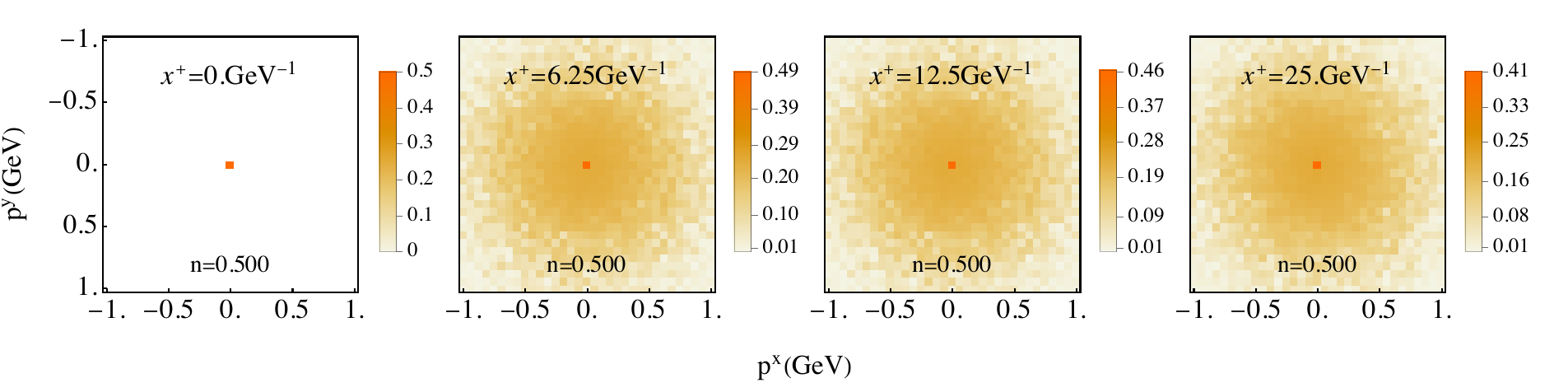}
  }
  \caption{ 
   The evolution of the transverse momentum distributions of (a) the quark in the $\ket{q}$ sector, (b) the quark in $\ket{qg}$ sector, and (c) the gluon in $\ket{qg}$ sector. 
   The interaction contains just the background interaction term $V(x^+)=V_{\mathcal{A}}(x^+)$, and phase factor is included.
    The initial state is a superposition of a $\ket{q}$ state with $p^+_Q=P^+=8.5 ~\GeV$, $\vec p_{\perp,Q}=\vec 0_\perp$, helicity $\lambda_Q=1/2$, color index $c_Q=1$, and a $\ket{qg}$ state with $p^+_q=0.5 ~\GeV$, $p^+_g=8~\GeV$, $\vec p_{\perp,q}=\vec p_{\perp,g}=\vec 0_\perp$, helicity $\lambda_q=1/2, \lambda_g=1$, color index $c_q=1, c_g=1$. The basis coefficient for each of the two is $1/\sqrt{2}$.
    From left to right, the transverse momentum distributions of the particle are shown at increasing light-front time instances. 
    The number at the bottom of each panel is the total probability of the plotted states.
    Parameters in the simulation: $m_g=0.1~\GeV$, $N_\perp=16$, $L_\perp=50~\GeV^{-1}$, $g^2\tilde\mu=0.018~\GeV^{3/2}$, $m_q=0.02~\GeV$. 
    The duration of each background field layer is $\tau=12.5~\GeV^{-1}$ and that of the each time step in the simulation is $\delta x^+=0.39~\GeV^{-1}$.
    For the rightmost panels, which are at the last of the evolution, the total evolution time is $L_\eta=25~\GeV^{-1}$; the value of the dimensionless quantity $Q_s a_\perp$ [$Q_s$ is defined in Eq.~\eqref{eq:Qs}] is $0.13$.
  }
  \label{fig:pT_VA_phase_8p5_g2dx3}
\end{figure*}
\begin{figure*}[tbp!]
  \centering
  \subfigure[\ the quark in $\ket{q}$  \label{fig:pT_VA_phase_8p5a_g2dx8}]
  {\includegraphics[width=.9\textwidth]
  {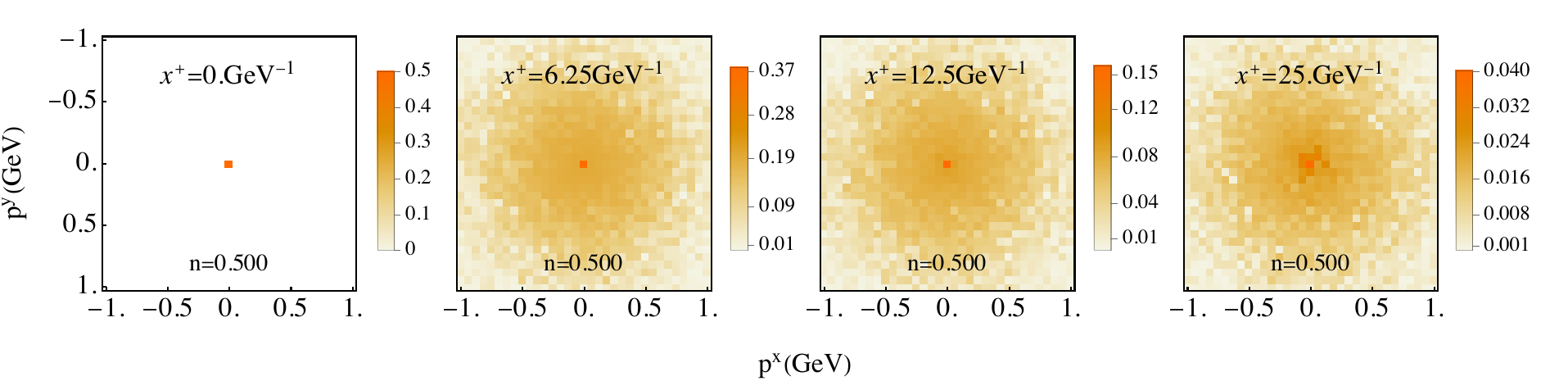}
  }
  \subfigure[\ the quark in $\ket{qg}$  \label{fig:pT_VA_phase_8p5b_g2dx8}]
  {\includegraphics[width=.9\textwidth]{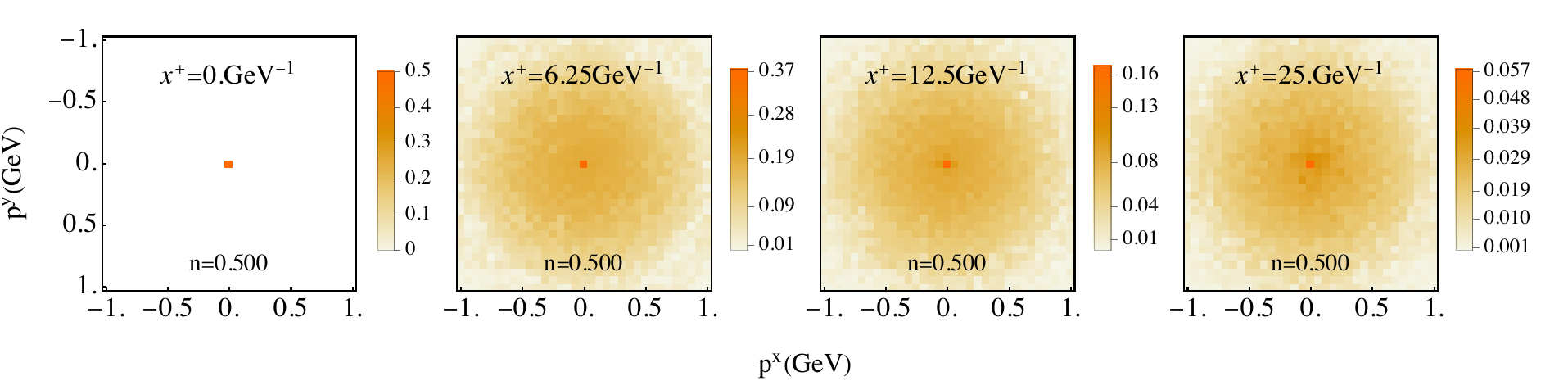}
  }
  \subfigure[\ the gluon in $\ket{qg}$  \label{fig:pT_VA_phase_8p5c_g2dx8}]
  {\includegraphics[width=.9\textwidth]{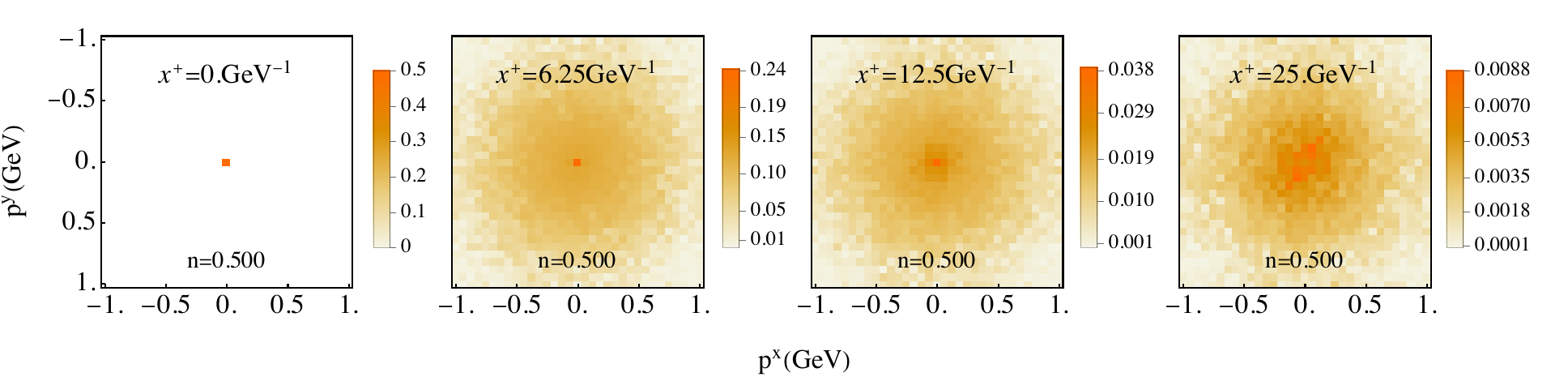}
  }
  \caption{
   The evolution of the transverse momentum distributions of (a) the quark in the $\ket{q}$ sector, (b) the quark in $\ket{qg}$ sector, and (c) the gluon in $\ket{qg}$ sector. 
   The interaction contains just the background interaction term $V(x^+)=V_{\mathcal{A}}(x^+)$, and phase factor is included.
    The initial state is a superposition of a $\ket{q}$ state with $p^+_Q=P^+=8.5 ~\GeV$, $\vec p_{\perp,Q}=\vec 0_\perp$, helicity $\lambda_Q=1/2$, color index $c_Q=1$, and a $\ket{qg}$ state with $p^+_q=0.5 ~\GeV$, $p^+_g=8~\GeV$, $\vec p_{\perp,q}=\vec p_{\perp,g}=\vec 0_\perp$, helicity $\lambda_q=1/2, \lambda_g=1$, color index $c_q=1, c_g=1$. The basis coefficient for each of the two is $1/\sqrt{2}$.
    From left to right, the transverse momentum distributions of the particle are shown at increasing light-front time instances. 
  The number at the bottom of each panel is the total probability of the plotted states.
  Parameters in those panels: $m_g=0.1~\GeV$, $N_\perp=16$, $L_\perp=50~\GeV^{-1}$, $g^2\tilde\mu=0.108~\GeV^{3/2}$, $m_q=0.02~\GeV$. 
  The duration of each background field layer is $\tau=12.5~\GeV^{-1}$ and that of the each time step in the simulation is $\delta x^+=0.39~\GeV^{-1}$.
  For the rightmost panels, which are at the last of the evolution, the total evolution time is $L_\eta=25~\GeV^{-1}$; the value of the dimensionless quantity $Q_s a_\perp$ [$Q_s$ is defined in Eq.~\eqref{eq:Qs}] is $0.78$.
  }
  \label{fig:pT_VA_phase_8p5_g2dx8}
\end{figure*}

We present the evolution of the quark state in transverse momentum space in two cases: one with a relatively weaker field with $g^2\tilde\mu=0.018~\GeV^{3/2}$ in Fig.~\ref{fig:pT_VA_phase_8p5_g2dx3}, and the other with a relatively stronger field with $g^2\tilde\mu=0.108~\GeV^{3/2}$ in Fig.~\ref{fig:pT_VA_phase_8p5_g2dx8}.
In both cases, the initial state is a superposition of a $\ket{q}$ state with $p^+_Q=P^+=8.5 ~\GeV$, $\vec p_{\perp,Q}=\vec 0_\perp$, helicity $\lambda_Q=1/2$, color index $c_Q=1$ and a $\ket{qg}$ state with $p^+_q=0.5 ~\GeV$, $p^+_g=8 ~\GeV$, $\vec p_{\perp,q}=\vec p_{\perp,g}=\vec 0_\perp$, helicity $\lambda_q=1/2, \lambda_g=1$, and color index $c_q=1, c_g=1$. 
The basis coefficient for each of the two is $1/\sqrt{2}$. The total evolution time of the presented results is $L_\eta=25~\GeV^{-1}$.
The typical transverse momentum that the particles obtained from the background field is characterized by the saturation scale $Q_s$, as defined in Eq.~\eqref{eq:Qs}.
In both simulations, the values of $Q_s$ are far below the UV cutoff of the grid $\lambda_{UV} = \pi/a_\perp$ so that the calculated result is close to the continuum limit and away from the lattice effects.
The values of the dimensionless quantity $Q_s a_\perp$ in the two cases are $0.13$ and $0.78$, respectively, both sufficiently smaller than $\pi$.
As we see in Figs.~\ref{fig:pT_VA_phase_8p5_g2dx3} and~\ref{fig:pT_VA_phase_8p5_g2dx8}, the majority of the occupied momentum modes are still away from the boundary of the transverse momentum lattice by the end of the evolution.

Under the interaction with the background field, both the initial $\ket{q}$ state and the initial $\ket{qg}$ state transfer to other momentum modes within their Fock sector.
This momentum transfer is more obvious with the stronger field in Fig.~\ref{fig:pT_VA_phase_8p5_g2dx8} compared to that in Fig.~\ref{fig:pT_VA_phase_8p5_g2dx3}.
The circular pattern resulting from the phase factor appears in the transverse momentum distribution. 
Because of its relatively small longitudinal momentum $p^+_q=0.5~\GeV$, compared to the total  $P^+=8.5 ~\GeV$ of the system, the quark in the $\ket{qg}$ has a more significant phase rotation from the phase factors. 
Thus, the circular pattern is most noticeable for the quark in the $\ket{qg}$ sector when the background field is weak, as in Fig.~\ref{fig:pT_VA_phase_8p5b_g2dx3}. 
By comparing the transverse momentum distribution of the gluon and that of the quark, one sees, especially with the stronger field in 
Fig.~\ref{fig:pT_VA_phase_8p5_g2dx8}, the effect of Casimir scaling; because $C_A > C_F$ the gluon gets a stronger momentum kick from the background field than the quark.

\begin{figure*}[tbp!]
  \centering
  \subfigure [\ $V_{I}(x^+)= V(x^+)$]
  {  \label{fig:color_evolution_VA_a}
    \includegraphics[width=0.8\textwidth]{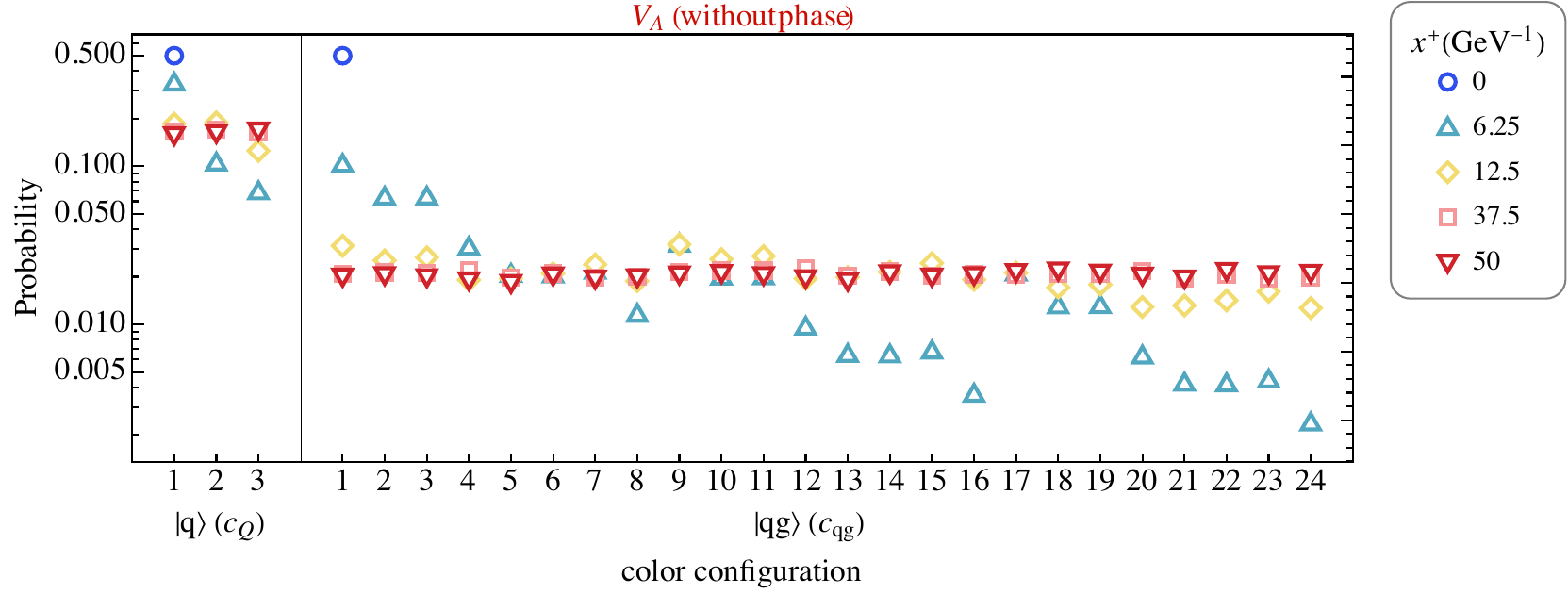}
  } 
  \subfigure [\ $V_{I}(x^+)=e^{i\frac{1}{2}P^-_{KE}x^+} V(x^+) e^{-i\frac{1}{2}P^-_{KE}x^+}$]
  {\label{fig:color_evolution_VA_b}
    \includegraphics[width=0.8\textwidth]{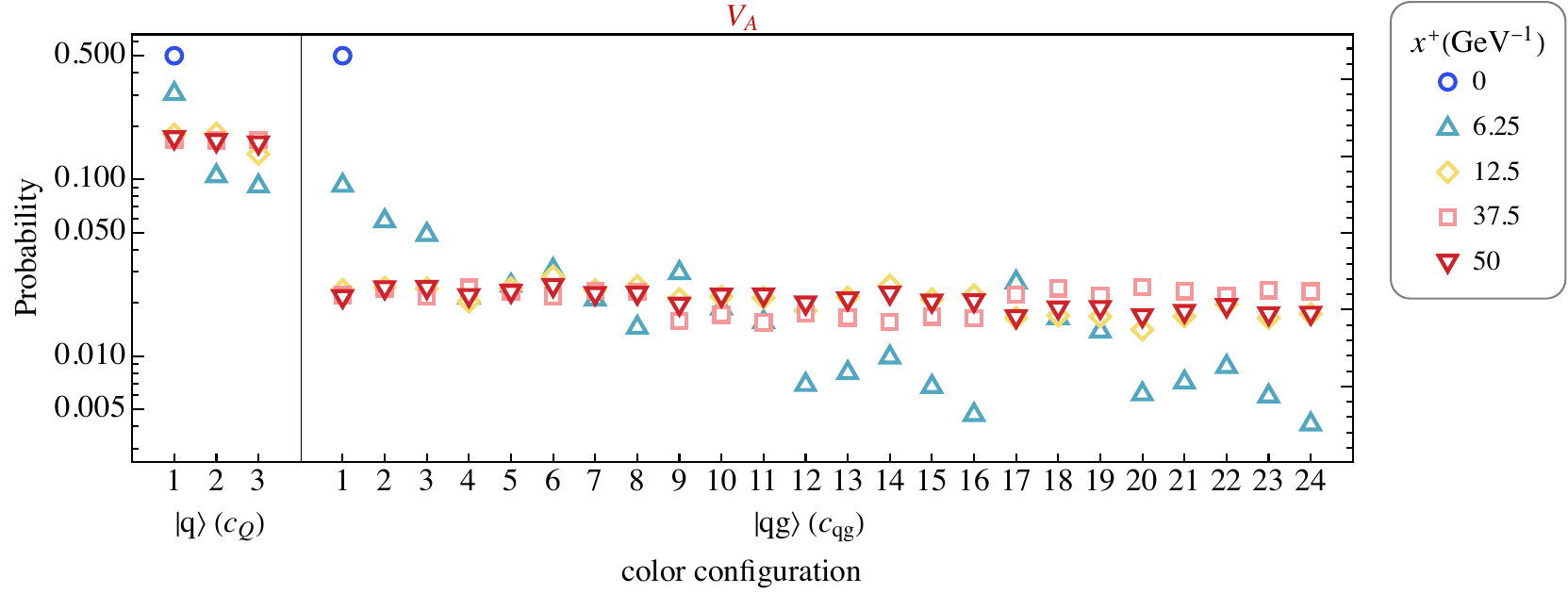}
  } 
  \caption{
  Evolution of the quark state in color space. 
  The interaction contains just the background interaction term $V(x^+)=V_{\mathcal{A}}(x^+)$, (a) without the phase factor and (b) with the phase factor.
  The color configuration in the $\ket{q}$ sector is labeled by the quark color index $c_Q=1,2,3$.
  The color configuration in the $\ket{qg}$ sector is labeled by the color index $c_{qg}=(c_q-1)8+c_g$, where the quark color index $c_q=1,2,3$ is the outer iterator and the gluon color index $c_g=1,\ldots,8$ is the inner iterator.
    The initial state is a superposition of a $\ket{q}$ state with $p^+_Q=P^+=4.25 ~\GeV$, $\vec p_{\perp,Q}=\vec 0_\perp$, helicity $\lambda_Q=1/2$, color index $c_Q=1$ and a $\ket{qg}$ state with $p^+_q=0.25 ~\GeV$, $p^+_g=4 ~\GeV$, $\vec p_{\perp,q}=\vec p_{\perp,g}=\vec 0_\perp$, helicity $\lambda_q=1/2, \lambda_g=1$, color index $c_q=1, c_g=1$. The basis coefficient for each of the two is $1/\sqrt{2}$.
    Parameters in these panels: $N_\perp=16$, $L_\perp=50~\GeV^{-1}$, $m_g=0.1~\GeV$, $g^2\tilde\mu=0.144~\GeV^{3/2}$, $m_q=0.02~\GeV$, $P^+=4.25~\GeV$, $K=8.5$.
    The duration of each background field layer is $\tau=12.5~\GeV^{-1}$, and that of each time step in the simulation is $\delta x^+=0.39~\GeV^{-1}$.
  }
  \label{fig:color_evolution_VA}
\end{figure*}

Then, we look at the evolution of the quark state in color space, as in Fig.~\ref{fig:color_evolution_VA}. 
The initial state is a superposition of a $\ket{q}$ state with color $c_Q=1$ and a $\ket{qg}$ state with color $c_q=1, c_g=1$. The basis coefficient for each of the two is $1/\sqrt{2}$.
The interaction $V_\mathcal{A}$ carries out a color rotation within the $\ket{q}$ and within the $\ket{qg}$ sector, separately.
In the two cases with and without the phase factor, as in Figs.~\ref{fig:color_evolution_VA_a} and  \ref{fig:color_evolution_VA_b}, all color states emerge during the evolution, and the state approaches a uniform color distribution in each Fock sector.

\subsubsection{Cross sections}

The interaction of a particle with the background field is usually quantified in terms of the cross section for scattering off the field. The study in Ref.~\cite{Li:2020uhl} calculated the cross section of a pure $\ket{q}$ state under the CGC background field.
In this work we study the cross section of a pure $\ket{qg}$ state.
These studies would get us prepared for calculating the cross section of a QCD eigenstate in the $\ket{q}+\ket{qg}$ Fock space in the future.

The cross section of a process is defined as the sum of the squares of the transition amplitudes,
\begin{align}\label{eq:defxs}
  \begin{split}
    \frac{\diff\sigma}{\diff^2 b}
    =&\sum_{\phi_f}
    {\left|  M(\phi_f; \psi_i)\right|}^2
    =\sum_{\phi_f}
    {\left|\expval{\phi_f|S|\psi_i}-\expval{\phi_f|\psi_i}\right|}^2
    \;.
  \end{split}
\end{align}
Here, $\psi_i$ stands for the initial state and $\phi_f$ the final state; $\sum_{\phi_f}$ sums over the phase space of the final state. 
The $S$ in the equation is the evolution operator from the initial state to the final state.

In the usual case corresponding to a physical scattering experiment, the time evolution happens over an infinite interval from $x^+=-\infty$ to $x^+=\infty$. For a finite-size target, this allows for an incoming quark to develop a cloud of gluons before the target and for the Fock states of the scattered particle to reorganize after the target through the $V_{qg}$ interaction. 
In our explicit numerical time-evolution procedure, such an infinite time evolution would not be feasible. 
Instead, we initialize our system in a specific Fock state at the time $x^+=0$ and study the evolution within the target color field for a finite time  $L_\eta$. 
Thus the calculation we are doing here does not actually correspond to scattering, and the quantity defined by Eq.~\eqref{eq:defxs} should not be interpreted as a usual cross section.
Studying a physical scattering process is possible with the same time-evolution algorithm. 
However, it requires using initial conditions at $x^+=0$ with a Fock state with a fully developed gluon cloud that corresponds to an incoming quark at $x^+=-\infty$, projecting out to similar scattering states at the end of the target.
In this paper, we focus on understanding the interaction within the target and leave the description of the correct asymptotic states to future work. 
Note that this issue did not concern the earlier tBLFQ calculation with the bare quark in Ref.~\cite{Li:2020uhl}, since in the absence of gluon radiation the time development between $x^+=\pm \infty$ and the target is trivial; thus, the results of that work could indeed be understood as quark-nucleus scattering cross sections.

In evaluating the total cross section, one should average over the color charge density $\rho$ of the target as in Eq.~\eqref{eq:chgcor},
\begin{align}\label{eq:crss_tot}
  \begin{split}
    \frac{\diff\sigma_{\mathrm{tot}}}{\diff^2 b}
    =&\expval{\sum_{\phi_f}{|  M(\phi_f; \psi_i)|}^2}
    \;.
  \end{split}
\end{align}
Here, the $\expval{\ldots}$ stands for a configuration average of the background field.
The total cross section includes a projection to the final state at the amplitude level, and a summation over all possible states at the cross section level~\cite{Mueller:1997ik, Kovchegov:1999kx,Dumitru:2002qt}. 
Using the unitarity of the $S$-matrix, i.e.,the optical theorem, the total cross section can also be expressed in terms of the expectation value of the diagonal elements of the scattering amplitude, i.e.,in terms of the imaginary part of the forward elastic amplitude. 

Since the background field interacts with both the quark and the gluon in the $\ket{qg}$ state, we first study their respective effects.
In the eikonal limit of $p^+=\infty$, both the single quark cross section $\diff\sigma_q/\diff^2 b$ and the single gluon cross section $\diff\sigma_{g}/\diff^2 b$ reduce to traces of Wilson lines and can be written in terms of the charge density $g^2\tilde{\mu}$, the interaction duration $L_\eta$, and the IR cutoff $m_g$ ~\cite{Dumitru:2002qt}.
The total cross section of a single quark interacting with the background field is (see Appendix~\ref{app:Wilsonline} for detailed derivations of Wilson line expectation values)
\begin{align}\label{eq:crss_eikonal_q}
  \begin{split}
    \frac{\diff\sigma_{q,\mathrm{tot}}}{\diff^2 b}\bigg|_{p^+=\infty}
    =&2\left[
      1- \frac{1}{N_c} \Re \expval{\Tr U_F(0,L_\eta; \vec x_\perp)}
      \right]\\
    =&2\bigg\{
    1-
    \exp\bigg[
    -\frac{C_F (g^2\tilde \mu)^2L_\eta}{8\pi m_g^2}
    \bigg]
    \bigg\}   
  \end{split}
\end{align}
and that of a single gluon is
\begin{align}\label{eq:crss_eikonal_g}
    \begin{split}
        \frac{\diff\sigma_{g,\mathrm{tot}}}{\diff^2 b}\bigg|_{p^+=\infty}
        =&2\left[
          1-\frac{1}{N_c^2-1} \Re \expval{\Tr U_A(0,L_\eta; \vec x_\perp)}
        \right]\\
        =&2\bigg\{
        1-
        \exp\bigg[
        -\frac{C_A (g^2\tilde \mu)^2L_\eta}{8\pi  m_g^2}
        \bigg]
        \bigg\}
        \;,
  \end{split}
\end{align}
where $C_F=(N_c^2-1)/(2N_c)=4/3$ and $C_A=N_c=3$. 
Here, one uses the representation in terms of the forward elastic amplitude, which in this case is the expectation value of a single Wilson line. 
Note that in the CGC picture, this means that the total cross section is the expectation value of a nonsinglet Wilson line operator, and as a consequence, it very strongly depends on the IR cutoff provided by $m_g$. 
Thinking differentially in terms of the momentum transfer from the target, it includes, besides the finite-$\vec{k}_\perp$ cross section, which results from the Fourier transform of the color singlet dipole operator, the part at $\vec{k}_\perp=\vec 0_\perp$ that is not singlet (see similar considerations in, e.g., Refs.~\cite{Gelis:2002ki,JalilianMarian:2005jf}).

Now let us look at the cross section of a $\ket{qg}$ state. 
The same background field would interact with the quark and the gluon, and the total cross section in the eikonal limit reads
\begin{multline}\label{eq:crss_eikonal_qg_full}
    \frac{\diff\sigma_{qg,\mathrm{tot}}}{\diff^2 b}\bigg|_{p^+=\infty}
    =2\bigg[
      1-\frac{1}{N_c(N_c^2-1)} \Re \langle\Tr U_F(0,L_\eta; \vec x_\perp)\\
      \otimes U_A(0,L_\eta; \vec y_\perp)\rangle
      \bigg]\\
      =2\Bigg\{
        1-
      \exp \bigg[
      - 
      \frac{g^4\tilde \mu^2 L_\eta}{8\pi m_g^2 }(C_F+C_A) 
      \bigg]
    \\ \times 
    f_{qg}
    \left[
    \frac{g^4
    \tilde{\mu}^2 L_\eta}{4\pi m_g}
    |\vec x_\perp-\vec y_\perp| K_1
    \left(
    m_g|\vec x_\perp-\vec y_\perp|
    \right)
    \right]
    \ 
    \Bigg\}
    \;.
\end{multline}
The calculation of this product of quark and gluon Wilson lines is discussed in Appendix~\ref{app:Wilsonline}. 
Here, $f_{qg}(\xi)$ is a correlation function between the quark and the gluon $f_{qg}(\xi)=[7\cos(\xi/2)+\cos(3\xi/2)]/8$ (see derivation in
Appendix~\ref{app:Wilsonline}). 
In the argument of $f_{qg}$ as in Eq.~\eqref{eq:crss_eikonal_qg_full}, $K_1$ is the modified Bessel function of the second kind, and $\vec x_\perp$ and $\vec y_\perp$ are the transverse coordinates of the quark and the gluon. 
Unlike the cross sections of the single particle, which are independent of the transverse coordinates, the quark-gluon cross section has a nontrivial dependence on their difference $|\vec x_\perp-\vec y_\perp|$. 
In the limit $|\vec x_\perp-\vec y_\perp| \to \infty$, the Wilson lines seen by the quark and the gluon become uncorrelated. 
In this limit, Bessel function $K_1$ approaches zero, and with $f(0)=1$, the quark-gluon cross section reduces to
\begin{align}\label{eq:crss_eikonal_qg}
  \begin{split}
    \frac{\diff\tilde\sigma_{qg,\mathrm{tot}}}{\diff^2 b}\bigg|_{p^+=\infty}
    = & 2\left\{
      1-
      \exp\left[
      -\frac{(C_F+C_A) (g^2\tilde\mu)^2L_\eta}{8\pi N_c m_g^2}
      \right]
      \right\}
    \;.
  \end{split}
\end{align}
This is just the product of the single quark and the single gluon cross sections, i.e., the case where the quark and the gluon interact with uncorrelated background fields separately.

\begin{figure}[tbp!]
  \centering
  \subfigure[\  $V(x^+)= V_{\mathcal{A},q}(x^+)$]
  {\includegraphics[width=0.4\textwidth]{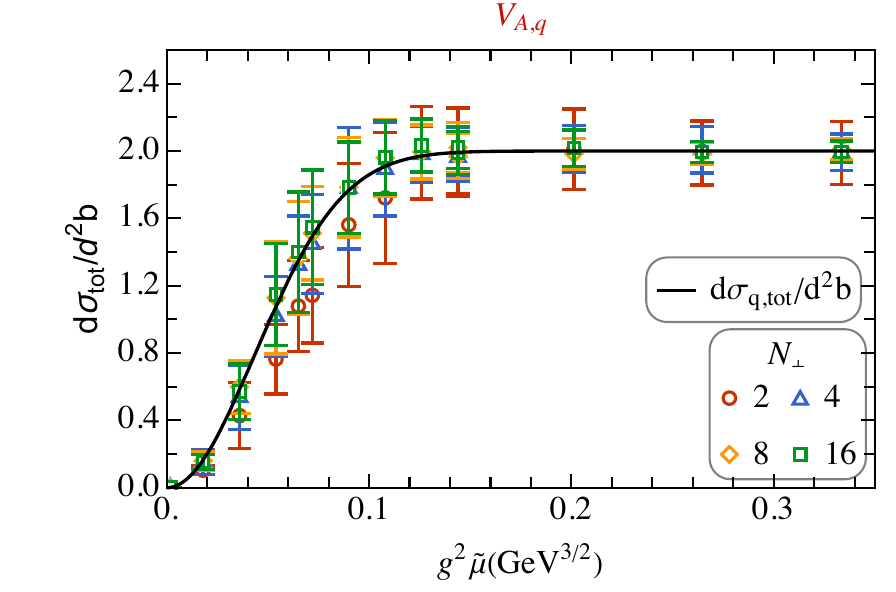}
  } 
  \qquad
  \subfigure[\  $V(x^+)= V_{\mathcal{A},g}(x^+)$]
  {\includegraphics[width=0.4\textwidth]{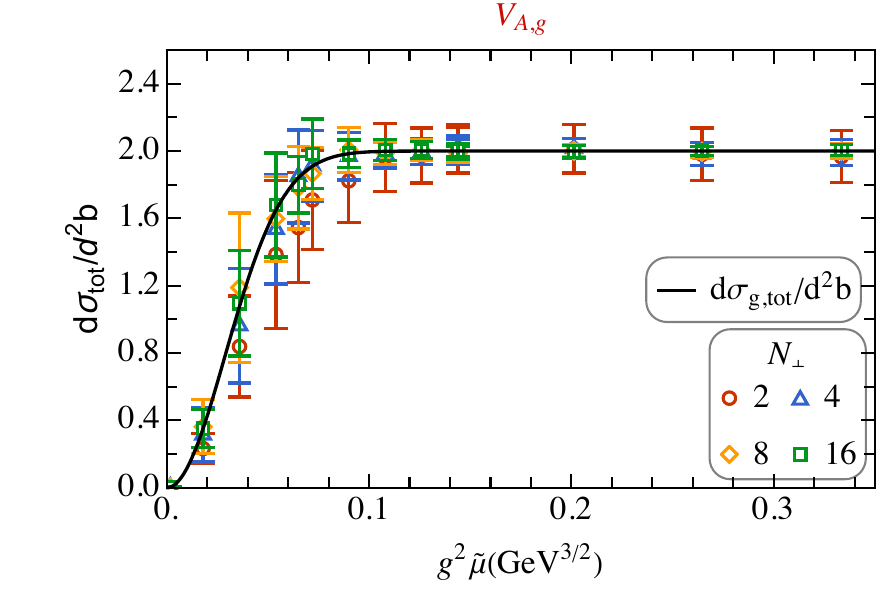}
  }
  \caption{
  The total cross sections of the $\ket{qg}$ state, as functions of $g^2
  \tilde\mu$, evaluated at various $N_\perp$.
  (a) The background field interacts with just the quark, i.e., $V(x^+)= V_{\mathcal{A},q}(x^+)$, and the solid line is the eikonal prediction as calculated from Eq.~\eqref{eq:crss_eikonal_q}. (b) The background field interacts with just the gluon, i.e., $V(x^+)= V_{\mathcal{A},g}(x^+)$, and the solid line is the eikonal prediction as calculated from Eq.~\eqref{eq:crss_eikonal_g}. 
   The initial state is a single quark-gluon state with $\vec p_{\perp,q} = \vec p_{\perp,g} = \vec 0_\perp$, light-front helicity $\lambda_q=1/2, \lambda_g=1$, and color $c_q=1, c_g=1$.
  The phase factor is not included in these simulations, i.e., $V_I(x^+)= V(x^+)$. 
  Each data point of the total cross section is calculated according to the definition in Eq.~\eqref{eq:crss_tot} by averaging over 50 configurations, and the standard deviation is taken as the uncertainty.
   Parameters in these simulations:  $N_\perp=2,4,8,16$, $L_\perp=50~\GeV^{-1}$, $N_\eta=4$, $L_\eta=50~\GeV^{-1}$, $m_g=0.1~\GeV$, $m_q=0.02~\GeV$.
    The duration of each background field layer is $\tau=12.5~\GeV^{-1}$ and that of each time step in the simulation is $\delta x^+=0.39~\GeV^{-1}$.
  }
  \label{fig:crss_qg_1}
\end{figure}

\begin{figure}[tbp!]
  \centering
\subfigure[\ \label{fig:crss_qg_2_diff} $V(x^+)= V_{\mathcal{A},q}(x^+)+V_{\mathcal{A}',g}(x^+)$]
  {\includegraphics[width=0.4\textwidth]{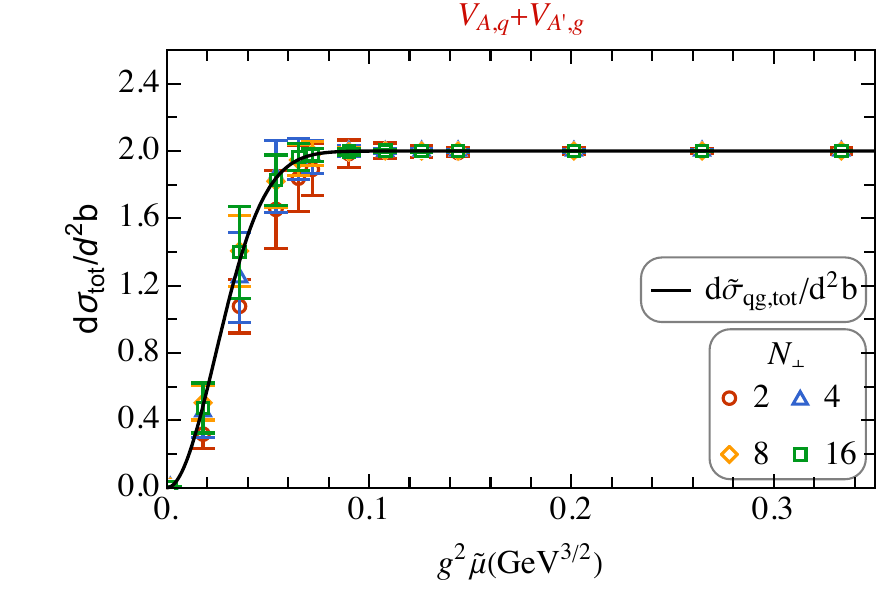}
  }
  \qquad
  \subfigure[\ \label{fig:crss_qg_2_same} $V(x^+)= V_{\mathcal{A},q}(x^+)+V_{\mathcal{A},g}(x^+)$]
  {\includegraphics[width=0.4\textwidth]{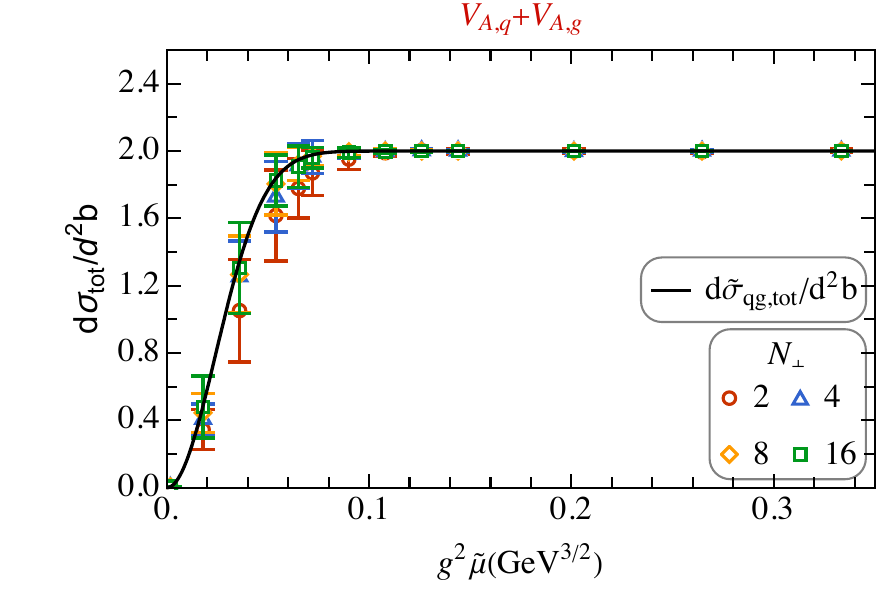}
  } 
\caption{
  The total cross sections of the $\ket{qg}$ state, as functions of $g^2
  \tilde\mu$, evaluated at various $N_\perp$.
  (a) Quark and gluon interact with different background fields, i.e., $V(x^+)= V_{\mathcal{A},q}(x^+)+V_{\mathcal{A}',g}(x^+) $, in which $\mathcal{A}$ and  $\mathcal{A}'$ are independently generated background fields for each simulation; (b) Quark and gluon interact with the same background field, i.e., $V(x^+)= V_{\mathcal{A},q}(x^+)+V_{\mathcal{A},g}(x^+) $.
  In both panels, the solid line is the uncorrelated eikonal prediction as calculated from Eq.~\eqref{eq:crss_eikonal_qg}.
   The initial state is a quark-gluon state with $\vec p_{\perp,q} = \vec p_{\perp,g} = \vec 0_\perp$, light-front helicity $\lambda_q=1/2, \lambda_g=1$, and color $c_q=1, c_g=1$.
  The phase factor is not included in these simulations, i.e., $V_I(x^+)= V(x^+)$. 
  Each data point of the total cross section is calculated according to the definition in Eq.~\eqref{eq:crss_tot}, by averaging over 50 configurations, and the standard deviation is taken as the uncertainty.
   Parameters in these simulations:  $N_\perp=2,4,8,16$, $L_\perp=50~\GeV^{-1}$, $N_\eta=4$, $L_\eta=50~\GeV^{-1}$, $m_g=0.1~\GeV$, $m_q=0.02~\GeV$.
    The duration of each background field layer is $\tau=12.5~\GeV^{-1}$ and that of each time step in the simulation is $\delta x^+=0.39~\GeV^{-1}$.
  }
  \label{fig:crss_qg_2}
\end{figure}

\begin{figure}[tbp!]
  \centering
  \includegraphics[width=0.4\textwidth]{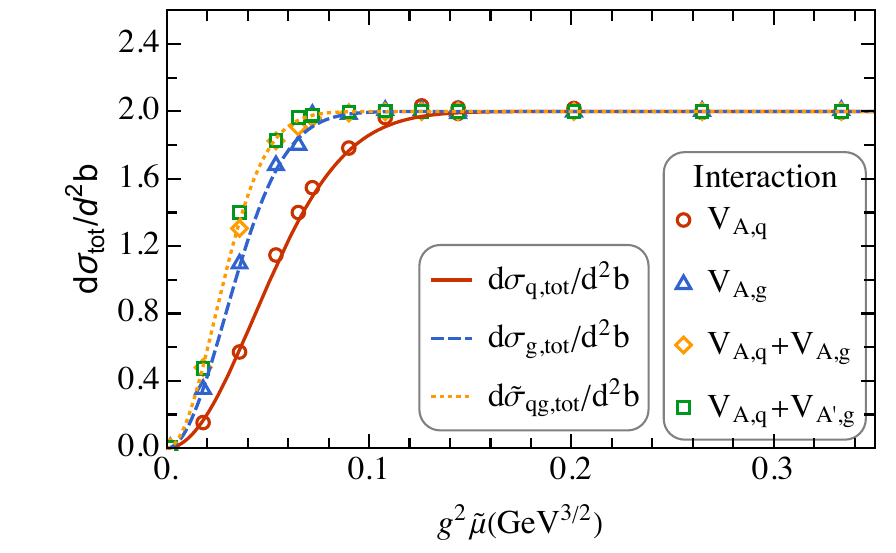}
      \caption{
  The total cross sections of the $\ket{qg}$ state, as functions of $g^2
  \tilde\mu$, with four different setups of the interaction,  $V(x^+)= V_{\mathcal{A},q}(x^+)$, $V(x^+)= V_{\mathcal{A},g}(x^+) $, $V(x^+)= V_{\mathcal{A},q}(x^+)+V_{\mathcal{A},g}(x^+) $, and  $V(x^+)= V_{\mathcal{A},q}(x^+)+V_{\mathcal{A}',g}(x^+) $.
  The data points in this figure are taken from the $N_\perp=16$ results of the four panels in Figs.~\ref{fig:crss_qg_1} and ~\ref{fig:crss_qg_2}, leaving out the uncertainties.
  The red solid, blue dashed, and yellow dotted lines are the eikonal predictions as calculated from Eqs.~\eqref{eq:crss_eikonal_q}, ~\eqref{eq:crss_eikonal_g} and~\eqref{eq:crss_eikonal_qg}, respectively.
   The initial state is a single quark-gluon state with $\vec p_{\perp,q} = \vec p_{\perp,g} = \vec 0_\perp$, light-front helicity $\lambda_q=1/2, \lambda_g=1$, and color $c_q=1, c_g=1$.
  The phase factor is not included in these simulations, i.e., $V_I(x^+)= V(x^+)$. 
   Parameters in these simulations:  $N_\perp=16$, $L_\perp=50~\GeV^{-1}$, $N_\eta=4$, $L_\eta=50~\GeV^{-1}$, $m_g=0.1~\GeV$, $m_q=0.02~\GeV$.
    The duration of each background field layer is $\tau=12.5~\GeV^{-1}$ and that of each time step in the simulation is $\delta x^+=0.39~\GeV^{-1}$.
  }
  \label{fig:crss_qg_3}
\end{figure}

We ran the simulations with a single $\ket{qg}$ initial state under the interaction with the background field at various $g^2\tilde\mu$, and we calculated the total cross sections according to Eq.~\eqref{eq:crss_tot}. 
Following the above discussions, we studied four different cases of the interaction: (1) the background field interacts with just the quark, i.e., $V(x^+)= V_{\mathcal{A},q}(x^+)$,  (2) the background field interacts with just the gluon, i.e., $V(x^+)= V_{\mathcal{A},g}(x^+)$; (3) the same background field interacts with both the quark and the gluon, i.e.,
 $V(x^+)= V_{\mathcal{A},q}(x^+)+V_{\mathcal{A},q}(x^+) = V_{\mathcal{A}}(x^+)$, (4) different background fields interact with the quark and the gluon, i.e., $V(x^+)= V_{\mathcal{A},q}(x^+)+V_{\mathcal{A}',g}(x^+) $,  where $\mathcal{A}$ and $\mathcal{A}'$ are independently generated background fields in the simulation.
We present the results in Figs.~\ref{fig:crss_qg_1},~\ref{fig:crss_qg_2}, and~\ref{fig:crss_qg_3}.
In these simulations, the initial state is a single quark-gluon state with $\vec p_{\perp,q} = \vec p_{\perp,g} = \vec 0_\perp$, light-front helicity $\lambda_q=1/2, \lambda_g=1$, and color $c_q=1, c_g=1$.
The phase factor is not included, which is equivalent to taking $P^+=\infty$.
As studied in Ref.~\cite{Li:2020uhl}, for the evolution of a single quark state with the chosen background field, the total cross sections at finite $P^+$ do not show noticeable differences from the $P^+=\infty$ case. 
We find it also true for the quark-gluon state by running simulations with various $P^+$.

Figure~\ref{fig:crss_qg_1} shows the calculated cross sections for the first two cases. 
The result of the background field interacting with just the quark (gluon) in the $\ket{qg}$  state agrees with the eikonal expectation of a single quark (gluon) separately scattering on the background field in Eq.~\eqref{eq:crss_eikonal_q} [Eq.~\eqref{eq:crss_eikonal_g}], as one would expect.
These calculations help check the correctness of our numerical calculations, and they might also be helpful to study processes involving a single quark or gluon.

Figure~\ref{fig:crss_qg_2} shows the calculated cross section for the latter two cases. 
In Fig.~\ref{fig:crss_qg_2_diff}, the total quark-gluon cross section of the quark and the gluon interacting with different background fields agrees with the eikonal prediction of the uncorrelated scattering in Eq.~\eqref{eq:crss_eikonal_qg}, as one would expect. 
However, even in the case where the quark and the gluon interact with the same background field, which is more likely to happen in a dressed quark scattering process, the total cross section agrees with this uncorrelated prediction in Eq.~\eqref{eq:crss_eikonal_qg} as well. 
In other words, the correlation between the quark and the gluon through interacting to the same background field is too small to be noticeable in the total cross section.
To see this quantitatively, at the strength $g^2\tilde\mu$ where the correlation is strong, the cross section is already close to its black disc limit of $\sigma_{tot}\to 2$, so the correlation is of little account (see more discussions in Appendix~\ref{app:Wilsonline}) .

To get an impression of the relative magnitude of the four cases discussed above, we put them together in Fig.~\ref{fig:crss_qg_3} for comparison. 
The cross section as a function of $g^2\tilde\mu$ saturates most rapidly for a $\ket{qg}$ state, second for a gluon state, and last for a quark state, also seen from their corresponding eikonal expectation in Eqs.~\eqref{eq:crss_eikonal_q}, \eqref{eq:crss_eikonal_g}, and \eqref{eq:crss_eikonal_qg}.

From the above results and discussions, the physical picture is that the interaction with the background field changes the distribution in transverse momentum space and color space. 
We also see that the cross section of a $\ket{qg}$ state agrees with the eikonal expectation, and the correlation between the quark and the gluon is significantly suppressed in the total cross section defined by Eq.~\eqref{eq:defxs}. 

\FloatBarrier
\begin{figure*}[tbph!]
    \centering
    \subfigure[\ Without background field]
    {\label{fig:Vfree_pro_evolution}
      \includegraphics[width=0.4\textwidth]{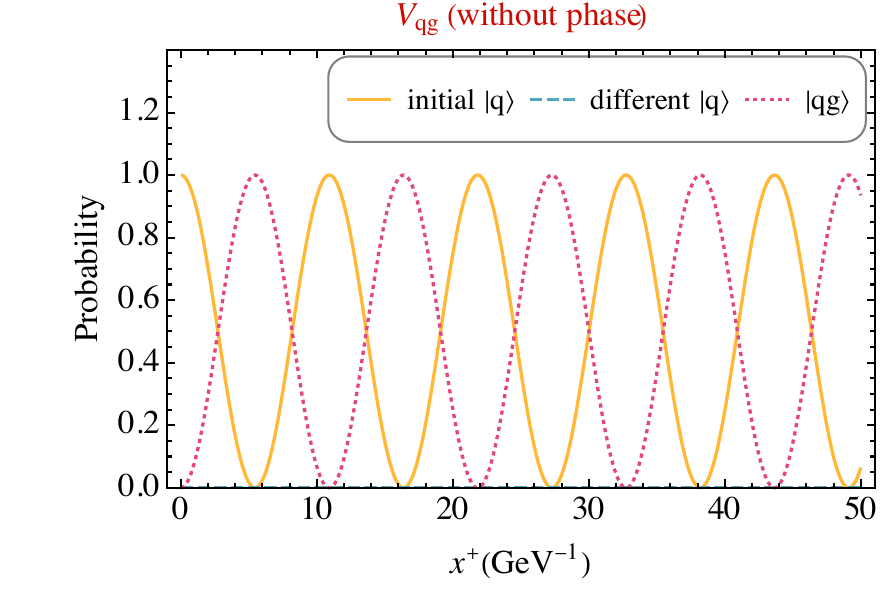}
      \includegraphics[width=0.4\textwidth]{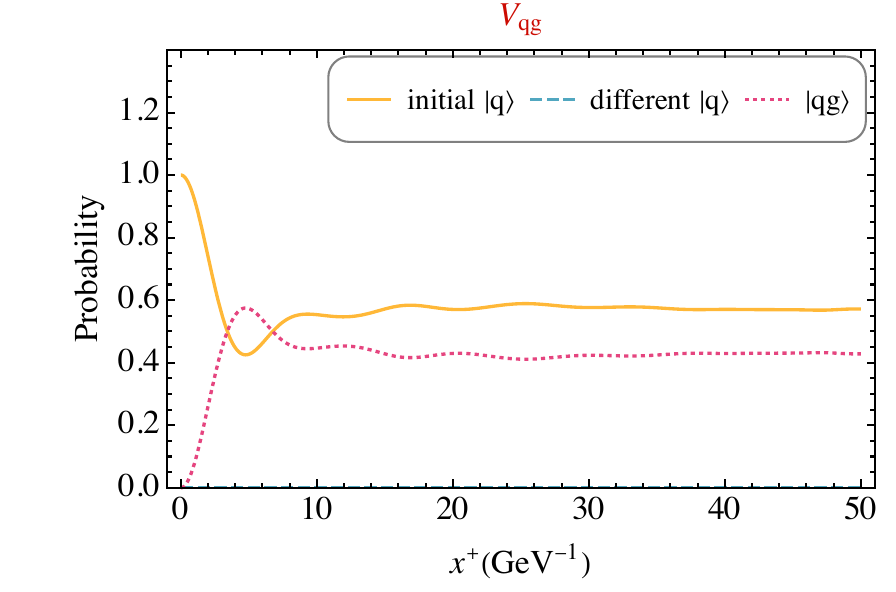}
    }
    \subfigure[\ With a weak  background field, $g^2\tilde \mu=0.018~\GeV^{3/2}$]
    {\label{fig:Vfull_pro_evolution_g2dx3}
      \includegraphics[width=0.4\textwidth]{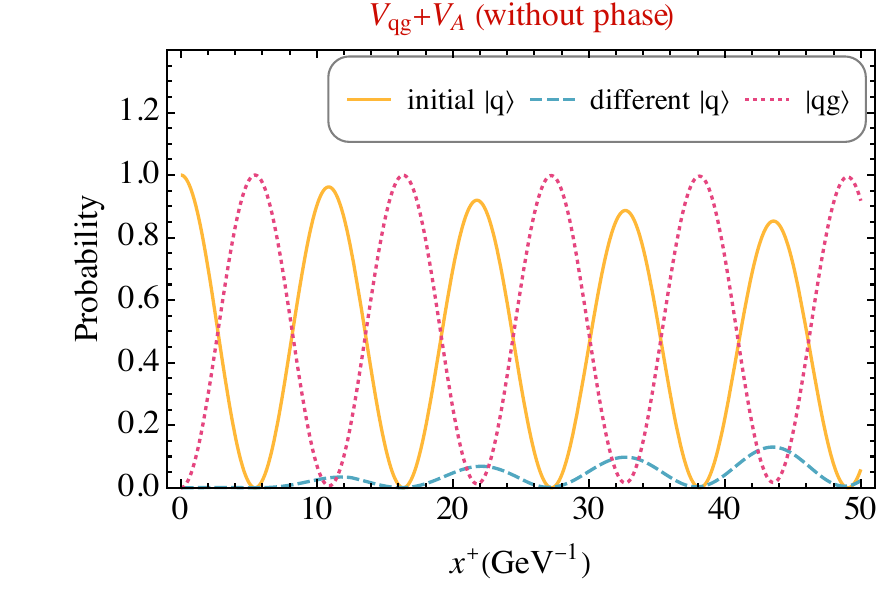}
      \includegraphics[width=0.4\textwidth]{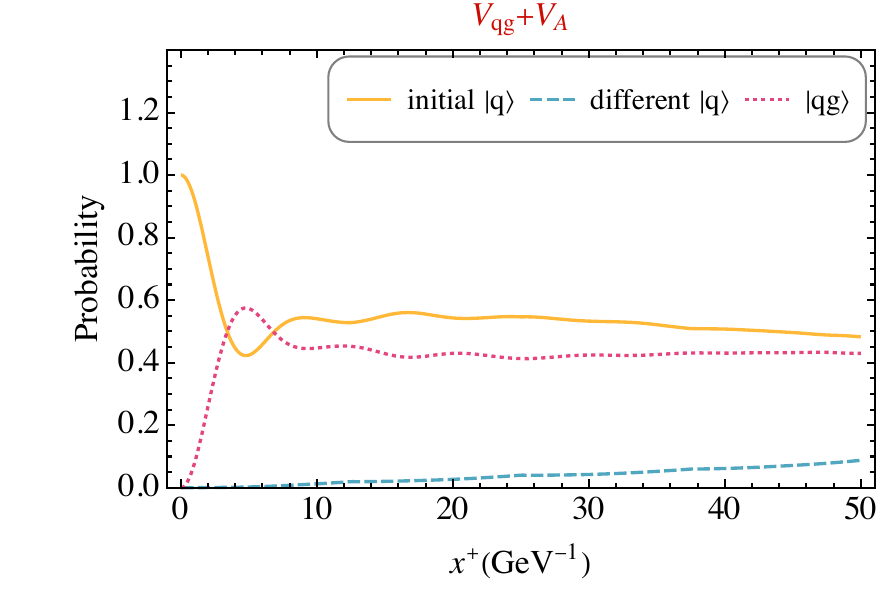}
    }
    \subfigure[\ With a stronger background field, $g^2\tilde \mu=0.144~\GeV^{3/2}$]
    {\label{fig:Vfull_pro_evolution_g2dx10}
      \includegraphics[width=0.4\textwidth]{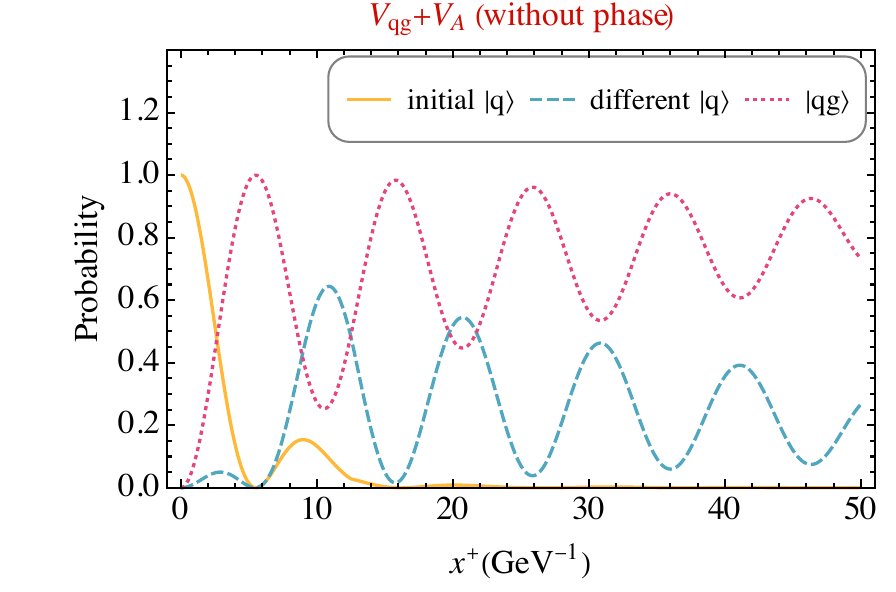}
      \includegraphics[width=0.4\textwidth]{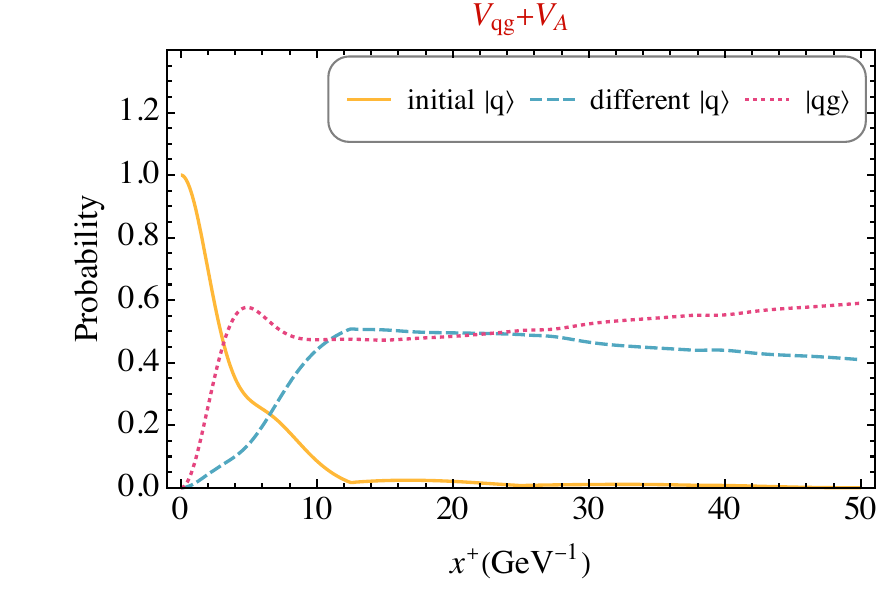}
    }
    \caption{
    The probability of the quark staying in its initial state and the transition probabilities to other states during the evolution.
   The initial state of the quark is a single quark state with $\vec p_{\perp,Q} = \vec 0_\perp$, $p_Q^+=P^+=8.5~\GeV$, light-front helicity $\lambda_Q=1/2$, and color $c_Q=1$. 
    The probability of the quark in its initial state is in the yellow solid line, that of the other states also in the $\ket{q}$ is in the blue dashed line, and that of the $\ket{qg}$ states is in the red dotted line.
  From top to bottom: (a) without background field, i.e., $g^2\tilde \mu=0$, (b) with a relatively weak background field, $g^2\tilde \mu=0.018~\GeV^{3/2}$, and (c) with a relatively strong background field, $g^2\tilde \mu=0.144~\GeV^{3/2}$. 
    The simulations in the left panels do not include the phase factors, and those in the right panels do.
    Parameters in these panels: $N_\perp=16$, $L_\perp=50~\GeV^{-1}$, $L_\eta=50~\GeV^{-1}$, $N_\eta=4$, $m_g=0.1~\GeV$, $m_q=0.02~\GeV$.
   The duration of each background field layer is $\tau=12.5~\GeV^{-1}$ and that of the each time step in the simulation is $\delta x^+=0.39~\GeV^{-1}$.
  }
    \label{fig:pro_evolve_4p25_8}
  \end{figure*}
  
  \begin{figure}[tbph!]
    \centering
    \subfigure[\ $V_{I}(x^+)=V(x^+)$]
    {\label{fig:Vqg_ppl_evolution}
    \includegraphics[width=0.4\textwidth]{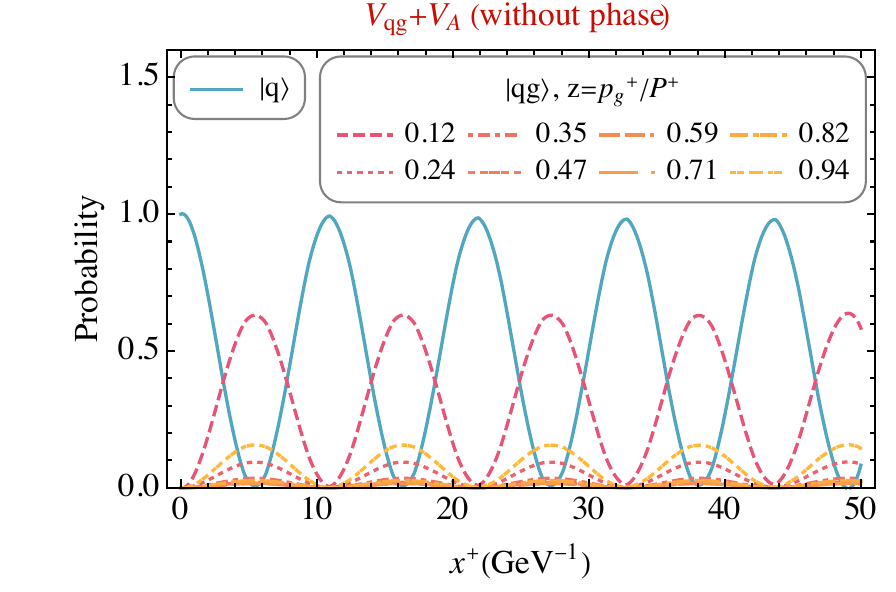}
    } 
    \subfigure[\ $V_{I}(x^+)=e^{i\frac{1}{2}P^-_{KE}x^+} V(x^+) e^{-i\frac{1}{2}P^-_{KE}x^+}$]
    {
    \includegraphics[width=0.4\textwidth]{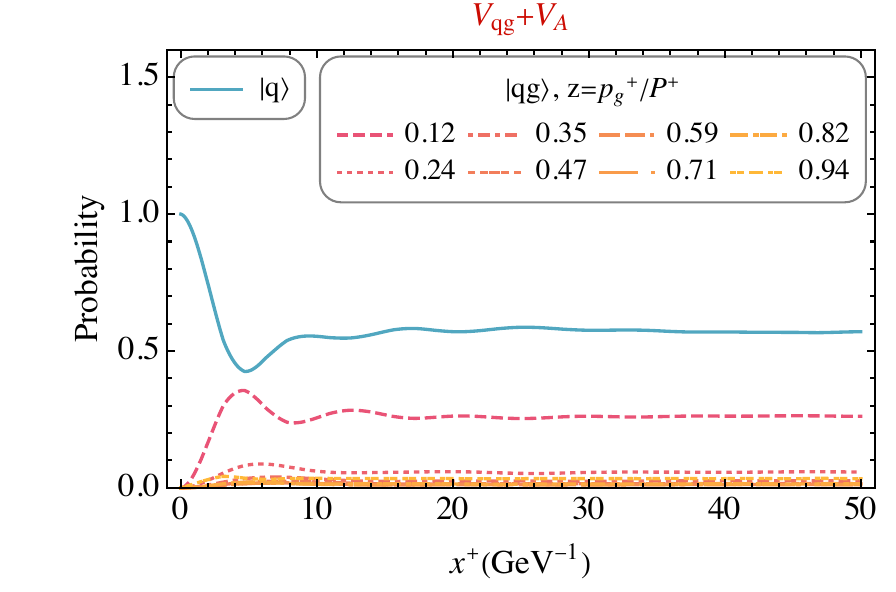}
    }
    \caption{
    The evolution of the probabilities of 
    different $p^+$ states, including the $\ket{q}$ sector and the K-segments of the $\ket{qg}$ sector characterized by the gluon longitudinal momentum fraction $z$.
    The interaction contains both the gluon emission/absorption term and the background interaction $V(x^+)=V_{qg}+V_{\mathcal{A}}(x^+)$, (a) without the phase factor and (b) with the phase factor.
    The initial state of the quark is a single quark state with $\vec p_{\perp,Q} = \vec 0_\perp$, $p_Q^+=P^+=8.5~\GeV$, light-front helicity $\lambda_Q=1/2$, and color $c_Q=1$. 
    Parameters in these panels: $N_\perp=16$, $L_\perp=50~\GeV^{-1}$, $L_\eta=50~\GeV^{-1}$, $N_\eta=4$, $m_g=0.1~\GeV$, $m_q=0.02~\GeV$, $K=8.5$, $g^2\tilde \mu=0.018~\GeV^{3/2}$, and the duration of each time step in the simulation is $\delta x^+=0.39~\GeV^{-1}$.
  }
    \label{fig:ppl_evolve_8p5_16}
  \end{figure}
  
  \begin{figure*}[tbp!]
    \centering
    \subfigure[\ the quark in $\ket{q}$  \label{fig:pT_Vfull_phase_8p5a}]
    {\includegraphics[width=.9\textwidth]
    {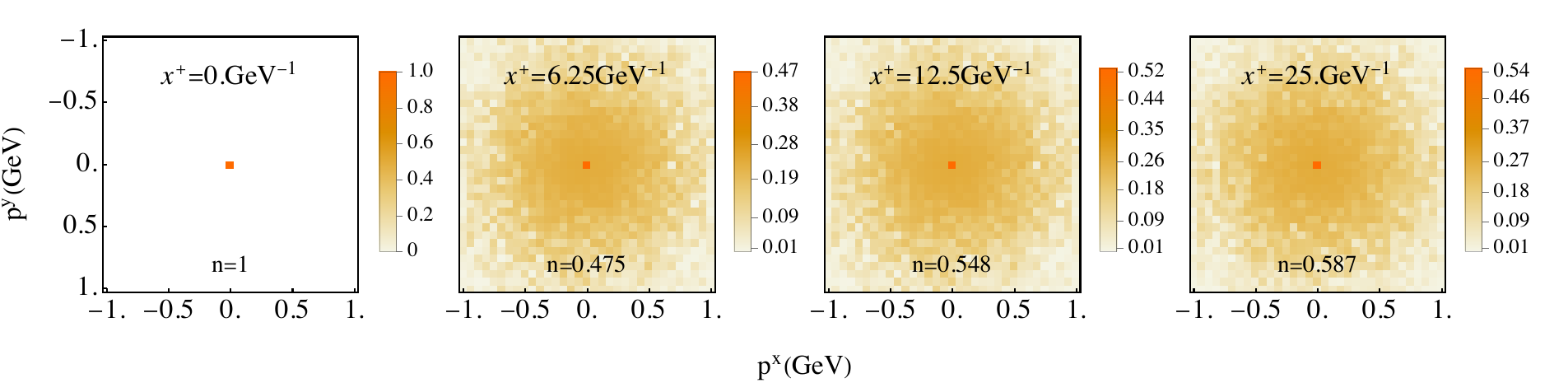}
    }
    \subfigure[\ the quark in $\ket{qg}$  \label{fig:pT_Vfull_phase_8p5b}]
    {\includegraphics[width=.9\textwidth]{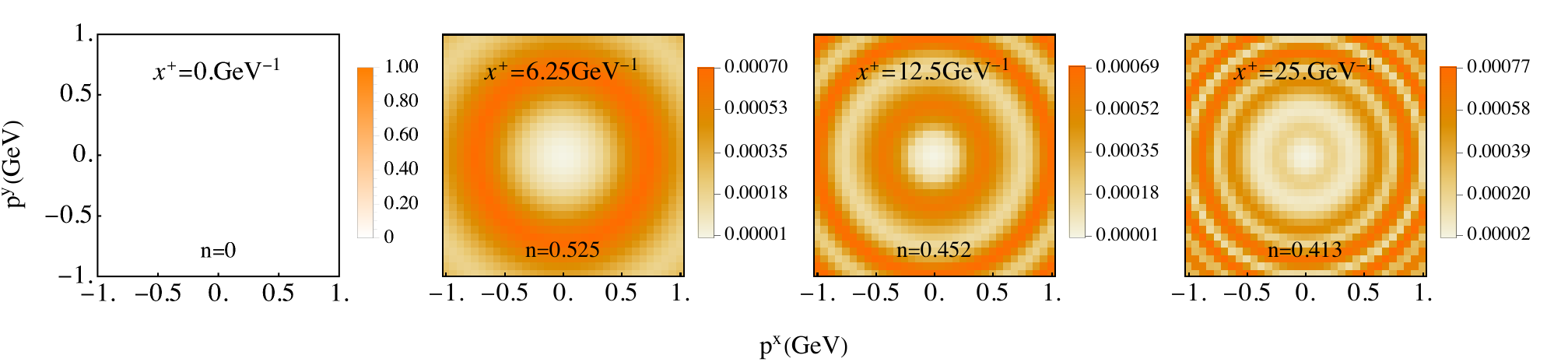}
    }
    \subfigure[\ the gluon in $\ket{qg}$  \label{fig:pT_Vfull_phase_8p5c}]
    {\includegraphics[width=.9\textwidth]{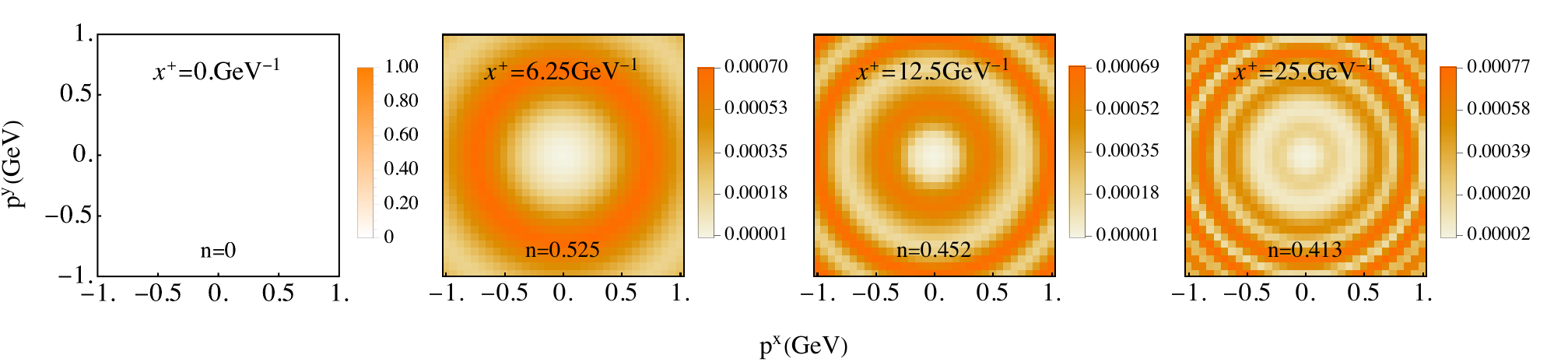}
    }
      \caption{ 
     The evolution of the transverse momentum distributions of (a) the quark in the $\ket{q}$ sector, (b) the quark in $\ket{qg}$ sector, and (c) the gluon in $\ket{qg}$ sector. 
     The interaction contains both the gluon emission/absorption and the background interaction term  $V(x^+)=V_{qg}+V_{\mathcal{A}}(x^+)$, and the phase factor is included.
     The initial state of the quark is a single quark state with $\vec p_{\perp,Q}=\vec 0_\perp$, light-front helicity $\lambda_Q=1/2$, color index $c_Q=1$.
      From left to right: the transverse momentum distributions of the particle are shown at increasing light-front time instances. 
     The number at the bottom of each panel is the total probability of the plotted states.
    Parameters in the simulation: $m_g=0.1~\GeV$, $N_\perp=16$, $L_\perp=50~\GeV^{-1}$, $g^2\tilde\mu=0.018~\GeV^{3/2}$, $m_q=0.02~\GeV$. 
    The duration of each background field layer is $\tau=12.5~\GeV^{-1}$ and that of the each time step in the simulation is $\delta x^+=0.39~\GeV^{-1}$.
    For the rightmost panels, which are at the last of the evolution, the total evolution time is $L_\eta=25~\GeV^{-1}$; the value of the dimensionless quantity $Q_s a_\perp$ [$Q_s$ is defined in Eq.~\eqref{eq:Qs}] is $0.13$.
    }
    \label{fig:pT_Vfull_phase_8p5}
  \end{figure*}
  
  \begin{figure*}[tbp!]
    \centering
    \subfigure[\ the quark in $\ket{q}$  \label{fig:pT_Vfull_phase_8p5a_g2d10}]
    {\includegraphics[width=.9\textwidth]
    {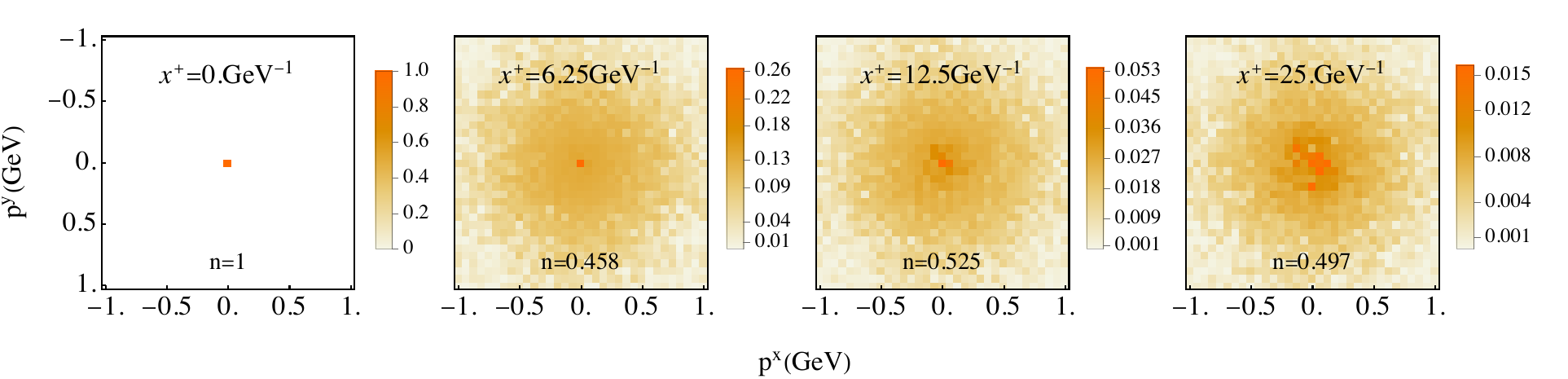}
    }
    \subfigure[\ the quark in $\ket{qg}$  \label{fig:pT_Vfull_phase_8p5b_g2d10}]
    {\includegraphics[width=.9\textwidth]{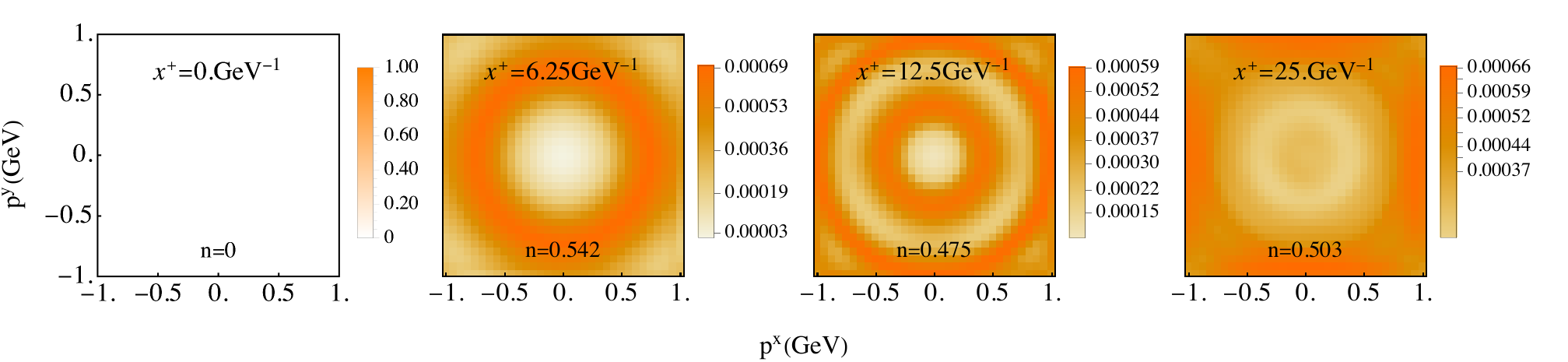}
    }
    \subfigure[\ the gluon in $\ket{qg}$  \label{fig:pT_Vfull_phase_8p5c_g2d10}]
    {\includegraphics[width=.9\textwidth]{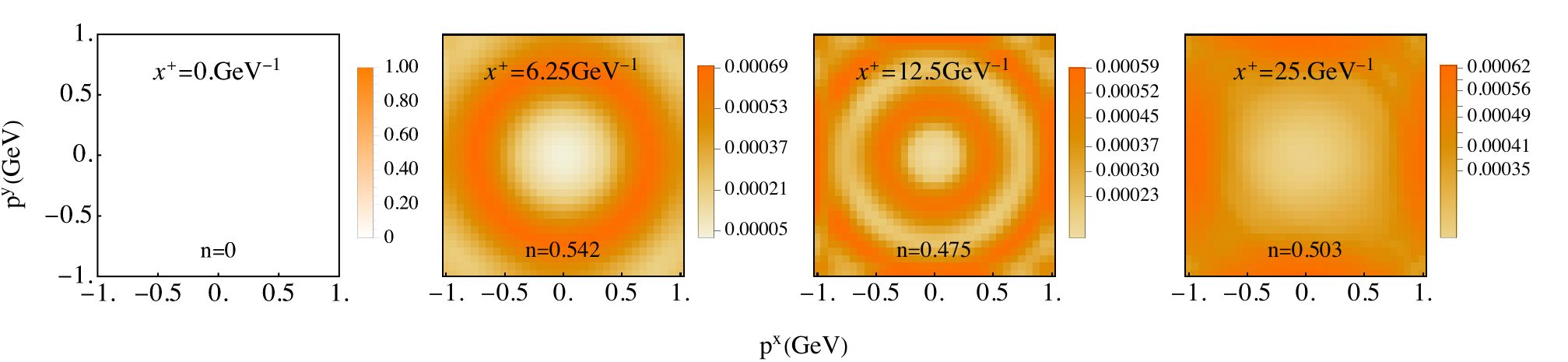}
    }
    \caption{   
    The evolution of the transverse momentum distributions of (a) the quark in the $\ket{q}$ sector, (b) the quark in $\ket{qg}$ sector, and (c) the gluon in $\ket{qg}$ sector. 
     The interaction contains both the gluon emission/absorption and the background interaction term  $V(x^+)=V_{qg}+V_{\mathcal{A}}(x^+)$, and phase factor is included.
     The initial state of the quark is a single quark state with $\vec p_{\perp,Q}=\vec 0_\perp$, light-front helicity $\lambda_Q=1/2$, color index $c_Q=1$.
      From left to right, the transverse momentum distributions of the particle are shown at increasing light-front time instances.  
     The number at the bottom of each panel is the total probability of the plotted states.
    Parameters in the simulation: $m_g=0.1~\GeV$, $N_\perp=16$, $L_\perp=50~\GeV^{-1}$, $g^2\tilde\mu=0.144~\GeV^{3/2}$, $m_q=0.02~\GeV$. 
    The duration of each background field layer is $\tau=12.5~\GeV^{-1}$ and that of the each time step in the simulation is $\delta x^+=0.39~\GeV^{-1}$.
    For the rightmost panels, which are at the last of the evolution, the total evolution time is $L_\eta=25~\GeV^{-1}$; the value of the dimensionless quantity $Q_s a_\perp$ [$Q_s$ is defined in Eq.~\eqref{eq:Qs}] is $1.04$.
  }
    \label{fig:pT_Vfull_phase_8p5_g2d10}
  \end{figure*}
  
  \begin{figure*}[tbp!]
    \centering
    \subfigure [\ $V_{I}(x^+)= V(x^+)$]
    {  \label{fig:color_evolution_Vfull_a}
      \includegraphics[width=0.8\textwidth]{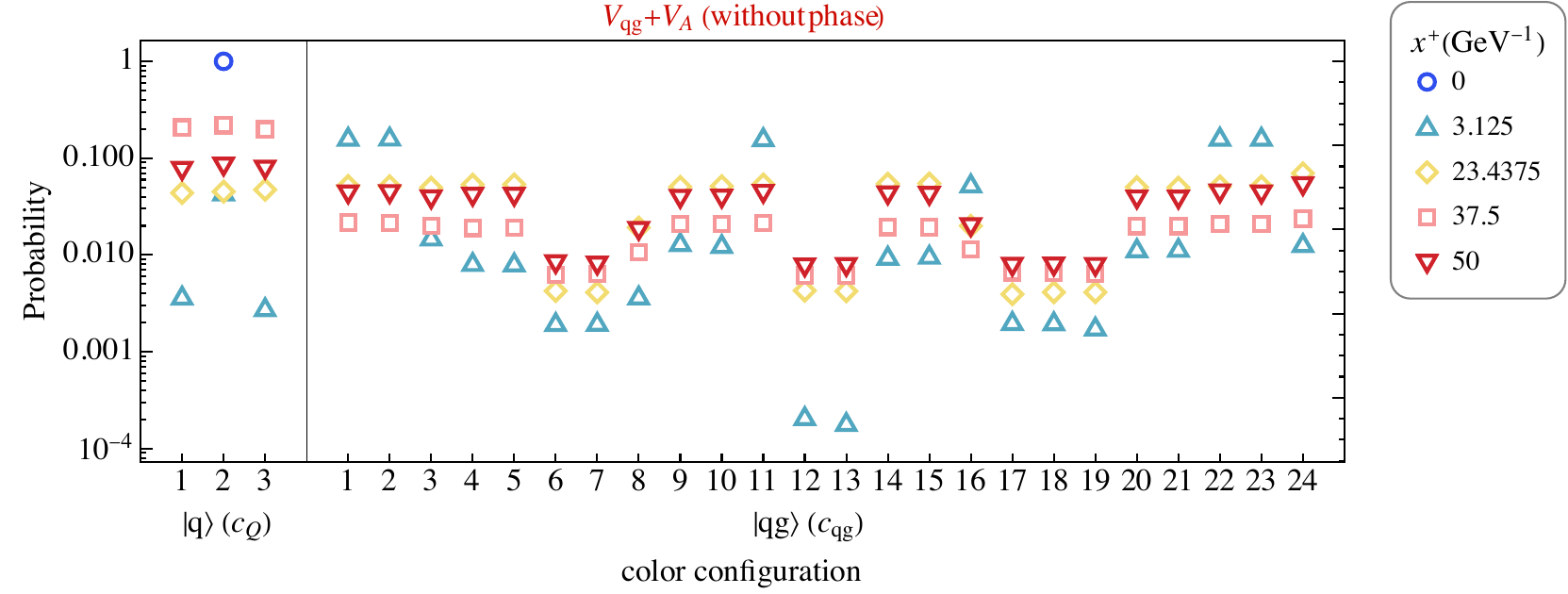}
    } 
    \subfigure [\ $V_{I}(x^+)=e^{i\frac{1}{2}P^-_{KE}x^+} V(x^+) e^{-i\frac{1}{2}P^-_{KE}x^+}$]
    {\label{fig:color_evolution_Vfull_b}
      \includegraphics[width=0.8\textwidth]{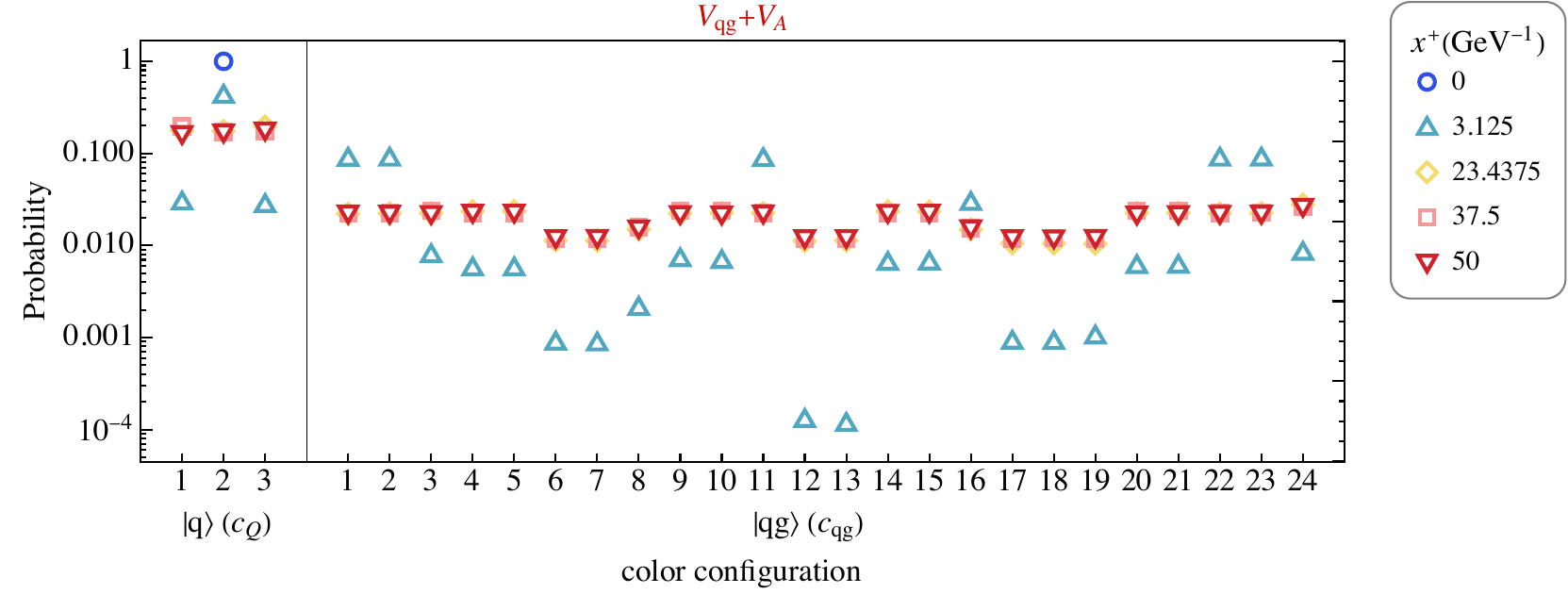}
    } 
      \caption{
    Evolution of the quark state in color space. 
    The interaction contains both the gluon emission/absorption and the background interaction term  $V(x^+)=V_{qg}+V_{\mathcal{A}}(x^+)$, (a) without the phase factor and (b) with the phase factor.
    The color configuration in the $\ket{q}$ sector is labeled by the quark color index $c_Q=1,2,3$.
    The color configuration in the $\ket{qg}$ sector is labeled by the color index $c_{qg}=(c_q-1)8+c_g$, where the quark color index $c_q=1,2,3$ is the outer iterator and the gluon color index $c_g=1,\ldots,8$ is the inner iterator.
    The initial state of the quark is a single quark state with $\vec p_{\perp,Q}=\vec 0_\perp$, light-front helicity $\lambda_Q=1/2$, color index $c_Q=2$.
      Parameters in these panels: $N_\perp=16$, $L_\perp=50~\GeV^{-1}$, $m_g=0.1~\GeV$, $g^2\tilde\mu=0.144~\GeV^{3/2}$, $m_q=0.02~\GeV$, $P^+=4.25~\GeV$, $K=8.5$.
     The duration of each background field layer is $\tau=12.5~\GeV^{-1}$, and that of each time step in the simulation is $\delta x^+=0.39~\GeV^{-1}$.
   }
    \label{fig:color_evolution_Vfull}
  \end{figure*}
  
  \subsection{Emission, absorption and background field}\label{sec:Vfull}
  Having studied the gluon emission/absorption and the background field separately in the Secs. ~\ref{sec:Vfree} and ~\ref{sec:VA}, we now put the two together to have the full interaction $V(x^+)=V_{qg}+V_{\mathcal{A}}(x^+)$. 
  
  We consider the initial state of the quark as a single quark state with $\vec p_{\perp,Q} = \vec 0_\perp$, light-front helicity $\lambda_Q=1/2$, and color $c_Q=1$. 
  The transition probabilities of the quark to other states are shown in Fig.~\ref{fig:pro_evolve_4p25_8}.
  The probability of the quark in its initial state is in the yellow solid line, that of the other $\ket{q}$ states is in the blue dashed line, and that of the $\ket{qg}$ states is in the red dotted line.
  When the background field is absent, the probabilities of the states in the $\ket{q}$ sector that are different from the initial state are always 0, as shown in Fig.~\ref{fig:Vfree_pro_evolution}.
  With the full interaction, the result shows the combined effects from the gluon emission/absorption and the interaction with the background field.
  When the background field is relatively weak, the result resembles the case with emission and absorption only, but different $\ket{q}$ states also emerge, see Fig.~\ref{fig:Vfull_pro_evolution_g2dx3}.
  With a stronger background field, the probability of different $\ket{q}$ states is larger, see Fig.~\ref{fig:Vfull_pro_evolution_g2dx10}.
  
  The evolution of the quark state in the $p^+$ phase space is shown in Fig.~\ref{fig:ppl_evolve_8p5_16}. The result is very similar to that without the background field in Fig.~\ref{fig:ppl_Vqg_evolve_8p5_16}, since the change in $p^+$ results from $V_{qg}$ and not from the background field interaction.
  
  The evolution of the quark state in the transverse momentum space is shown in Fig.~\ref{fig:pT_Vfull_phase_8p5} at $g^2\tilde\mu=0.018~\GeV^{3/2}$ and in Fig.~\ref{fig:pT_Vfull_phase_8p5_g2d10} at $g^2\tilde\mu=0.144~\GeV^{3/2}$. 
  Circular patterns appear as a result of the phase rotation, similar to those in the cases with the gluon emission/absorption in Fig.~\ref{fig:pT_Vfree_phase_8p5} and those with the background field in Figs.~\ref{fig:pT_VA_phase_8p5_g2dx3} and~\ref{fig:pT_VA_phase_8p5_g2dx8}.
  In addition, transitioning to other momentum modes in both the $\ket{q}$ and the $\ket{qg}$ sectors appear, resulting from the interaction with the background field.
  This effect is more obvious with the stronger field in Fig.~\ref{fig:pT_Vfull_phase_8p5_g2d10} compared to that in Fig.~\ref{fig:pT_Vfull_phase_8p5}.
  
  The evolution of the quark state in the color phase space is shown in Fig.~\ref{fig:color_evolution_Vfull}. 
  The initial state is a bare quark with color index $c_Q=2$.
  The $V_{qg}$ interaction allows the transition to six of the $\ket{qg}$ color states, as we have seen in Fig.~\ref{fig:color_evolution_Vfree}.
  The $V_\mathcal{A}$ interaction allows the color transitions within the $\ket{q}$ sector and within the $\ket{qg}$ sector, as we have seen in Fig.~\ref{fig:color_evolution_VA}.
  As a result, all color states emerge during the evolution in Fig.~\ref{fig:color_evolution_Vfull}.
  Similar to the evolution with just the $V_{qg}$ interaction, the probabilities of those states oscillate in the simulation without the phase factor, as in Fig.~\ref{fig:color_evolution_Vfull_a}, and the oscillation is suppressed when the phase factor is restored as in Fig.~\ref{fig:color_evolution_Vfull_b}.
  
\begin{figure}[ht!]
    \centering
    \subfigure [\ $V_{I}(x^+)= V(x^+)$]
    {
      \includegraphics[width=0.4\textwidth]{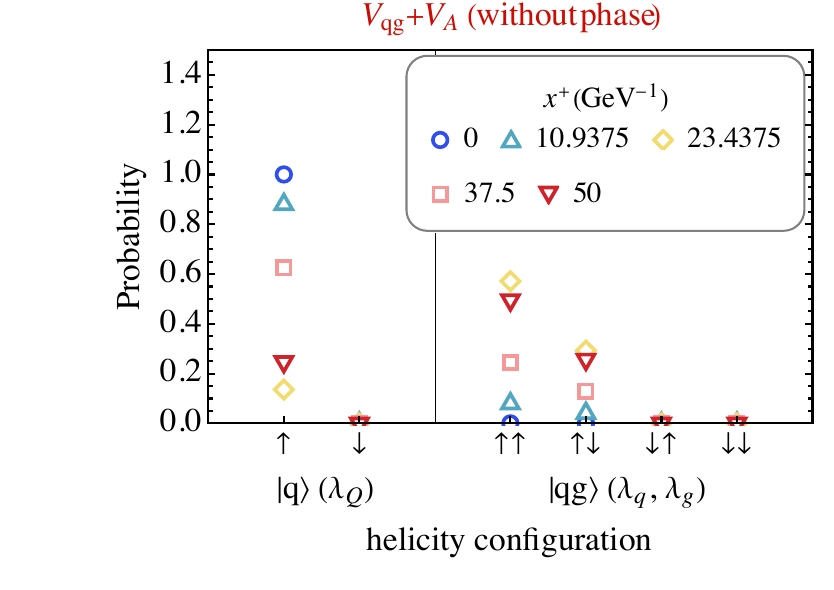}
    } 
    \qquad
    \subfigure [\ $V_{I}(x^+)=e^{i\frac{1}{2}P^-_{KE}x^+} V(x^+) e^{-i\frac{1}{2}P^-_{KE}x^+}$]
    {
      \includegraphics[width=0.4\textwidth]{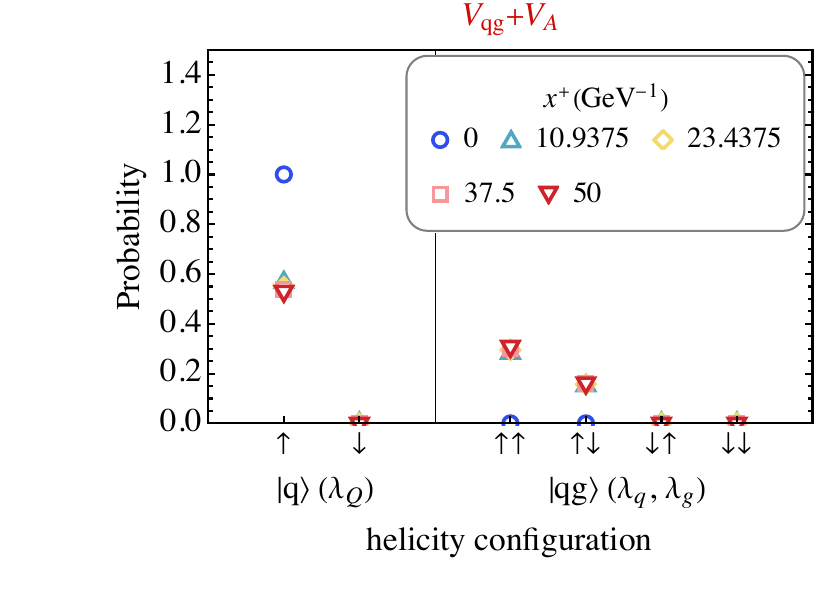}
    } 
    \caption{
    Evolution of the quark state in  light-front helicity phase space.
    The interaction contains both the gluon emission/absorption and the background interaction term  $V(x^+)=V_{qg}+V_{\mathcal{A}}(x^+)$, (a) without the phase factor and (b) with the phase factor.
    The helicity configuration in the $\ket{q}$ sector is labeled by the quark helicity $\lambda_Q=\uparrow,\downarrow$.
    The helicity configuration in the $\ket{qg}$ sector is labeled by the quark helicity $\lambda_q=\uparrow,\downarrow$ and the gluon helicity $\lambda_g=\uparrow,\downarrow$.
      The initial state of the quark is a single quark state with $\vec p_{\perp,Q}=\vec 0_\perp$, helicity $\lambda_Q=1/2$, color $c_Q=2$.
      Parameters in these panels: $N_\perp=16$, $L_\perp=50~\GeV^{-1}$, $m_g=0.1~\GeV$, $g^2\tilde\mu=0.144~\GeV^{3/2}$, $m_q=0.02~\GeV$, $P^+=4.25~\GeV$, $K=8.5$.
      The duration of each background field layer is $\tau=12.5~\GeV^{-1}$, and that of each time step in the simulation is $\delta x^+=0.39~\GeV^{-1}$.
  }
    \label{fig:spin_evolution_Vfull}
  \end{figure}

  The evolution of the quark state in helicity space is shown in Fig.~\ref{fig:spin_evolution_Vfull}. 
  The result is very similar to that in the $V_{qg}$ evolution in Fig.~\ref{fig:spin_evolution_Vfree}, since the change in helicity results from $V_{qg}$ and not from the background field interaction.

  To sum up this section, we have studied the evolution with the full interaction $V_{qg}+V_{\mathcal{A}}$, where the former is in charge of gluon emission/absorption, and the latter term controls the transitions within each of the $\ket{q}$ and the $\ket{qg}$ sector. 
  By adjusting the relative magnitude of the two, one could access different physics regimes.
  From the nonperturbative time evolution, we investigate the combined effects from the full interaction in the quark phase space, including the longitudinal momentum, the transverse momentum, helicity, and color spaces. 
  By adjusting the strength of the background field, one is able to change the relative importance of the gluon emission and absorption, and the color decoherence and momentum broadening due to the background field. 
  Our results overall are consistent with the expectations from having the two different kinds of interactions separately.
  
  \FloatBarrier
  
  \section{Conclusions and outlook}\label{sec:summary}

  In this work, we developed a numerical implementation of the time-evolution Hamiltonian formalism, tBLFQ, for the interactions of a $\ket{q} + \ket{qg}$ system with a target color field.
  Our formulation enables us to access the wave function of the quark at any intermediate time during the evolution, and to continuously tune the relative importance of the interaction with the target field, and of the gluon emissions and absorptions, without taking any parametric limits.

  We carried out explicit time evolutions of the quark as a quantum state inside the background color field. 
  Our calculation enables us to access explicitly the time evolution of the transverse and longitudinal momentum, color, and helicity of the scattering partons.
  The light-front Hamiltonian of our system consists of three parts: the kinetic energy term, which leads to a phase rotation of the state, the interaction with the background field, and the gluon emission/absorption.
  We studied these effects both individually and in combination. 
  The simulations were done for three different cases: the gluon emission/absorption alone, the interaction with the background field alone, and the full interaction that combines the previous two terms. We also compared the processes with and without the phase rotation from the kinetic energy term. Overall, in the limiting cases, the results correspond qualitatively and quantitatively to what one could expect based on general physical arguments, or explicit calculations. We therefore believe that our numerical method is now well tested and robust to be applied to several different physical situations. 
  
   In this paper we have focused on developing and testing the numerical method. In the future, as discussed in the Introduction, our goal is to apply this numerical method to different physical situations, such as jet quenching in a hot plasma and a high-energy scattering with subeikonal effects. 
   In this work, we use a single $\ket{q}$ state or a single $\ket{qg}$ state with definite momentum to study the dynamical process in a simplified yet clean picture. 
   Specific physical applications require initial conditions that are matched to the studied physical system, and calculations of the physical observables that are of interest. 
   For the case of high-energy scattering, one needs as an initial condition a dressed quark state formulated in a way that is consistent with our truncation of the Fock space. In addition to the perturbative calculation of this state, another possibility  would be to solve the eigenvalue equation with the QCD Hamiltonian in our truncated Fock space. 
  In this work, we take the background field of the nucleus as the MV model, and keep the dominant field component at high energy ($\mathcal{A}^-$) in our calculation. For the purposes of understanding subeikonal effects and the role of spin in high-energy scattering, it would be interesting to generalize this to a background field with transverse components~\cite{Cougoulic:2020tbc}. 
  In a separate physical situation from that of high-energy scattering, our calculation provides a systematic way to study the interactions of an energetic parton in a colored medium, which is the situation in jet quenching, when a highly energetic parton interacts with a colored medium. 
  Many calculations of jet quenching are done in the approximation of independent static scattering centers. 
  We hope that our formulation would provide for a way to generalize this and enable an understanding of jet quenching in a more general nonperturbatively strong gluonic field configuration, such as the one provided by the pre-equilibrium glasma fields in the initial stage of a heavy ion collision.

\section*{Acknowledgements}
T. Lappi and M. Li are supported by the Academy of Finland, Project No. 321840 (T. L.) and  under the European Union’s Horizon 2020 Research and Innovation Programme by the European Research Council (ERC, Grant Agreement No. ERC-2015-CoG-681707) and by the STRONG-2020 project (Grant Agreement No 824093). The content of this article does not reflect the official opinion of the European Union and responsibility for the information and views expressed therein lies entirely with the authors. Computing resources from CSC-IT Center for Science in Espoo, Finland and from the Finnish Grid and Cloud Infrastructure (persistent identifier \texttt{urn:nbn:fi:research-infras-2016072533}) were used in this work. 
X. Zhao is supported by new faculty startup funding by the Institute of Modern Physics, Chinese Academy of Sciences, Key Research Program of Frontier Sciences, Chinese Academy of Sciences, Grant No. ZDB-SLY-7020, by the Natural Science Foundation of Gansu Province, China, Grant No. 20JR10RA067 and by the Strategic Priority Research Program of the Chinese Academy of Sciences, Grant No. XDB34000000.

\FloatBarrier
\appendix
\section{Conventions}
\label{app:convention}
The light-front coordinates are defined as \( (x^+, x^-, x^1, x^2) \), where \(x^+=x^0+ x^3\) is the light-front time,  \(x^-=x^0-
x^3\) the longitudinal coordinate, and  \(\vec x_\perp=(x^1, x^2)\) the transverse coordinates.
In this paper, we also use ``$x$" and ``$y$" as the transverse indices, and they should be understood the same as the indices ``$1$" and ``$2$" introduced here.
The nonvanishing elements of the metric tensor are
\begin{align}
  g^{+-}=g^{-+}=2\;, \
  g_{+-}=g_{-+}=\frac{1}{2}\;,\
  g^{11}=g^{22}=-1\;.
\end{align}
The Dirac matrices are four unitary traceless $4 \times 4$ matrices,
\begin{align}
  \begin{split}
  \gamma^0=\beta=
  \begin{pmatrix}
    0&-i\\
    i&0
  \end{pmatrix},
  \quad
  \gamma^+=
  \begin{pmatrix}
    0 & 0\\
    2i & 0
  \end{pmatrix},\\
  \gamma^-=
  \begin{pmatrix}
    0&-2i\\
    0&0
  \end{pmatrix},
  \quad
  \gamma^i=
  \begin{pmatrix}
    -i\hat{\sigma}^i&0\\
    0&i\hat{\sigma}^i
  \end{pmatrix}
  ,
  \end{split}
\end{align}
where,
\begin{align}
  \hat{\sigma}^1=\sigma^2=
  \begin{pmatrix}
    0&-i\\
    i&0
  \end{pmatrix},
       \quad
       \hat{\sigma}^2=-\sigma^1=
       \begin{pmatrix}
         0&-1\\
         -1&0
       \end{pmatrix}.
\end{align}

\section{The light-front Hamiltonian}
\subsection{Derivation of the light-front QCD Hamiltonian with a background field}\label{app:Hamiltonian}
In this section, we derive the light-front QCD Hamiltonian according to Ref.~\cite{Brodsky:1997de} but with an additional background field. 
The QCD Lagrangian with a background field is given in Eq.~\eqref{eq:Lagrangian},
\begin{align}
  \mathcal{L}=-\frac{1}{4}{F^{\mu\nu}}_a F^a_{\mu\nu}+\overline{\Psi}(i\gamma^\mu D_\mu - m_q)\Psi\;.
\end{align}

The equation of motion for the gauge field gives the color-Maxwell equation, 
\begin{align}\label{eq:colorM_v2}
  \partial_\lambda F_s^{\lambda\kappa}=gJ_s^\kappa \;,
\end{align}
with the current density $J_s^\kappa\equiv f^{sac}F^{\kappa\mu}_a C^c_\mu+\overline{\Psi}\gamma^\kappa T^s\Psi$. 
In the light-cone gauge of $A_a^+=\mathcal{A}_a^+=0$, the $\kappa=+$ component of Eq.~\eqref{eq:colorM_v2} does not contain time derivatives and can be written as
\begin{align}
  g J_a^+=\partial_\lambda F_a^{\lambda +}=-\partial^+\partial_-C^-_a-\partial^+\partial_i C^i_a
  \;.
\end{align}
By disregarding the zero modes~\cite{Pauli:1995dt}, one inverts the equation to
\begin{align}\label{eq:Aconstrain_v2}
    \frac{1}{2}A^-_a=-g\frac{1}{{(\partial^+)}^2}J^+_a
    -\frac{1}{\partial^+}\partial_i C^i_a- \frac{1}{2}\mathcal{A}^-_a
    \;.
  \end{align}
We define the free solution $\tilde{A}_a^\mu$ such that $\lim_{g\to 0}A_a^\mu=\tilde{A}_a^\mu$. According to Eq.~\eqref{eq:Aconstrain_v2}, the free field reads
\begin{align}\label{eq:tilde_A}
  \tilde{A}^\mu_a=(0,\tilde{A}^-_a,A^i_a), \quad \text{with}\ \
  \frac{1}{2}\tilde{A}^-_a\equiv\frac{1}{2}A^-_a+g\frac{1}{{(\partial^+)}^2}J^+_a
  \;.
\end{align}

The equation of motion for the fermion field gives the color-Dirac equation,
\begin{align}\label{eq:cDirac2}
  [i\gamma^\mu(\partial_\mu +ig\bm C_\mu)-m_q]\Psi=0
  \;.
\end{align}
We now separate the dynamical component of the fermion field by introducing projectors $\Lambda^\pm=\gamma^0\gamma^\pm/2$. The projected spinors are thereby $\Psi_\pm=\Lambda^\pm \Psi$, and we  obtain a coupled set of spinor equations from Eq.~\eqref{eq:cDirac2},
\begin{align}
  2i\partial_+\Psi_+=(-i\alpha^i \bm D_i+m_q\beta)\Psi_-+2g\bm C_+\Psi_+ \label{eq:psi_1}\;,\\
  2i\partial_-\Psi_-=(-i\alpha^i \bm D_i+m_q\beta)\Psi_++2g\bm C_-\Psi_- \label{eq:psi_2}\;.
\end{align}
Equation~\eqref{eq:psi_2} does not contain time derivatives and can be written as a constraint relation,
\begin{align}\label{eq:psi_constraint}
    \Psi_-=\frac{1}{2i\partial_-}(m_q\beta-i\alpha^i \bm D_i)\Psi_+
    \;.
  \end{align}
By substituting Eq.~\eqref{eq:psi_constraint} into Eq.~\eqref{eq:psi_1}, we get
\begin{align}\label{eq:d-psi+}
  2iD_+\Psi_+=(m_q\beta-i\alpha^i \bm D_i)\frac{1}{2i\partial_-}(m_q\beta-i\alpha^i \bm D_i)\Psi_+
  \;.
\end{align}
In analogy to the free solution $\tilde{A}$, we define the free spinor $\tilde{\Psi}=\tilde{\Psi}_++\tilde{\Psi}_-$ with 
\begin{align}\label{eq:free_spinor}
  \tilde{\Psi}_+=\Psi_+,\quad\tilde{\Psi}_-=\frac{1}{2i\partial_-}(m_q\beta-i\alpha^i \partial_i)\Psi_+
  \;.
\end{align}
It is also easy to see that $\tilde{\Psi}_\pm=\Lambda^\pm \tilde{\Psi}$.
The conjugate momenta are 
\begin{align}
\Pi_{A^s_\kappa}^\lambda=-F_s^{\lambda\kappa},
\quad
\Pi_{\Psi}^\lambda=\frac{i}{2}\overline{\Psi}\gamma^\lambda,
\quad 
\Pi_{\overline{\Psi}}^\lambda=\frac{i}{2}\gamma^\lambda\Psi
\;.
\end{align}

We now turn to the construction of the canonical Hamiltonian density through a Legendre transformation,
\begin{align}\label{eq:Legendre_H}
  \begin{split}
    \mathcal{P}_+=& (\partial_+ A^s_\kappa)\Pi_{A^s_\kappa}^+ +(\partial_+ \Psi)\Pi_{\Psi}^++(\partial_+ \overline{\Psi})\Pi_{\overline{\Psi}}^+-\mathcal{L}\\
    =&-F_s^{+\kappa}\partial_+ A^s_\kappa+\frac{i}{2}[\overline{\Psi}\gamma^+\partial_+ \Psi+h.c.]
    +\frac{1}{4}{F^{\mu\nu}}_a F^a_{\mu\nu}
    \;,
  \end{split}
\end{align}
It is convenient to add a total derivative $-\partial_\kappa(F_s^{\kappa+}A^s_+)$ to the Hamiltonian $P^-=2 P_+$, 
\begin{align}\label{eq:Legendre_Pmn}
  \begin{split}
    P^-=&2 \int\diff x_+\diff^2 x_\perp\ \mathcal{P}_+\\
    =&\int\diff x^-\diff^2 x_\perp\ -F_s^{+\kappa}\partial_+ A^s_\kappa+\frac{i}{2}[\overline{\Psi}\gamma^+\partial_+ \Psi+h.c.]\\
    &+\frac{1}{4}{F^{\mu\nu}}_a F^a_{\mu\nu}-\partial_\kappa(F_s^{\kappa+}A^s_+)
    \;.
  \end{split}
\end{align}
We eliminate the light-front time derivatives of the fields by applying the equations of motions in Eqs.~\eqref{eq:colorM_v2} and~\eqref{eq:cDirac2}, and rewrite the full light-front Hamiltonian in
terms of only the ‘‘tilde’’ variables defined in Eqs.~\eqref{eq:tilde_A} and~\eqref{eq:free_spinor}.
We introduce the current density of free fields solution $\tilde{J}^\mu_a$ in analogy to $J^\mu_a$, 
as $\tilde J_s^\mu\equiv f^{sac}F^{\mu\kappa}_a \tilde C^c_\kappa+\overline{\tilde \Psi}\gamma^\mu T^s\tilde \Psi$, and notice that their ``+" components are the same,
\begin{align}
  \begin{split}
    J_s^+
    = f^{sac}\partial^+C^i_a C^c_i+\overline{\tilde{\Psi}}\gamma^+T^s\tilde{\Psi}
    =\tilde{J}^+_s
    \;.
  \end{split}
\end{align}
Finally, we get the light-front Hamiltonian with the background field as 
\begin{align}
  \begin{split}
    P^-&=
    \int\diff x^-\diff^2 x_\perp
    \bigg\{
    \\
    &
    -\frac{1}{2}C^j_a{(i\nabla)}^2_\perp C_j^a
    +\frac{1}{2}\overline{\tilde{\Psi}}\gamma^+\frac{m_q^2-\nabla_\perp^2}{2i\partial_-}\tilde{\Psi}\\
    &
    -g f^{abc}\partial^i C^j_a C_i^bC_j^c
    +g \tilde J^+_a\tilde{A}^a_+
    +g \tilde J^+_a\mathcal{A}^a_+
    +g\overline{\tilde{\Psi}}\gamma^i \bm{C}_i\tilde{\Psi}\\
    &-\frac{1}{2}g^2 \tilde J^+_a\frac{1}{{(\partial^+)}^2}\tilde J^+_a
    +\frac{g^2}{4}f^{abc}C^i_b C^j_c f^{a e f}C_i^e C_j^f\\
    &
    +\frac{g^2}{2}\overline{\tilde{\Psi}}\gamma^i \bm{C}_i\frac{\gamma^+}{2i\partial_-}\gamma^j \bm{C}_j\tilde{\Psi}
\bigg\}   
\;.
  \end{split}
\end{align}
The two terms in the first line are the kinetic energy for the gauge field, the background field, and the fermion field. 
The four terms in the second line can be written collectively as $g\tilde J^\mu_a C^a_\mu$, which include the three-gluon interaction and the vertex interaction; the latter is responsible for the gluon emission and quark-antiquark-pair-production processes. 
The two terms in the third line are the instantaneous-gluon interaction and the four-gluon interaction respectively. 
The last line contains the instantaneous-fermion interaction. 
For each interaction involving the gluon field, it also involves the background field. 
Since we are interested in the interactions introduced by the background field to the quark but not the dynamics of the background field itself, we thereby neglect the kinetic energy of the background field and its self interaction in this work. 
In the text, we drop the tilde on all variables to simplify the notations, but their meanings are not changed.

\subsection{Spin and polarization}\label{app:spinor}
We use the following spinor representation. The $u$, $v$ spinors are defined as,
\begin{equation}
  \begin{split}
    &u(p,\lambda=\frac{1}{2})=\frac{1}{\sqrt{p^+}}{(p^+,0,im_q,ip^x-p^y)}^\intercal \;,\\
    &u(p,\lambda=-\frac{1}{2})=\frac{1}{\sqrt{p^+}}{(0,p^+,-ip^x-p^y,im_q)}^\intercal \;,\\
    &\bar{u}(p,\lambda=\frac{1}{2})=\frac{1}{\sqrt{p^+}}(m_q,p^x- i p^y,-ip^+,0)\;,\\
    &\bar{u}(p,\lambda=-\frac{1}{2})=\frac{1}{\sqrt{p^+}}(-p^x - i p^y,m_q,0,-ip^+)\;,
  \end{split}
\end{equation}
and
\begin{equation}
  \begin{split}
    &v(p,\lambda=\frac{1}{2})=\frac{1}{\sqrt{p^+}}{(p^+,0,-im_q, ip^x-p^y)}^\intercal \;,\\
    &v(p,\lambda=-\frac{1}{2})=\frac{1}{\sqrt{p^+}}{(0,p^+,-ip^x-p^y, -im_q)}^\intercal \;,\\
    &\bar{v}(p,\lambda=\frac{1}{2})=\frac{1}{\sqrt{p^+}}(-m_q, p^x- i p^y,-ip^+,0)\;,\\
    &\bar{v}(p,\lambda=-\frac{1}{2})=\frac{1}{\sqrt{p^+}}(-p^x - i p^y, -m_q,0,-ip^+)\;.
  \end{split}
\end{equation}

The polarization vectors for gluon are defined as 
\begin{align}
e(k,\lambda=\pm 1)
=\left( 0,\frac{2\bm{\epsilon}^\perp_\lambda\cdot \vec{k}_\perp}{k^+},\bm{\epsilon}^\perp_\lambda
\right)
\;,
\end{align}
where $\bm{\epsilon}^\perp_\pm=(1,\pm i)/\sqrt{2}$. 

The spinor-polarization vector contraction part $\bar{u}(p_Q,\lambda_Q)$ $ \gamma^\mu u(p_q,\lambda_q)\epsilon_\mu(p_g,\lambda_g)$ and its complex conjugates are summarized in Table~\ref{tab:uue} for different helicity configurations. 
\begin{table*}[ht]
  \caption{Spinor-polarization vector contraction for different helicity configurations. For any transverse two-dimensional vector, $\vec p_\perp = (p^x, p^y)$, define $p^R\equiv p^x + i p^y$, and $p^L\equiv p^x-i p^y$.
  As defined in Sec.~\ref{sec:int_basis_qg}, $z\equiv p_g^+/p_Q^+$ is the longitudinal momentum fraction of the gluon, and  $\vec \Delta_m$ is the relative center-of-mass momentum defined in Eq.~\eqref{eq:Dm}.
  }\label{tab:uue}
  \begin{tabular}{c c c}
    \hline\hline\\[-3mm]
  \begin{tabular}{c}Helicity configurations  \\ ($\lambda_Q, \lambda_q, \lambda_g$)\end{tabular}
    & ~~~~~
    $\bar{u}(p_Q,\lambda_Q) 
    \gamma^\mu
    u(p_q,\lambda_q)
    \epsilon_\mu(p_g,\lambda_g)$ ~~~~~~~~~~
    & ~~~~~
    $\bar{u}(p_q,\lambda_q) 
    \gamma^\mu
    u(p_Q,\lambda_Q)
    \epsilon_\mu^*(p_g,\lambda_g)$
    ~~~
    \\
   % [1mm]\\
    \hline\\[-2mm]
    $\uparrow\uparrow\uparrow$    
    & $ \dfrac{\sqrt{2(1-z)}}{z(1-z)} \Delta_m^R$
    & $ \dfrac{\sqrt{2(1-z)}}{z(1-z)} \Delta_m^L$
\\[3mm]
\\
$\uparrow\uparrow\downarrow$    
& $ \dfrac{\sqrt{2(1-z)}}{z(1-z)} (1-z)\Delta_m^L$
& $ \dfrac{\sqrt{2(1-z)}}{z(1-z)} (1-z)\Delta_m^R$
\\[3mm]
\\
$\uparrow\downarrow\uparrow$    
& $\dfrac{\sqrt{2(1-z)}}{z(1-z)} (z^2 m_q)$
& $\dfrac{\sqrt{2(1-z)}}{z(1-z)} (-z^2 m_q)$
\\[3mm]
\\
$\uparrow\downarrow\downarrow$    
& 0
& 0
\\[3mm]
\\
$\downarrow\uparrow\uparrow$    
& 0
& 0
\\[3mm]
\\
$\downarrow\uparrow\downarrow$    
& $\dfrac{\sqrt{2(1-z)}}{z(1-z)} (-z^2 m_q)$
& $\dfrac{\sqrt{2(1-z)}}{z(1-z)} (z^2 m_q)$
\\[3mm]
\\
$\downarrow\downarrow\uparrow$    
& $ \dfrac{\sqrt{2(1-z)}}{z(1-z)} (1-z)\Delta_m^R$
& $ \dfrac{\sqrt{2(1-z)}}{z(1-z)} (1-z)\Delta_m^L$
\\[3mm]
\\
$\downarrow\downarrow\downarrow$    
& $ \dfrac{\sqrt{2(1-z)}}{z(1-z)} \Delta_m^L$
& $ \dfrac{\sqrt{2(1-z)}}{z(1-z)} \Delta_m^R$
\\[2mm]\\
\hline\hline
\end{tabular}
\end{table*}

\subsection{Quantization in a discrete space}\label{app:modes}
We consider that the system is contained in a box of finite volume $\Omega=2L{(2L_\perp)}^2$. 
We have introduced two artificial length parameters, $L$ in the longitudinal direction and $L_\perp$ in transverse directions. In the longitudinal direction, $-L\le x^- \le L$, we impose periodic boundary conditions for bosons and  antiperiodic boundary conditions for fermions such that the longitudinal momentum space is discretized as,
\begin{align}
    &p^+ = 
    \begin{dcases}
        &\frac{2\pi}{L}k^+, ~\text{with}~ k^+ = \frac{1}{2}, \frac{3}{2}, \ldots,\infty 
        ~\text{for fermions}\;, \\
        &\frac{2\pi}{L}k^+, ~\text{with}~ k^+ = 1,2, \ldots,\infty
        ~\text{for bosons}\;.
    \end{dcases}
\end{align}
In the transverse dimension,  $-L_\perp \le x^1, x^2 \le L_\perp$, we impose the periodic boundary conditions and discretize the space into $2 N_\perp\times 2 N_\perp$ grids.
The corresponding momentum space is also discrete with periodic boundary conditions,
\begin{align}
   & p^i = \frac{2\pi}{2L_\perp} k^i,  ~\text{with}~ k^1, k^2 = -N_\perp, -N_\perp+1 \ldots, N_\perp-1 \;.
\end{align}

The conversion of the integration is
  \begin{align}
    \int\frac{\diff^2 \vec p_\perp}{{(2\pi)}^2} \to \frac{1}{{(2L_\perp)}^2}\sum_{k_1, k_2}\;,
    \qquad
    \int \diff^2 \vec r_\perp \to a_\perp^2\sum_{n_1, n_2}\;.
  \end{align}
 The Dirac delta is converted to the Kronecker delta as follows:
  \begin{align}
    \begin{split}
    &\int \diff^2 \vec r_\perp e^{-i \vec p_\perp\cdot \vec x_\perp}
    = {(2\pi)}^2\delta^2(\vec p_\perp)\ \\
    &\to
    \sum_{n_1, n_2}  a^2 e^{-i (n_1 k_1 + n_2 k_2)\pi/N_\perp} = {(2L_\perp)}^2 \delta_{k_1,0}\delta_{k_2,0}\;,
    \end{split}
  \end{align}
  and
  \begin{align}
    \begin{split}
      &\int \diff^2 \vec p_\perp e^{i \vec p_\perp\cdot \vec x_\perp}={(2\pi)}^2\delta^2(\vec r_\perp)\ \\
      & \to
      \sum_{k_1, k_2} \frac{1}{{(2L_\perp)}^2}  e^{i (n_1 k_1 + n_2 k_2)\pi/N_\perp} =  \frac{1}{a_\perp^2}\delta_{n_1,0}\delta_{n_2,0}\;.
    \end{split}
  \end{align}
  The (inverse-)Fourier transformation becomes
  \begin{align}
    \begin{split}
    f(n_1, n_2)= & \frac{1}{ {(2 L_\perp)}^2} \sum_{k_1, k_2} e^{i(n_1 k_1 + n_2 k_2)\pi/N_\perp}\tilde f(k_1, k_2),\\
    \tilde f(k_1, k_2)  = & \sum_{n_1, n_2} a^2 e^{-i(n_1 k_1 + n_2 k_2)\pi/N_\perp}  f(n_1, n_2)\;.
    \end{split}
  \end{align}

The mode expansion for field operators on such a discrete momentum basis is
\begin{align}
  \begin{split}
 & \Psi^{\text{Box}}_c(x)=\sum_{\bar{\alpha}} \frac{1}{\sqrt{ p^+ 2L (2L_\perp)^2}}\\
 &\quad \quad \times [b_{\bar{\alpha},c} u(p,\lambda)
   e^{-ip\cdot x}+d^\dagger_{\bar{\alpha},c} v(p,\lambda) e^{ip\cdot x}]\;,   
  \end{split}
  \end{align}
\begin{align}
    \begin{split}
 &A^{\text{Box}}_{\mu,a}(x)=\sum_{\bar{\alpha}} \frac{1}{\sqrt{ p^+ 2L (2L_\perp)^2}}\\
  &\quad \quad \times [a_{\bar{\alpha},a}\epsilon_\mu(p,\lambda)e^{-ip\cdot x}+a^\dagger_{\bar{\alpha},a}\epsilon_\mu^*(p,\lambda)e^{ip\cdot x}]   \;,   
    \end{split}
\end{align}
  where $p\cdot x=p^+ x^-/2 -\vec p_\perp\cdot \vec x_\perp$ is the 3-product for the spatial components of $p^\mu$ and $x^\mu$.
Each single particle state is specified by five quantum numbers, $\bar{\alpha}=\{ k^+, k^1, k^2, \lambda \}$ and $c$ ($a$), where $\lambda$ is the light-front helicity, and $c$ ($a$) is the color index.
Note that this is the same with the basis number $\beta = \{\bar{\alpha},c\}$ defined in our basis representation.
The creation operators $b^{\dagger}_{\bar{\alpha},c}$, $d^{\dagger}_{\bar{\alpha},c}$ and $a^{\dagger}_{\bar{\alpha},a}$
create quarks, antiquarks, and gluons with their corresponding quantum numbers, respectively. 
They obey the following commutation and anticommutation relations:
\begin{align}
    \begin{split}
 & \{b_{\bar{\alpha},c},b^{\dagger}_{\bar{\alpha}',c'}\}=\{d_{\bar{\alpha},c},d_{\bar{\alpha}',c'}^{\dagger}\}
=\delta_{\bar{\alpha},\bar{\alpha}'}
  \delta_{c,c'}\;,\\
 & [a_{\bar{\alpha},a},a^{\dagger}_{\bar{\alpha}',a'}]=\delta_{\bar{\alpha},\bar{\alpha}'}
  \delta_{a,a'}  \;.  
    \end{split}
\end{align}
The fields obey the standard equal-light-front-time commutation relations, and here we write it out  for the dynamical fields:
  \begin{multline}
      \{\Psi^{\text{Box}}_{+,c}(x), \Psi^{\dagger \text{Box}}_{+,c'}(y)\}_{x^+=y^+}
  =\\
  \Lambda^+ \delta(x^- - y^-) \delta^2(\vec x_\perp - \vec y_\perp)\delta_{c,c'}\;,  
  \end{multline}
  in which $ \Lambda^+ =\gamma^0\gamma^+/2$ is the same light-front projector introduced in Appendix.~\ref{app:Hamiltonian}, and
  \begin{multline}
      [A^{\text{Box}}_{i, a}(x), A^{\dagger \text{Box}}_{j,b}(y)]_{x^+=y^+}
      =\\-\frac{i}{4} \epsilon(x^- - y^-) \delta^2(\vec x_\perp - \vec y_\perp)\delta_{i,j} \delta_{a,b}
      \;,
    \end{multline}
with $i,j=1,2$, and $\epsilon(x)$ is the sign function.

A single quark basis state is defined as 
\begin{align}
  \ket{\beta_{q}(k_q^+, k^1_q, k^2_q, \lambda_q, c_q)}
  =b^\dagger_{k_q^+, k^1_q, k^2_q, \lambda_q, c_q}\ket{0}\;.
\end{align}
The basis in the transverse coordinate space that is related to it through Fourier transformation is defined as
\begin{multline}
  \ket{\bar{\beta}_{q}(k_q^+, n^1_q, n^2_q, \lambda_q, c_q)}
  = \\
  \sum_{k^1_q, k^2_q} 
  e^{i(n^1_q k^1_q + n^2_q k^2_q)\pi/N_\perp}
  b^\dagger_{k_q^+, k^1_q, k^2_q, \lambda_q, c_q}\ket{0}\;.    
\end{multline}
We define single gluon basis states as
\begin{align}
  \ket{\beta_{g}(k_g^+, k^1_g, k^2_g, \lambda_g, c_g)}
  =a^\dagger_{k_g^+, k^1_g, k^2_g, \lambda_g, c_g}\ket{0}\;,
\end{align}
and
\begin{multline}
  \ket{\bar{\beta}_{g}(k_g^+, n^1_g, n^2_g, \lambda_g, c_g)}
  = \\
  \sum_{k^1_g, k^2_g} 
  e^{i(n^1_g k^1_g + n^2_g k^2_g)\pi/N_\perp}
  a^\dagger_{k_g^+, k^1_g, k^2_g, \lambda_g, c_g}\ket{0}\;.    
\end{multline}

\section{Momentum transfer on the periodic lattice}\label{app:periodic}
In our calculation, we are working on a discretized transverse lattice in coordinate space. This means that we have a finite momentum space lattice with periodic boundary conditions. The interaction matrix elements for gluon emission and absorption depend on momentum differences between particles. The periodicity of the lattice means that there are several ways to calculate the sum or difference between two momentum vectors, depending on which copy of the periodic lattice one uses. 
One option for resolving this ambiguity would be to fully embrace the lattice discretization and write down the light-front Hamiltonian in discrete space, imposing periodic boundary conditions. This would, however, lead to the presence of high transverse momentum, low-energy fermion doubler modes~\cite{Nielsen:1981hk,Nielsen:1980rz,Faessler:2004yf}. 
Since the dynamics of our system crucially depends on the phase factors determined by the single particle light-cone energies, which we treat in momentum space for numerical efficiency as discussed in Sec.-~\ref{sec:basistimeevol}, such doublers would be unacceptable. 
Making the doubler modes energetic e.g. by adding a Wilson term~\cite{Wilson:1974sk} would be an interesting avenue of investigation for the future. Here, we instead pursue an approach where we evaluate the interaction matrix elements in the continuum. 
We then choose a physically motivated prescription for calculating the value of the transverse momentum differences that these matrix elements depend on from the momenta of the three partons participating in the splitting or merging. 
Our prescription is based on the principle that we want to maintain an accurate description of the physics at small momenta near the center of the Brillouin zone, acknowledging that the interactions of the modes near the transverse UV cutoff will in any case be affected by the discretization and cannot be treated as accurately as the low momentum ones.  
In this Appendix, we specify how this treatment of the periodicity is defined. 

Due to the periodicity a particle with a transverse momentum quantum number $k_l^i \  (i=x,y)$ is equivalent to that with $k_l^i\pm 2 N_\perp$. 
On the lattice, which is inside the fundamental Brillouin zone, the allowed modes are $k_l^i = -N_\perp, -N_\perp+1,\ldots, N_\perp -1$.
This artifact in potential brings ambiguities when dealing with large momentum modes near the boundaries. 
The concrete question is, for example, how to deal with the situation when the two momenta of particles in the $\ket{qg}$ state, each within the fundamental Brillouin zone, add up to a momentum for the $\ket{q}$ state outside it. 
In such a situation, one must decide which copy of the quark and gluon momenta to use to calculate the momentum difference in the matrix elements of the splitting/merging processes. 

For simplicity, we discuss the one-dimensional case in the following. The same procedure is applied separately to both the $x$ and the $y$ dimensions in the transverse plane.
Consider the transition between a $\ket{q}$ state and a $\ket{qg}$ state. 
To distinguish the two quarks in the initial and in the final states, we use ``$Q$'' to denote the quark in the $\ket{q}$ state and ``$q$'' to denote the quark in the $\ket{qg}$ state.
In addition, we use ``$g$'' to denote the gluon in the $\ket{qg}$ state.
The transverse momentum quantum numbers of the three particles are $k_Q$, $k_q$, and $k_g$, respectively, and each of them is within the lattice range $[-N_\perp, N_\perp -1]$. 
The total momentum of the $\ket{qg}$ state, $k_q +k_g$, is therefore in the range of $[-2N_\perp, 2N_\perp -2]$, exceeding the fundamental Brillouin zone.
Considering the periodic boundary condition, the momentum conservation is satisfied if either $k_q + k_g=k_Q$ or $ k_q + k_g=k_Q\pm 2 N_\perp$. 
As a consequence, a process on the lattice specified by the values of $k_q, k_g, k_Q$ could represent two or more different physical processes.

The transition amplitude depends on the transferred momenta, $\Delta_q\equiv k_q-k_{tot}$ and $\Delta_g\equiv k_g-k_{tot}$, where $k_{tot}$ is the total momentum quantum number.
The ambiguity rises when choosing $k_{tot}$ either as $k_q + k_g$ or as $k_Q$, with $k_q,k_g$ and  $k_Q$ always in the fundamental Brillouin zone.
We resolve this ambiguity by making consistent choices in matching the physical process and the process calculated on the lattice.
Let us first look into each of the three momentum-conserved cases of the $qg\leftrightarrow Q$ transition, i.e., $k_q + k_g=k_Q$, $ k_q + k_g=k_Q+ 2 N_\perp$, and $ k_q + k_g=k_Q- 2 N_\perp$, separately.

 \begin{figure}[tbp!]
    \centering
    \subfigure [\ Particles on the lattice ]
    {
      \includegraphics[width=0.45\textwidth]{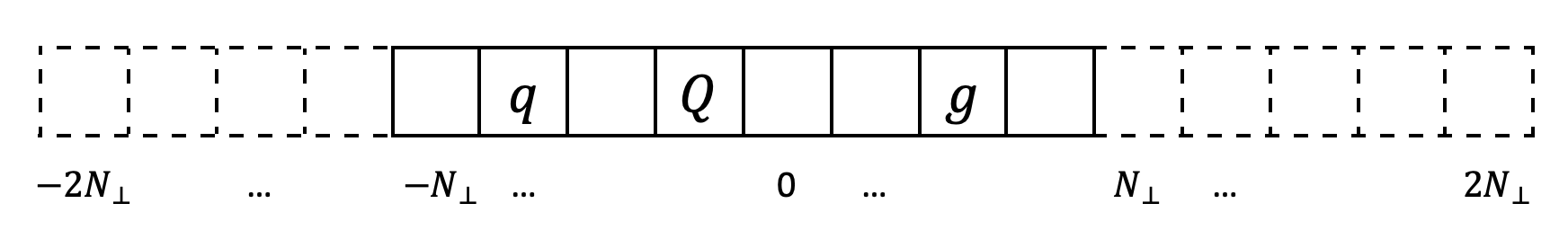}
    } 
    \subfigure [\ After applying the periodic boundary condition ]
    {
      \includegraphics[width=0.45\textwidth]{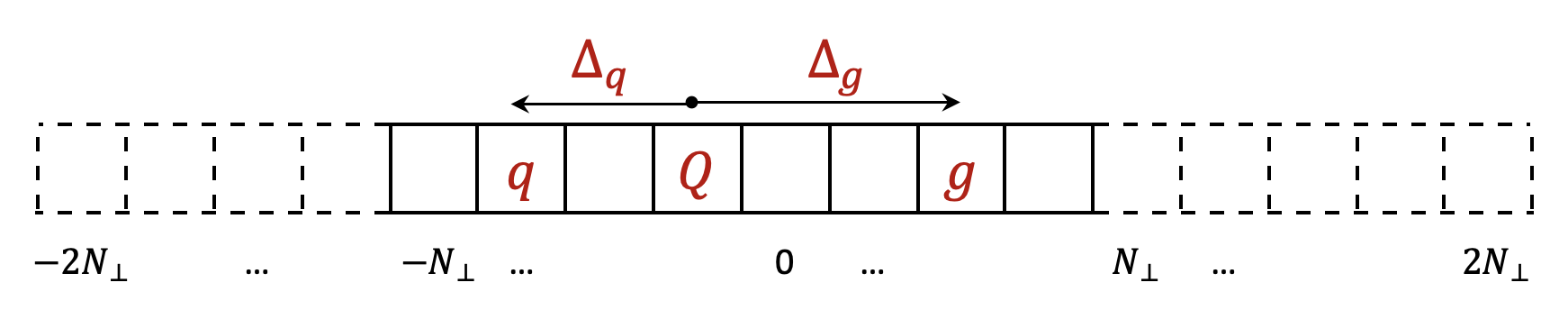}
    } 
    \caption{ 
    An example of a quark (denoted as ``$q$'') and a gluon (denoted as ``$g$'') transferring into/from a quark (denoted as ``$Q$") with their momenta satisfying $k_q + k_g = k_Q$. This is a nonproblematic case when we do not need to worry about the periodicity. 
   The grids inside the fundamental Brillouin zone, i.e., those with momentum numbers in the range of $[-N_\perp, N_\perp -1]$, are in solid lines.
   The grids outside this range are in dashed lines.
   In (a), particles are marked at their momentum quantum numbers assigned on the lattice. 
   In (b), particles are marked at their momentum quantum numbers used to calculate the transferred momenta $\Delta_q$ and $\Delta_g$. 
    }
    \label{fig:qgQ_pt_1}
  \end{figure}

  \begin{figure}[tbp!]
    \centering
    \subfigure [\ Particles on the lattice ]
    {\label{fig:qgQ_pt_2a}
      \includegraphics[width=0.45\textwidth]{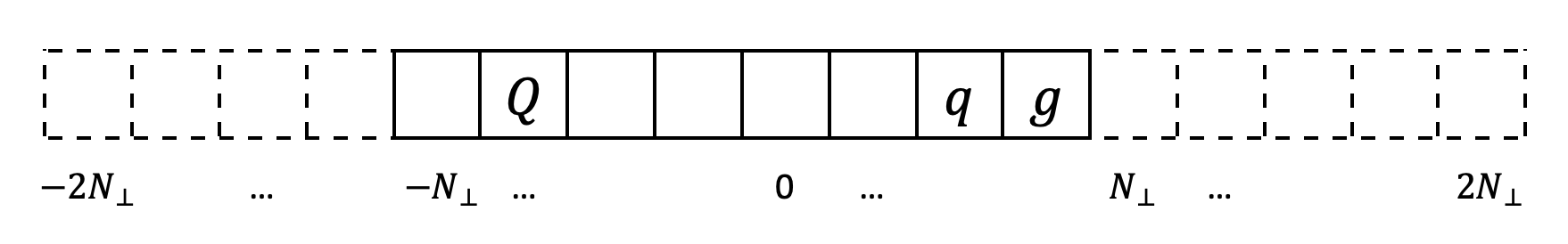}
    } 
    \subfigure [\ After applying the periodic boundary condition according to our chosen prescription]
    {\label{fig:qgQ_pt_2b}
      \includegraphics[width=0.45\textwidth]{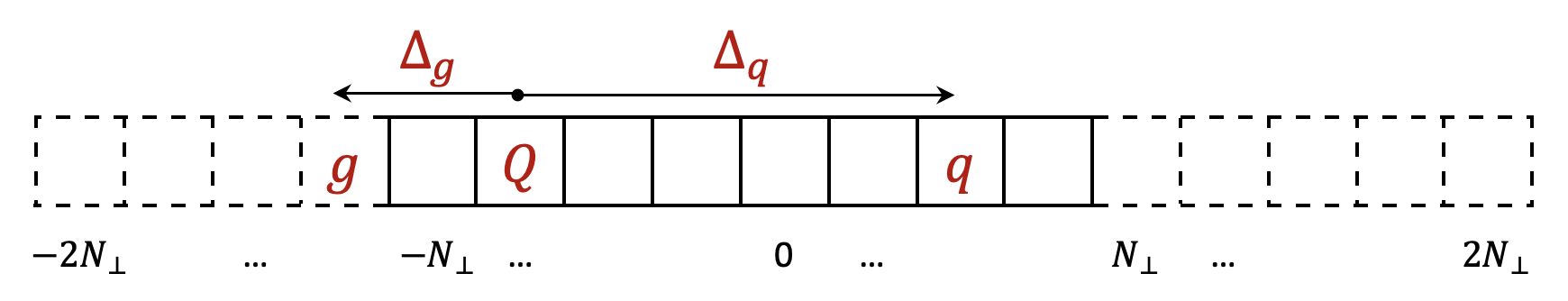}
    } 
    \subfigure [\ An alternative choice of applying the periodic boundary condition ]
    {\label{fig:qgQ_pt_2c}
      \includegraphics[width=0.45\textwidth]{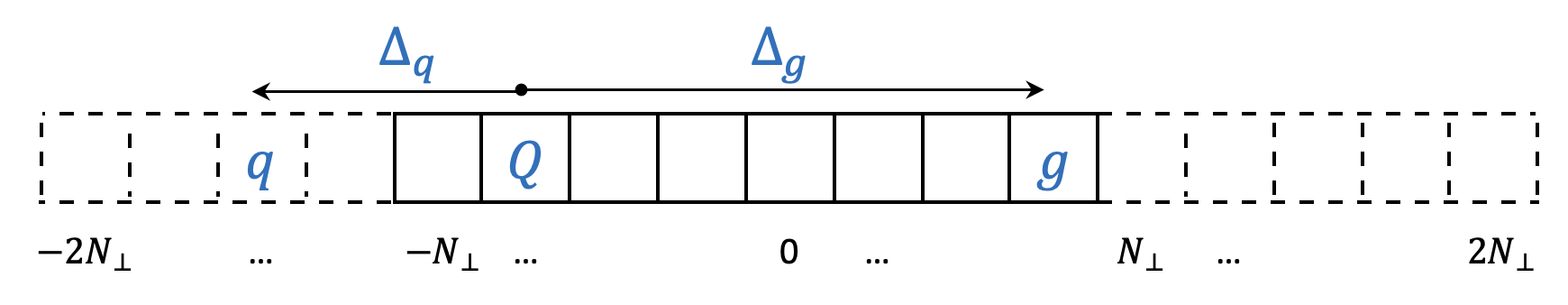}
    } 
     \subfigure [\ Another choice of applying the periodic boundary condition ]
    {\label{fig:qgQ_pt_2d}
      \includegraphics[width=0.45\textwidth]{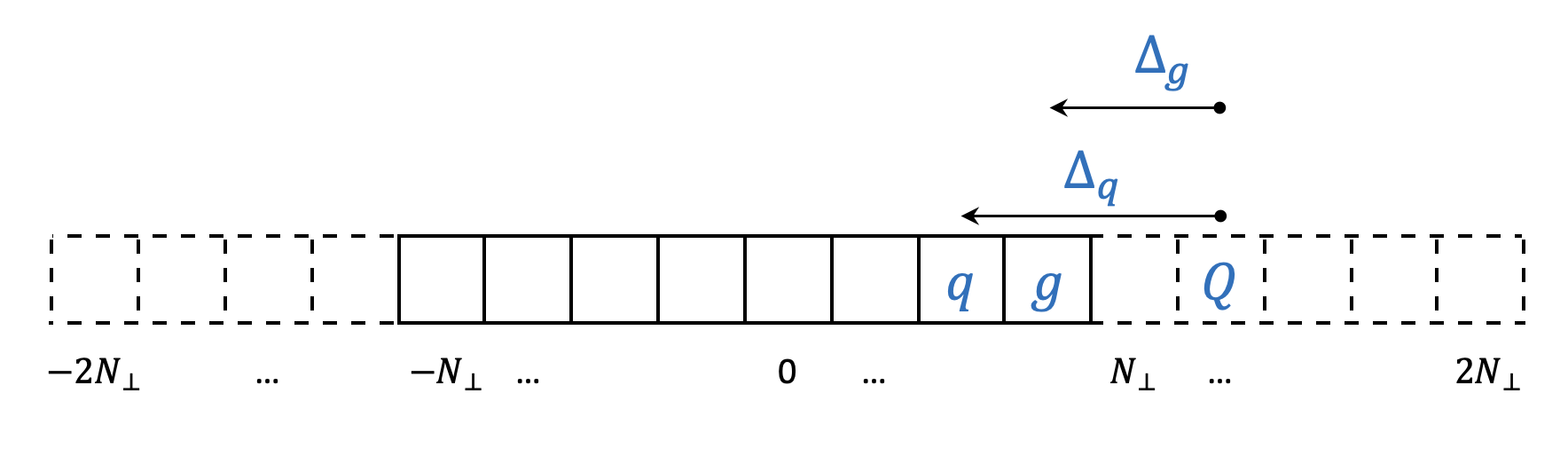}
    } 
    \caption{
    An example of a quark (denoted as ``$q$") and a gluon (denoted as ``$g$") transferring into/from a quark (denoted as ``$Q$") with their momenta satisfying $k_q + k_g = k_Q+2N_\perp$.
    The grids inside the fundamental Brillouin zone, i.e., those with momentum numbers in the range of $[-N_\perp, N_\perp -1]$, are in solid lines.
  The grids outside this range are in dashed lines.
  In (a), particles are marked at their momentum quantum numbers assigned on the lattice. 
  In (b), particles are marked at their momentum quantum numbers used to calculate the transferred momenta $\Delta_q$ and $\Delta_g$. 
    In (c) and (d), two other choices in applying the periodic boundary conditions are shown.
    }
    \label{fig:qgQ_pt_2}
  \end{figure}

\begin{figure}[tbp!]
    \centering
    \subfigure [\ Particles on the lattice ]
    {\label{fig:qgQ_pt_3a}
      \includegraphics[width=0.45\textwidth]{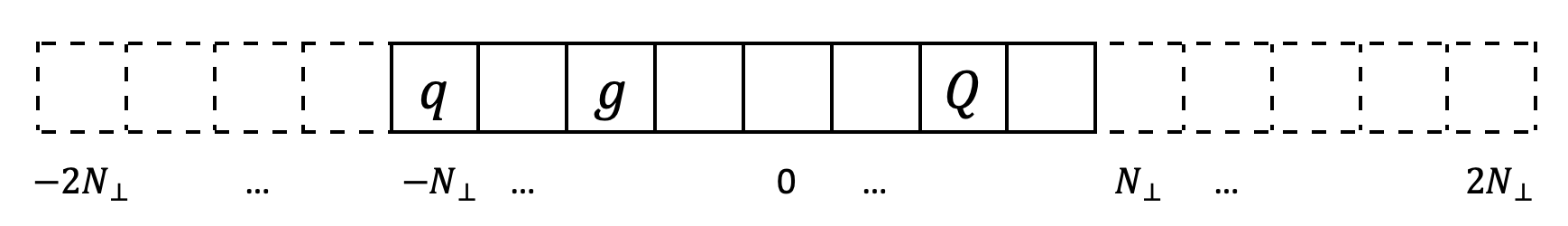}
    } 
    \subfigure [\ After applying the periodic boundary condition ]
    {\label{fig:qgQ_pt_3b}
      \includegraphics[width=0.45\textwidth]{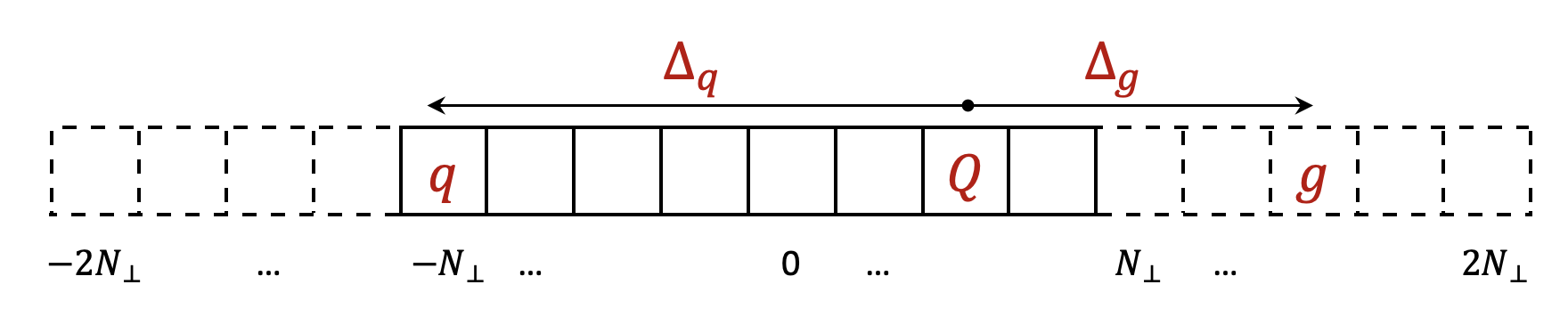}
    } 
     \subfigure [\ An alternative choice of applying the periodic boundary condition ]
    {\label{fig:qgQ_pt_3c}
      \includegraphics[width=0.45\textwidth]{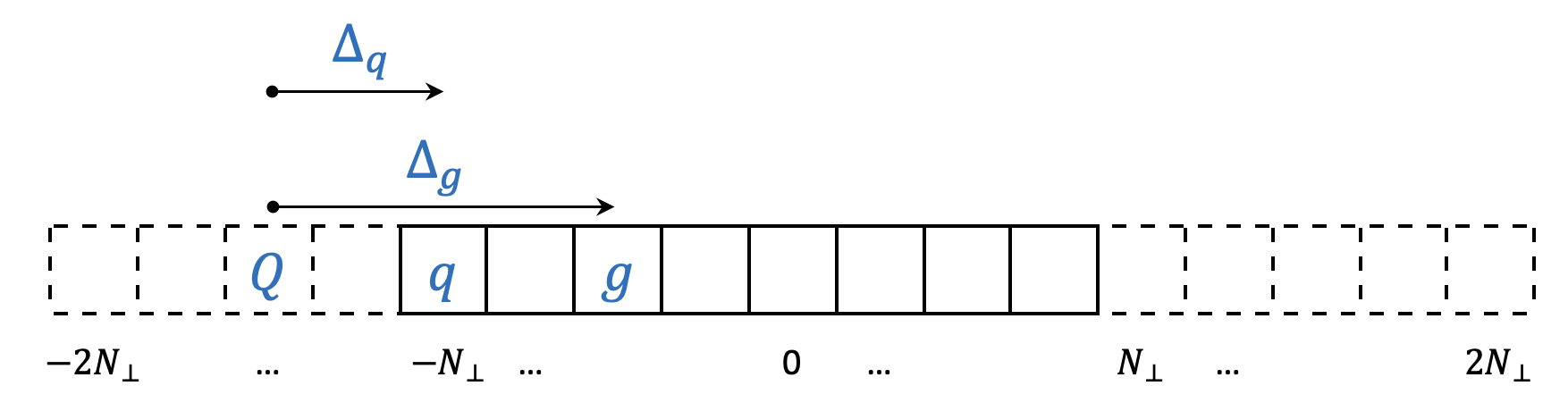}
    } 
    \subfigure [\ Another choice of applying the periodic boundary condition ]
    {\label{fig:qgQ_pt_3d}
      \includegraphics[width=0.45\textwidth]{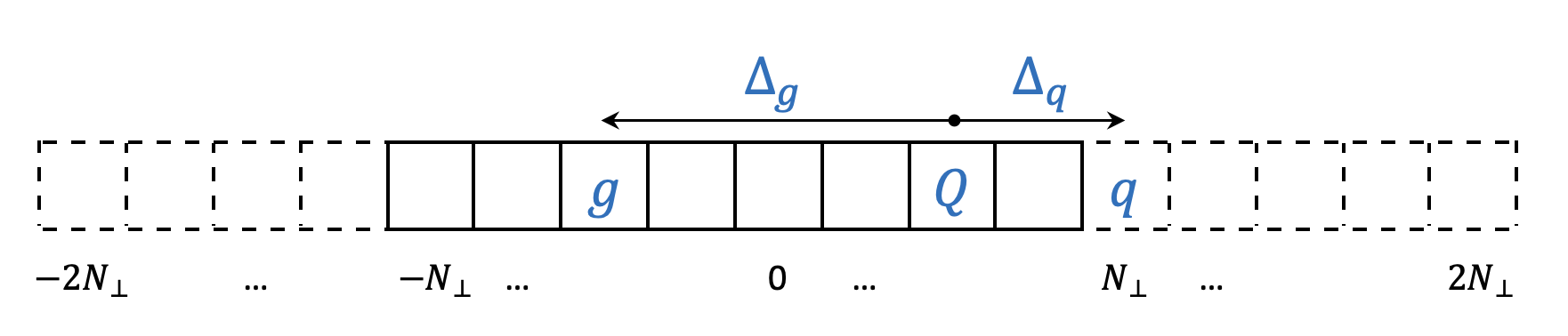}
    } 
    \caption{
    An example of a quark (denoted as ``$q$") and a gluon (denoted as ``$g$") transferring into/from a quark (denoted as ``$Q$") with their momenta satisfying $k_q + k_g = k_Q-2N_\perp$.
    The grids inside the fundamental Brillouin zone, i.e., those with momentum numbers in the range of $[-N_\perp, N_\perp -1]$, are in solid lines.
    The grids outside this range are in dashed lines.
    In (a), particles are marked at their momentum quantum numbers assigned on the lattice. 
    In (b), particles are marked at their momentum quantum numbers used to calculate the transferred momenta $\Delta_q$ and $\Delta_g$. 
    In (c) and (d), two other choices in applying the periodic boundary conditions are shown.
    }
    \label{fig:qgQ_pt_3}
  \end{figure}

\begin{enumerate}
  \item $k_q + k_g = k_Q$\\
 Since the sum $k_q + k_g$ is already inside the lattice range $[-N_\perp, N_\perp -1]$, we take $k_{tot}=k_q + k_g = k_Q$ directly and calculate the transferred momenta as $\Delta_q\equiv k_q-k_{tot}$ and $\Delta_g\equiv k_g-k_{tot}$.
 An example is shown in Fig.~\ref{fig:qgQ_pt_1}.
  \item $ k_q + k_g = k_Q+ 2 N_\perp$\\
  In this situation, the sum $k_q + k_g$ exceeds the positive boundary of the lattice.
  This could happen when both $k_q$ and $k_g$ are large and positive, as the example illustrated in Fig.~\ref{fig:qgQ_pt_2a}.
  There could be more than one choice in applying the periodic boundary conditions to the momentum quantum numbers.
  We choose to bring the gluon to the opposite direction as $k_g\to k_g-2N_\perp$.
  Therefore we calculate the transferred momenta as $\Delta_q\equiv k_q-k_Q$ and $\Delta_g\equiv k_g-(k_q + k_g)$.
  This prescription is shown in Fig.~\ref{fig:qgQ_pt_2b}. 
  The corresponding physical process is a quark and a gluon carrying large but opposite momenta transforming into/from a quark carrying a small momentum.
 There could be alternative ways in applying the periodic boundary conditions, as shown in Figs.~\ref{fig:qgQ_pt_2c} and \ref{fig:qgQ_pt_2d}.  
The process shown in Fig.~\ref{fig:qgQ_pt_2c} is obtained by bringing the quark $q$ one period below, $k_q\to k_q-2N_\perp$. In this interpretation, a quark and a gluon, carrying opposite momentum, transfer into/from a quark carrying a small momentum.
Differently, the process shown in Fig.~\ref{fig:qgQ_pt_2d} is obtained by bringing the quark $Q$ one period above,  $k_Q\to k_Q+2N_\perp$. In this interpretation, a quark and a gluon, each carrying a positive momentum, transfer into/from a quark carrying a larger positive momentum.
  \item $ k_q + k_g = k_Q- 2 N_\perp$\\
  This is very similar to the previous situation where $ k_q + k_g = k_Q+ 2 N_\perp$.
  This could happen when both $k_q$ and $k_g$ are large and negative, as the example illustrated in Fig.~\ref{fig:qgQ_pt_3a}.
  We choose to bring the gluon to the opposite direction as $k_g\to k_g+2N_\perp$ [see Fig.~\ref{fig:qgQ_pt_3b}].
  Therefore, we calculate the transferred momenta as $\Delta_q\equiv k_q-k_Q$ and $\Delta_g\equiv k_g-(k_q + k_g)$.
  Two alternative ways in applying the periodic boundary conditions are shown in Figs.~\ref{fig:qgQ_pt_3c} and \ref{fig:qgQ_pt_3d}.  
\end{enumerate}

Our choices for all three cases discussed above can be summarized into one as 
\[\Delta_q= k_q-k_Q, \qquad \Delta_g= k_g-(k_q + k_g)\;.\]
It is generalized to the two-dimensional transverse space in Eq.~\eqref{eq:Dq_Dg}. 
With this prescription, we could, on the lattice of one fundamental Brillouin zone, maintain the interpretation of back-to-back splitting and merging, which is physically the most significant process in the $qg\leftrightarrow Q$ transition.

\section{The fourth-order Runge-Kutta method}\label{app:RK4}
In solving initial value problems for ordinary differential equations, the Runge-Kutta method takes ``trial" steps between the beginning and the ending points, then uses the values at those ``trial" points to compute the ``real" step across the whole interval.
The fourth-order Runge-Kutta (RK4) method uses symmetrization to cancel out errors up to $O((\delta x^+)^4)$.
In this section, we write out the RK4 simulation in Eq.~\eqref{eq:M_qg} explicitly.
Note that Eq.~\eqref{eq:M_qg} is in the matrix form of the basis representation, and here we take its operator form, $V_{qg,I}$ in place of $\mathcal{V}_{qg,I}$,
\begin{multline}
\ket{\psi; x^+ + \delta x^+}_I\\
  =
  U_{RK4} 
  \left(
  -\frac{i}{2} 
    V_{qg,I}; x^+, x^+ + \delta x^+
  \right)
  \ket{\psi; x^+}_I 
  \;,
\end{multline}
where $V_{qg,I}(x^+) =  
e^{i\frac{1}{2}P^-_{KE}x^+}
V_{qg} 
e^{-i\frac{1}{2}P^-_{KE}x^+} $.
The $U_{RK4}$ operation consists of a sequence of evaluations,
\begin{align}
  \begin{split}
  & k_1 = \delta x^+ \left[-i 
  V_{qg,I}(x^+) / 2\right] \ket{\psi;x^+}_I
  \;,\\
   & \ket{\psi; x^+ + 
   \delta x^+/2 }^1_I
  =\ket{\psi;x^+}_I + \frac{1}{2}k_1
  \;,\\
  & k_2 = \delta x^+\left[-i 
  V_{qg,I}(x^+  +
  \delta x^+/2) / 2\right] \ket{\psi;x^+ +
  \delta x^+/2}^1_I
  \;,\\
  & \ket{\psi; x^+ +
  \delta x^+/2}^2_I
  =\ket{\psi;x^+}_I + \frac{1}{2}k_2
  \;,\\
  & k_3 = \delta x^+ \left[-i
  V_{qg,I}(x^+  + 
  \delta x^+/2) / 2\right] \ket{\psi;x^+ + 
  \delta x^+/ 2 }^2_I
  \;,\\
  &\ket{\psi; x^+ + \delta x^+}^1_I
  =\ket{\psi; x^+}_I + k_3
  \;,\\
  & k_4 = \delta x^+ \left[-i 
  V_{qg,I}(x^+  +  \delta x^+)/ 2 \right] \ket{\psi;x^+ + \delta x^+}^1_I
  \;,\\
  &\ket{\psi; x^+ + \delta x^+}_I
  =\ket{\psi; x^+}_I + \frac{k_1}{6}+ \frac{k_2}{3}+ \frac{k_3}{3}+ \frac{k_4}{6}
  \;.
  \end{split}
\end{align}
Here, the states with the superscript 1 or 2 are the ``trial" states that are evaluated at the midpoint and the end point.

\begin{figure}[ht]
  \centering
\includegraphics[width=0.4\textwidth]{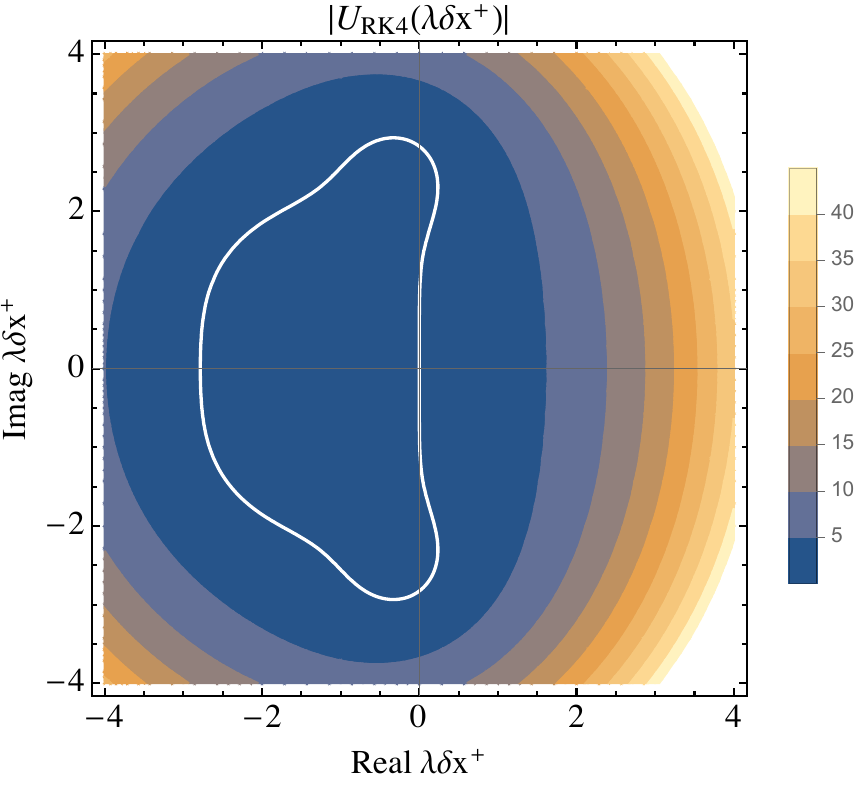}
  \caption{$|U_{RK4} (\lambda \delta x^+) |$ plotted as a function $\lambda \delta x^+$ on the complex plane.  The contour $|U_{RK4} (\lambda \delta x^+) |=1$ is plotted in the white line, and it is the stability boundary for the RK4 algorithm.}
  \label{fig:RK4_stability}
\end{figure}

In the high-energy limit of $P^+\to\infty$, $V_{qg,I}(x^+) $ loses its dependence on the light-front time and reduces to $V_{qg}$. In this case, we could write $U_{RK4}$ in a collective form. By defining $\lambda \equiv - i/2 V_{qg}$, the Runge-Kutta algorithm reduces to
\begin{align}
  \begin{split}
  U_{RK4}&  (\lambda \delta x^+) \\
  =&1
  +\lambda \delta x^+
  +\frac{1}{2} (\lambda \delta x^+)^2
  +\frac{1}{6} (\lambda \delta x^+)^3
  +\frac{1}{24} (\lambda \delta x^+)^4
  \;.
  \end{split}
\end{align}
To see the stability of this method, we can plot $|U_{RK4} (\lambda \delta x^+) |$ in the complex plane of $\lambda \delta x^+$. The stability boundary defined by the contour $|U_{RK4} (\lambda \delta x^+) |=1$ is shown in Fig.~\ref{fig:RK4_stability}. Note that in this case $\lambda$ is effectively purely imaginary, and one sees from the plot that the method is very close to unitary for a large range of $\Im[\lambda\delta x^+]$.

\section{The eikonal limit of the Wilson line}\label{app:Wilsonline}
In this Appendix, we derive the Wilson line of a quark-gluon state in the eikonal limit and discuss its behavior with regard to the total scattering cross section. 

To begin with, consider a quark or a gluon propagating through the background of a classical color field. 
In the eikonal limit, the momentum of the particle is approximated as $P^\mu=(P^+\approx\sqrt{s},P^-=0,P_\perp=0)$ and likewise for the background field $P_{\mathcal{A}}^\mu=(P_{\mathcal{A}}^+=0,P_{\mathcal{A}}^- \approx\sqrt{s},P_{\mathcal{A},\perp}=0)$. 
In such circumstances, the interaction Hamiltonian in the interaction picture is equivalent to that in the Schr\"odinger picture, $V_I(x^+)=V(x^+)$, since the phase factor $e^{\pm i/2P^- x^+}$ reduces to 1.
The evolution of the quark interacting with the background field for a finite distance in light-front time, $x^+=[0,L_\eta]$, is written in terms of a fundamental Wilson line,
\begin{align}
  U_F(0,L_\eta; \vec x_\perp)\equiv\mathcal{T}_+\exp\bigg(
  -i g\int_{0}^{L_\eta}\diff x^+\mathcal{A}_a^-(\vec x_\perp, x^+)t^a
  \bigg)\;,
\end{align}
where $\mathcal{A}_\mu = \sum_a t^a \mathcal{A}_\mu^a$ and $t^a$ are the $\mathrm{SU}(3)$ generators in the fundamental representation. 
Similarly, the evolution of the gluon in the eikonal limit is described by the adjoint Wilson line,
\begin{align}
  U_A(0,L_\eta; \vec x_\perp)\equiv\mathcal{T}_+\exp\bigg(
  -i g\int_{0}^{L_\eta}\diff x^+\mathcal{A}_a^-(\vec x_\perp, x^+)T^a
  \bigg)\;,
\end{align}
where $\mathcal{A}_\mu = \sum_a T^a \mathcal{A}_\mu^a$ and $(T^a)_{bc}=-if^{abc}$ are the $\mathrm{SU}(3)$ generators in the adjoint representation. 

Next, we consider the scattering of a quark-gluon state, in which the quark and the gluon interact with the same background field simultaneously. 
The scattering amplitude is simply the tensor product of the quark and the gluon Wilson lines,
\begin{align}
U_{qg}(0,L_\eta; \vec x_\perp,\vec y_\perp)=
U_F(0,L_\eta; \vec x_\perp)\otimes U_A(0,L_\eta; \vec y_\perp)  
\;.
\end{align}

Physical observables such as the cross section could be determined from the Wilson line averaged over the background field configurations, which is essentially the scattering amplitude.
Note that the dimension of the Wilson line is the same as that of the particle's color space. In calculating the total scattering cross section of the particle state $l$, one should sum over the final color states and average over the initial color states,
\begin{equation}\label{eq:crss_trace}
    \frac{\diff\sigma_l}{\diff^2 b}
    =
    \expval{\frac{1}{N_l}\sum_{f=1}^{N_l}\sum_{i=1}^{N_l}
    {\left|U_l |_{f i}-\delta_{f i}\right|}^2 }
    =
     2
    \left[
      1-\Re \expval{\Tr U_l}
    \right]
    \;,
\end{equation}
where $U_l$ is the Wilson line, $i, f$ the color indices, and $N_l$ the dimension of the color space of particle $l$. The trace $\Tr $ is over the color indices.
The $\expval{\ldots}$ stands for a configuration average of the background field.

The configuration average of a single Wilson line in the $R$ representation (e.g., the fundamental and the adjoint representations) is,
  \begin{align}\label{eq:Wilsonline_R}
    \begin{split}
    \expval{U_R(0,L_\eta; \vec x_\perp)_{\beta\alpha}}
    =&\bar U_R(0,L_\eta; \vec x_\perp)\delta_{\beta\alpha}\\
      =&\exp \bigg[
      \frac{-g^4\tilde \mu^2 L_\eta}{8\pi m_g^2 }C_R 
      \bigg]\delta_{\beta\alpha}\;,      
    \end{split}
  \end{align}
  where $C_R$ is the second-order Casimir invariant in the $R$ representation. 
In deriving the above expression of the Wilson line, one uses the correlation relation of the color sources in Eq.~\eqref{eq:chgcor} to contract the multiple color sources in the time-ordered exponential. 
After the contraction, each term in the Wilson line is then recollected into an exponential, as in Eq.~\eqref{eq:Wilsonline_R}. A detailed calculation can be found in Ref.~\cite{Fukushima:2007dy}.

For a quark, the corresponding Wilson line is in the fundamental representation, $C_F=(N_c^2-1)/(2N_c)=4/3$, so
  \begin{align}\label{eq:Wilsonline_F}
    \begin{split}
      \bar U_F(0,L_\eta; \vec x_\perp)
      =&\exp \bigg[
      \frac{- g^4\tilde \mu^2 L_\eta}{6\pi m_g^2 }
      \bigg] 
      \;.
    \end{split}
  \end{align}
For a gluon, the corresponding Wilson line is in the adjoint representation, $C_A=N_c = 3$, so
  \begin{align}\label{eq:Wilsonline_A}
    \begin{split}
      \bar U_A(0,L_\eta; \vec x_\perp)
      =&\exp \bigg[
      \frac{-3 g^4\tilde \mu^2 L_\eta}{8\pi m_g^2 }
      \bigg]
      \;.
    \end{split}
  \end{align}
For a quark-gluon state, the scattering amplitude is a two-point function of the Wilson lines. Here,  we make the derivation for a more general case, the tensor product of two Wilson lines with one in the $R_1$ representation and the other in the $R_2$ representation. 
The tensor element reads explicitly as
\begin{widetext}
\begin{align}\label{eq:UU}
  \begin{split}
    &\expval{U_{R_1}(0,L_\eta; \vec x_\perp)_{\beta_1\alpha_1}U_{R_2}(0,L_\eta; \vec y_\perp)_{\beta_2\alpha_2}}\\
    &=
    \sum_{n=0}^\infty
    {(-ig)}^n
    \int
    \left(
    \prod_{i=1}^n \diff^2 z_{i\perp}
    G_0(\vec x_\perp - \vec z_{i\perp})
    \right)
    \int_{0}^{L_\eta}\diff z_1^+
    \int_{z_1^+}^{L_\eta}d z_2^+ \cdots \int_{z_{n-1}^+}^{L_\eta}\diff z_n^+
    \sum_{n=0}^\infty
    {(-ig)}^n
    \int
    \left(
    \prod_{i=1}^n \diff^2 w_{i\perp}    G_0(\vec y_\perp - \vec w_{i\perp})
    \right)
    \\
    & \quad \quad 
    \int_{0}^{L_\eta}\diff w_1^+
    \int_{w_1^+}^{L_\eta}d w_2^+ \cdots \int_{w_{n-1}^+}^{L_\eta}\diff w_n^+
    \mathcal{T}_+
    \langle
      \rho_{a_1}(z_1^+, z_{1\perp})\rho_{a_2}(z_2^+, z_{2\perp})\cdots \rho_{a_n}(z_n^+, z_{n\perp})
    \rho_{b_1}(w_1^+, w_{1\perp})\rho_{b_2}(w_2^+, w_{2\perp})\cdots \rho_{b_n}(w_n^+, w_{n\perp})
    \rangle\\
    &  \quad \quad 
    t^{a_1}_{R_1} t^{a_2}_{R_1}\cdots t^{a_n}_{R_1}
    \bigg|_{\beta_1\alpha_1}
    t^{b_1}_{R_2} t^{b_2}_{R_2}\cdots t^{b_n}_{R_2}
    \bigg|_{\beta_2\alpha_2}
    \;.
  \end{split}
\end{align}  
\end{widetext}
where $t_R$ is the $\mathrm{SU}(3)$ generator in the $R$ representation. 
As in calculating the single Wilson line, we use the correlation relation of the color sources in Eq.~\eqref{eq:chgcor} to contract the multiple color sources. The difference is that here the contraction happens not only along each Wilson line but also between the two. 
We rewrite Eq.~\eqref{eq:UU} into the integral equation form~\cite{Fukushima:2007dy},
\begin{widetext}
\begin{align}\label{eq:UU_integral}
  \begin{split}
    &\expval{U_{R_1}(0,L_\eta; \vec x_\perp)_{\beta_1\alpha_1}U_{R_2}(0,L_\eta; \vec y_\perp)_{\beta_2\alpha_2}}\\
    & =
    \expval{U_{R_1}(0,L_\eta; \vec x_\perp)_{\beta_1\alpha_1}}
    \expval{U_{R_2}(0,L_\eta; \vec y_\perp)_{\beta_2\alpha_2}}
+ (-ig)^2
    \int_{0}^{L_\eta}\diff z_1^+
    \int_{0}^{L_\eta}\diff z_2^+
    \expval{U_{R_1}(z_1^+,L_\eta; \vec x_\perp)}_{\beta_1\lambda_1}
    \expval{U_{R_2}(z_2^+,L_\eta; \vec y_\perp)}_{\beta_2\lambda_2}
    \\
    &  \quad \quad \int \diff^2 z_{1\perp} G_0(\vec x_{\perp}-\vec z_{1\perp})
    \int \diff^2 z_{2\perp} G_0(\vec y_{\perp}-\vec z_{2\perp})
    \expval{\rho_a(\vec{z}_{1\perp},z_1^+)\rho_b(\vec{z}_{2\perp},z_2^+)}
    t^{a}_{R_1;\lambda_1\gamma_1} t^{b}_{R_2;\lambda_2\gamma_2}
%    \\ &\quad \quad \times
    \expval{U_{R_1}(0,z_1^+; \vec x_\perp)_{\gamma_1\alpha_1}U_{R_2}(0,z_2^+; \vec y_\perp)_{\gamma_2\alpha_2}}\\
    & =
    \bar U_{R_1}(0,L_\eta; \vec x_\perp)
    \bar U_{R_2}(0,L_\eta; \vec y_\perp)
    \bigg(
    \delta_{\beta_1\alpha_1}
    \delta_{\beta_2\alpha_2}      
    -g^4
    \int_{0}^{L_\eta}\diff z^+
    t^{a}_{R_1;\beta_1\gamma_1} t^{a}_{R_2;\beta_2\gamma_2}
    \tilde{\mu}^2(z^+)
    L(\vec x_\perp, \vec y_\perp)
    \frac{\expval{U_{R_1}(0,z^+; \vec x_\perp)_{\gamma_1\alpha_1}U_{R_2}(0,z^+; \vec y_\perp)_{\gamma_2\alpha_2}}}{\bar U_{R_1}(0,z^+; \vec x_\perp)
    \bar U_{R_2}(0,z^+; \vec y_\perp)}
    \bigg)
    \;.
  \end{split}
\end{align}
\end{widetext}
The solution is 
\begin{align}\label{eq:UU_solution}
 \begin{split}
    &\expval{U_{R_1}(0,L_\eta; \vec x_\perp)_{\beta_1\alpha_1}U_{R_2}(0,L_\eta; \vec y_\perp)_{\beta_2\alpha_2}}\\
    =&
    \bar U_{R_1}(0,L_\eta; \vec x_\perp)
    \bar U_{R_2}(0,L_\eta; \vec y_\perp)
\\
&\times 
    \exp\Bigg[
-g^4 
\frac{m_g|\vec x_\perp-\vec y_\perp| K_1\left(m_g|\vec x_\perp-\vec y_\perp|\right)}{4\pi m_g^2}
\\
&
\times\int_{0}^{L_\eta}\diff z^+
\tilde{\mu}^2(z^+)
t^{a}_{R_1} \otimes
t^{a}_{R_2}
\Bigg]_{\beta_1\beta_2;\alpha_1\alpha_2}
    \;.
  \end{split}
\end{align}
For a quark-gluon state scattering on a background field with constant $\tilde\mu$, Eq.~\eqref{eq:UU_solution} becomes, 
\begin{align}\label{eq:UqUg_solution}
 \begin{split}
   &\expval{U_F(0,L_\eta; \vec x_\perp)_{\beta_1\alpha_1}U_A(0,L_\eta; \vec y_\perp)_{\beta_2\alpha_2}}\\
    =&\exp \bigg[
  - 
  \frac{g^4\tilde \mu^2 L_\eta}{8\pi m_g^2 }(C_F+C_A) 
  \bigg]\\
  &\times
  \exp\bigg[
-
\frac{g^4
\tilde{\mu}^2 L_\eta}{4\pi m_g}
|\vec x_\perp-\vec y_\perp| K_1\left(m_g|\vec x_\perp-\vec y_\perp|\right)
\\ 
&\times
t^{a} \otimes T^{a}
\bigg]
_{\beta_1\beta_2;\alpha_1\alpha_2}
    \;.
  \end{split}
\end{align}

\begin{figure}[tbp!]
  \centering
     \includegraphics[width=0.4\textwidth]{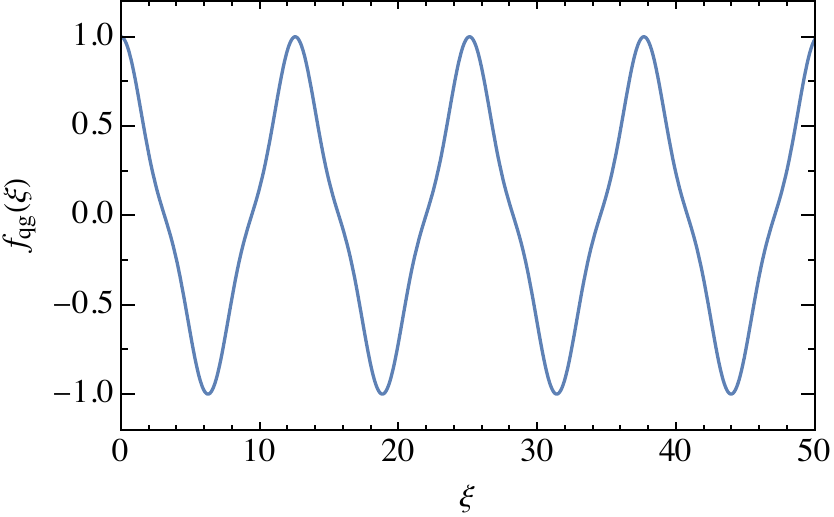}
  \caption{ 
  The function $f_{qg}(\xi)$ as a correlation between the quark and the gluon Wilson lines according to Eq.~\eqref{eq:f_qg_xi}.}
 \label{fig:f_qg_xi}
\end{figure}
\begin{figure}[tbp!]
  \centering
     \includegraphics[width=0.4\textwidth]{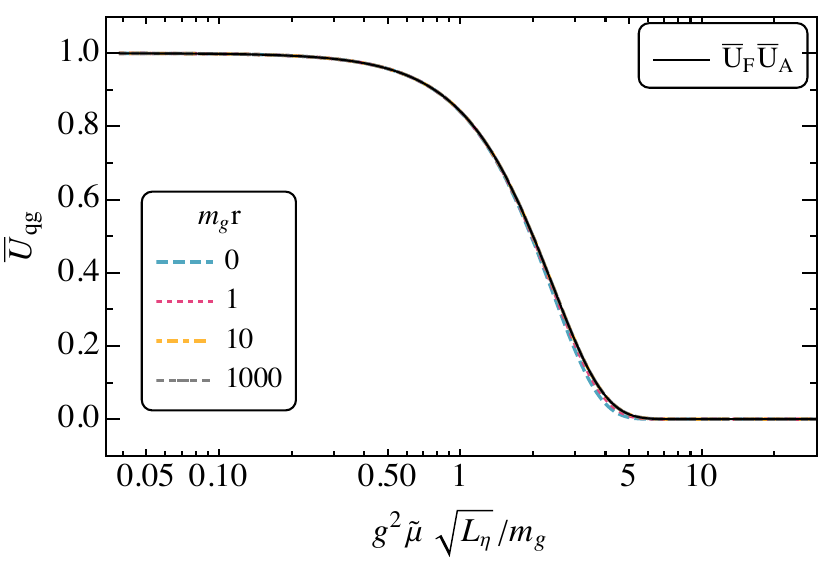}
  \caption{
    The product of the quark and the gluon Wilson lines $\bar U_{qg}$ plotted as a function of the dimensionless quantity $g^2\tilde{\mu} \sqrt{L_\eta}/ m_g$ at various $m_g r$, where $r=|\vec x_\perp -\vec y_\perp|$, according to Eq.~\eqref{eq:UqUg_ave}.
    The part without the quark-gluon correlation,
    $\bar{U}_F \bar{U}_A 
    = \exp [ - g^4\tilde \mu^2 L_\eta/(8\pi m_g^2)(C_F+C_A)]$ is plotted in the solid line.  }
  \label{fig:UqUg_Uqg}
\end{figure}

In calculating the total cross section, it is the real part of the trace of the averaged Wilson line that matters as in Eq.~\eqref{eq:crss_trace}, so we are interested in the following expression:
\begin{align}\label{eq:UqUg_ave}
  \begin{split}
    &\bar U_{qg}(0,L_\eta; \vec x_\perp,\vec y_\perp) \\
    =&\frac{1}{N_c(N_c^2-1)}
    \Re \Tr
    \expval{U_F(0,L_\eta; \vec x_\perp)
    U_A(0,L_\eta; \vec y_\perp)
    }\\
    =&
    \bar U_F(0,L_\eta; \vec x_\perp)
    \bar U_A(0,L_\eta; \vec y_\perp)\\
    &\times f_{qg}
    \left[
    \frac{g^4
    \tilde{\mu}^2 L_\eta}{4\pi m_g}
    |\vec x_\perp-\vec y_\perp| K_1\left(m_g|\vec x_\perp-\vec y_\perp|\right)
    \right]
    \;.
  \end{split}
\end{align}
In the second equation, $\bar{U}_F \bar{U}_A$ is the product of the averaged Wilson lines of the quark and the gluon  [see Eqs.~\eqref{eq:Wilsonline_F} and \eqref{eq:Wilsonline_A}].
Here, $f_{qg}$ is the contribution from the correlation between the quark's and the gluon's Wilson lines and acts as a correction factor to the $\bar{U}_F \bar{U}_A$ term. 
The functional form of $f_{qg}$ reads
\begin{align}\label{eq:f_qg_xi}
  \begin{split}
    f_{qg}(\xi)\equiv&
    \frac{1}{N_c(N_c^2-1)}
    \Re \Tr
    \exp(-\xi t^{a} \otimes T^{a})    \\ 
    =&\frac{1}{8}[7\cos(\xi/2)+\cos(3\xi/2)]\;,
  \end{split}
\end{align} 
and it is plotted in Fig.~\ref{fig:f_qg_xi}. 
It is a periodic function with a period of $4\pi$ and oscillates between $1$ and $-1$.
In Eq.~\eqref{eq:UqUg_ave}, this term depends on the dimensionless quantity $g^2\tilde{\mu} \sqrt{L_\eta}/ m_g$, just as the $\bar{U}_F \bar{U}_A$ term, but it also depends on the separation between the quark and the gluon, $r\equiv|\vec x_\perp-\vec y_\perp|$. 
The smaller the value of $r$ is, the faster $f_{qg}$ deviates from $1$ as a function of $g^2\tilde{\mu} \sqrt{L_\eta}/ m_g$, suggesting that the correlation is stronger when the quark and the gluon are closer.
One could also see this in the limit of infinite separation where $r=\infty$, the correlation becomes $f_{qg}(0)=1$.

The contribution from $f_{qg}(\xi)$ as a correction to the $\bar{U}_F \bar{U}_A$ term inside $\bar U_{qg}$ is actually very small.
Even in the strongest correlation case where $m_g r=0$, the first node of $f_{qg}=0$ occurs at $g^2\tilde{\mu} \sqrt{L_\eta}/ m_g = 2\pi$, where the value of $\bar{U}_F \bar{U}_A$ already reduces to $0.0011$. 
We present the plots of $\bar{U}_F \bar{U}_A$ and the correlated Wilson line $\bar U_{qg}$ as functions of the dimensionless quantity $g^2\tilde{\mu} \sqrt{L_\eta}/ m_g$ at various $m_g r$ in Fig.~\ref{fig:UqUg_Uqg}. 
Those $\bar U_{qg}$ curves barely deviate from $\bar{U}_F \bar{U}_A$, even in the zero separation case.
Indeed, the influence from the correlation function $f_{qg}$ is not very noticeable in $\bar U_{qg}$ as we have discussed.
From here, one could expect that $f_{qg}$ has little influence in the total cross section of a $\ket{qg}$ state interacting with the background field.

\bibliographystyle{JHEP-2modlong}
\bibliography{qA}

\providecommand{\href}[2]{#2}\begingroup\raggedright\begin{thebibliography}{10}

\bibitem{Gelis:2010nm}
F.~Gelis, E.~Iancu, J.~Jalilian-Marian and R.~Venugopalan, {\it {The Color
  Glass Condensate}},
  \href{http://dx.doi.org/10.1146/annurev.nucl.010909.083629}{{\em Ann. Rev.
  Nucl. Part. Sci.} {\bf 60} (2010) 463}
  [\href{http://arXiv.org/abs/1002.0333}{{\tt arXiv:1002.0333 [hep-ph]}}].
%%CITATION = ARXIV:1002.0333;%%

\bibitem{Casalderrey-Solana:2011ule}
J.~Casalderrey-Solana and E.~Iancu, {\it {Interference effects in
  medium-induced gluon radiation}},
  \href{http://dx.doi.org/10.1007/JHEP08(2011)015}{{\em JHEP} {\bf 08} (2011)
  015} [\href{http://arXiv.org/abs/1105.1760}{{\tt arXiv:1105.1760 [hep-ph]}}].

\bibitem{Mehtar-Tani:2011lic}
Y.~Mehtar-Tani, C.~A. Salgado and K.~Tywoniuk, {\it {The radiation pattern of a
  QCD antenna in a dilute medium}},
  \href{http://dx.doi.org/10.1007/JHEP04(2012)064}{{\em JHEP} {\bf 04} (2012)
  064} [\href{http://arXiv.org/abs/1112.5031}{{\tt arXiv:1112.5031 [hep-ph]}}].

\bibitem{Armesto:2012qa}
N.~Armesto, H.~Ma, M.~Martinez, Y.~Mehtar-Tani and C.~A. Salgado, {\it
  {Interference between initial and final state radiation in a QCD medium}},
  \href{http://dx.doi.org/10.1016/j.physletb.2012.09.039}{{\em Phys. Lett. B}
  {\bf 717} (2012) 280} [\href{http://arXiv.org/abs/1207.0984}{{\tt
  arXiv:1207.0984 [hep-ph]}}].

\bibitem{Armesto:2013fca}
N.~Armesto, H.~Ma, M.~Martinez, Y.~Mehtar-Tani and C.~A. Salgado, {\it
  {Coherence Phenomena between Initial and Final State Radiation in a Dense QCD
  Medium}},  \href{http://dx.doi.org/10.1007/JHEP12(2013)052}{{\em JHEP} {\bf
  12} (2013) 052} [\href{http://arXiv.org/abs/1308.2186}{{\tt arXiv:1308.2186
  [hep-ph]}}].

\bibitem{Kajantie:2019nse}
K.~Kajantie, L.~D. McLerran and R.~Paatelainen, {\it {Gluon Radiation from a
  classical point particle II: dense gluon fields}},
  \href{http://dx.doi.org/10.1103/PhysRevD.101.054012}{{\em Phys. Rev. D} {\bf
  101} (2020) 054012} [\href{http://arXiv.org/abs/1911.12738}{{\tt
  arXiv:1911.12738 [hep-ph]}}].

\bibitem{Kajantie:2019hft}
K.~Kajantie, L.~D. McLerran and R.~Paatelainen, {\it {Gluon Radiation from a
  Classical Point Particle}},
  \href{http://dx.doi.org/10.1103/PhysRevD.100.054011}{{\em Phys. Rev. D} {\bf
  100} (2019) 054011} [\href{http://arXiv.org/abs/1903.01381}{{\tt
  arXiv:1903.01381 [nucl-th]}}].

\bibitem{Altinoluk:2020oyd}
T.~Altinoluk, G.~Beuf, A.~Czajka and A.~Tymowska, {\it {Quarks at
  next-to-eikonal accuracy in the CGC I: Forward quark-nucleus scattering}},
  \href{http://arXiv.org/abs/2012.03886}{{\tt arXiv:2012.03886 [hep-ph]}}.

\bibitem{Chirilli:2021lif}
G.~A. Chirilli, {\it {High-energy Operator Product Expansion at sub-eikonal
  level}},  \href{http://arXiv.org/abs/2101.12744}{{\tt arXiv:2101.12744
  [hep-ph]}}.

\bibitem{Accardi:2012qut}
A.~Accardi {\em et.~al.}, {\it {Electron Ion Collider: The Next QCD Frontier}:
  {Understanding the glue that binds us all}},
  \href{http://dx.doi.org/10.1140/epja/i2016-16268-9}{{\em Eur. Phys. J. A}
  {\bf 52} (2016) 268} [\href{http://arXiv.org/abs/1212.1701}{{\tt
  arXiv:1212.1701 [nucl-ex]}}].

\bibitem{Kovchegov:2015pbl}
Y.~V. Kovchegov, D.~Pitonyak and M.~D. Sievert, {\it {Helicity Evolution at
  Small-x}},  \href{http://dx.doi.org/10.1007/JHEP01(2016)072}{{\em JHEP} {\bf
  01} (2016) 072} [\href{http://arXiv.org/abs/1511.06737}{{\tt arXiv:1511.06737
  [hep-ph]}}].
\newblock [Erratum: JHEP 10, 148 (2016)].

\bibitem{Kovchegov:2017lsr}
Y.~V. Kovchegov, D.~Pitonyak and M.~D. Sievert, {\it {Small-$x$ Asymptotics of
  the Gluon Helicity Distribution}},
  \href{http://dx.doi.org/10.1007/JHEP10(2017)198}{{\em JHEP} {\bf 10} (2017)
  198} [\href{http://arXiv.org/abs/1706.04236}{{\tt arXiv:1706.04236
  [nucl-th]}}].

\bibitem{Kovchegov:2018znm}
Y.~V. Kovchegov and M.~D. Sievert, {\it {Small-$x$ Helicity Evolution: an
  Operator Treatment}},
  \href{http://dx.doi.org/10.1103/PhysRevD.99.054032}{{\em Phys. Rev. D} {\bf
  99} (2019) 054032} [\href{http://arXiv.org/abs/1808.09010}{{\tt
  arXiv:1808.09010 [hep-ph]}}].

\bibitem{Jalilian-Marian:2019kaf}
J.~Jalilian-Marian, {\it {Rapidity loss, spin and angular asymmetries in
  scattering of a quark from color field of a proton (nucleus)}},
  \href{http://arXiv.org/abs/1912.08878}{{\tt arXiv:1912.08878 [hep-ph]}}.
%%CITATION = ARXIV:1912.08878;%%

\bibitem{Adamiak:2021ppq}
D.~Adamiak, Y.~V. Kovchegov, W.~Melnitchouk, D.~Pitonyak, N.~Sato and M.~D.
  Sievert, {\it {First analysis of world polarized DIS data with small-$x$
  helicity evolution}},  \href{http://arXiv.org/abs/2102.06159}{{\tt
  arXiv:2102.06159 [hep-ph]}}.

\bibitem{Blaizot:2012fh}
J.-P. Blaizot, F.~Dominguez, E.~Iancu and Y.~Mehtar-Tani, {\it {Medium-induced
  gluon branching}},  \href{http://dx.doi.org/10.1007/JHEP01(2013)143}{{\em
  JHEP} {\bf 01} (2013) 143} [\href{http://arXiv.org/abs/1209.4585}{{\tt
  arXiv:1209.4585 [hep-ph]}}].

\bibitem{CasalderreySolana:2012ef}
J.~Casalderrey-Solana, Y.~Mehtar-Tani, C.~A. Salgado and K.~Tywoniuk, {\it {New
  picture of jet quenching dictated by color coherence}},
  \href{http://dx.doi.org/10.1016/j.physletb.2013.07.046}{{\em Phys. Lett. B}
  {\bf 725} (2013) 357} [\href{http://arXiv.org/abs/1210.7765}{{\tt
  arXiv:1210.7765 [hep-ph]}}].

\bibitem{Blaizot:2013vha}
J.-P. Blaizot, F.~Dominguez, E.~Iancu and Y.~Mehtar-Tani, {\it {Probabilistic
  picture for medium-induced jet evolution}},
  \href{http://dx.doi.org/10.1007/JHEP06(2014)075}{{\em JHEP} {\bf 06} (2014)
  075} [\href{http://arXiv.org/abs/1311.5823}{{\tt arXiv:1311.5823 [hep-ph]}}].

\bibitem{Mehtar-Tani:2019ygg}
Y.~Mehtar-Tani and K.~Tywoniuk, {\it {Improved opacity expansion for
  medium-induced parton splitting}},
  \href{http://dx.doi.org/10.1007/JHEP06(2020)187}{{\em JHEP} {\bf 06} (2020)
  187} [\href{http://arXiv.org/abs/1910.02032}{{\tt arXiv:1910.02032
  [hep-ph]}}].

\bibitem{Barata:2021byj}
J.~a. Barata, F.~Dom\'\i{}nguez, C.~Salgado and V.~Vila, {\it {A modified
  in-medium evolution equation with color coherence}},
  \href{http://dx.doi.org/10.1007/JHEP05(2021)148}{{\em JHEP} {\bf 05} (2021)
  148} [\href{http://arXiv.org/abs/2101.12135}{{\tt arXiv:2101.12135
  [hep-ph]}}].

\bibitem{Zhao:2013cma}
X.~Zhao, A.~Ilderton, P.~Maris and J.~P. Vary, {\it {Scattering in
  Time-Dependent Basis Light-Front Quantization}},
  \href{http://dx.doi.org/10.1103/PhysRevD.88.065014}{{\em Phys. Rev.} {\bf
  D88} (2013) 065014} [\href{http://arXiv.org/abs/1303.3273}{{\tt
  arXiv:1303.3273 [nucl-th]}}].
%%CITATION = ARXIV:1303.3273;%%

\bibitem{Hu:2019hjx}
B.~Hu, A.~Ilderton and X.~Zhao, {\it {Scattering in strong electromagnetic
  fields: Transverse size effects in time-dependent basis light-front
  quantization}},  \href{http://dx.doi.org/10.1103/PhysRevD.102.016017}{{\em
  Phys. Rev. D} {\bf 102} (2020) 016017}
  [\href{http://arXiv.org/abs/1911.12307}{{\tt arXiv:1911.12307 [nucl-th]}}].

\bibitem{Chen:2017uuq}
G.~Chen, X.~Zhao, Y.~Li, K.~Tuchin and J.~P. Vary, {\it {Particle distribution
  in intense fields in a light-front Hamiltonian approach}},
  \href{http://dx.doi.org/10.1103/PhysRevD.95.096012}{{\em Phys. Rev.} {\bf
  D95} (2017) 096012} [\href{http://arXiv.org/abs/1702.06932}{{\tt
  arXiv:1702.06932 [nucl-th]}}].
%%CITATION = ARXIV:1702.06932;%%

\bibitem{Li:2020uhl}
M.~Li, X.~Zhao, P.~Maris, G.~Chen, Y.~Li, K.~Tuchin and J.~P. Vary, {\it
  {Ultrarelativistic quark-nucleus scattering in a light-front Hamiltonian
  approach}},  \href{http://dx.doi.org/10.1103/PhysRevD.101.076016}{{\em Phys.
  Rev. D} {\bf 101} (2020) 076016} [\href{http://arXiv.org/abs/2002.09757}{{\tt
  arXiv:2002.09757 [nucl-th]}}].

\bibitem{McLerran:1993ni}
L.~D. McLerran and R.~Venugopalan, {\it {Computing quark and gluon distribution
  functions for very large nuclei}},
  \href{http://dx.doi.org/10.1103/PhysRevD.49.2233}{{\em Phys. Rev.} {\bf D49}
  (1994) 2233} [\href{http://arXiv.org/abs/hep-ph/9309289}{{\tt
  arXiv:hep-ph/9309289 [hep-ph]}}].
%%CITATION = HEP-PH/9309289;%%

\bibitem{McLerran:1993ka}
L.~D. McLerran and R.~Venugopalan, {\it {Gluon distribution functions for very
  large nuclei at small transverse momentum}},
  \href{http://dx.doi.org/10.1103/PhysRevD.49.3352}{{\em Phys. Rev.} {\bf D49}
  (1994) 3352} [\href{http://arXiv.org/abs/hep-ph/9311205}{{\tt
  arXiv:hep-ph/9311205 [hep-ph]}}].
%%CITATION = HEP-PH/9311205;%%

\bibitem{McLerran:1994vd}
L.~D. McLerran and R.~Venugopalan, {\it {Green's functions in the color field
  of a large nucleus}},  \href{http://dx.doi.org/10.1103/PhysRevD.50.2225}{{\em
  Phys. Rev.} {\bf D50} (1994) 2225}
  [\href{http://arXiv.org/abs/hep-ph/9402335}{{\tt arXiv:hep-ph/9402335
  [hep-ph]}}].
%%CITATION = HEP-PH/9402335;%%

\bibitem{Brodsky:1997de}
S.~J. Brodsky, H.-C. Pauli and S.~S. Pinsky, {\it {Quantum chromodynamics and
  other field theories on the light cone}},
  \href{http://dx.doi.org/10.1016/S0370-1573(97)00089-6}{{\em Phys. Rept.} {\bf
  301} (1998) 299} [\href{http://arXiv.org/abs/hep-ph/9705477}{{\tt
  arXiv:hep-ph/9705477 [hep-ph]}}].
%%CITATION = HEP-PH/9705477;%%

\bibitem{McLerran:1998nk}
L.~D. McLerran and R.~Venugopalan, {\it {Fock space distributions, structure
  functions, higher twists and small x}},
  \href{http://dx.doi.org/10.1103/PhysRevD.59.094002}{{\em Phys. Rev.} {\bf
  D59} (1999) 094002} [\href{http://arXiv.org/abs/hep-ph/9809427}{{\tt
  arXiv:hep-ph/9809427 [hep-ph]}}].
%%CITATION = HEP-PH/9809427;%%

\bibitem{krasnitz2003gluon}
A.~Krasnitz, Y.~Nara and R.~Venugopalan, {\it Gluon production in the color
  glass condensate model of collisions of ultrarelativistic finite nuclei},
  {\em Nuclear Physics A} {\bf 717} (2003) 268.

\bibitem{Dumitru:2002qt}
A.~Dumitru and J.~Jalilian-Marian, {\it {Forward quark jets from protons
  shattering the colored glass}},
  \href{http://dx.doi.org/10.1103/PhysRevLett.89.022301}{{\em Phys. Rev. Lett.}
  {\bf 89} (2002) 022301} [\href{http://arXiv.org/abs/hep-ph/0204028}{{\tt
  arXiv:hep-ph/0204028 [hep-ph]}}].
%%CITATION = HEP-PH/0204028;%%

\bibitem{Fukushima:2007dy}
K.~Fukushima and Y.~Hidaka, {\it {Light projectile scattering off the color
  glass condensate}},
  \href{http://dx.doi.org/10.1088/1126-6708/2007/06/040}{{\em JHEP} {\bf 06}
  (2007) 040} [\href{http://arXiv.org/abs/0704.2806}{{\tt arXiv:0704.2806
  [hep-ph]}}].

\bibitem{Lappi:2007ku}
T.~Lappi, {\it {Wilson line correlator in the MV model: Relating the glasma to
  deep inelastic scattering}},
  \href{http://dx.doi.org/10.1140/epjc/s10052-008-0588-4}{{\em Eur. Phys. J. C}
  {\bf 55} (2008) 285} [\href{http://arXiv.org/abs/0711.3039}{{\tt
  arXiv:0711.3039 [hep-ph]}}].

\bibitem{MSD2}
A.~Askar and A.~S. Cakmak, {\it Explicit integration method for the
  time-dependent schrodinger equation for collision problems},  {\em The
  Journal of Chemical Physics} {\bf 68} (1978) 2794.

\bibitem{press2007numerical}
W.~H. Press, {\em Numerical recipes 3rd edition: The art of scientific
  computing}.
\newblock Cambridge university press, 2007.

\bibitem{Soper:1972xc}
D.~E. Soper, {\it {Infinite-momentum helicity states}},
  \href{http://dx.doi.org/10.1103/PhysRevD.5.1956}{{\em Phys. Rev. D} {\bf 5}
  (1972) 1956}.

\bibitem{Curtright:2015iba}
T.~L. Curtright and C.~K. Zachos, {\it {Elementary results for the fundamental
  representation of SU(3)}},
  \href{http://dx.doi.org/10.1016/S0034-4877(15)30040-9}{{\em Rept. Math.
  Phys.} {\bf 76} (2015) 401} [\href{http://arXiv.org/abs/1508.00868}{{\tt
  arXiv:1508.00868 [math.RT]}}].

\bibitem{Rabi:1937dgo}
I.~I. Rabi, {\it {Space Quantization in a Gyrating Magnetic Field}},
  \href{http://dx.doi.org/10.1103/PhysRev.51.652}{{\em Phys. Rev.} {\bf 51}
  (1937) 652}.

\bibitem{fox2006quantum}
M.~Fox, {\em Quantum Optics: An Introduction}.
\newblock Oxford Master Series in Physics. OUP Oxford, 2006.

\bibitem{Mueller:1997ik}
A.~H. Mueller, {\it {General issues in small x and diffractive physics}},
  \href{http://dx.doi.org/10.1007/s100500050027}{{\em Eur. Phys. J.} {\bf A1}
  (1998) 19} [\href{http://arXiv.org/abs/hep-ph/9710531}{{\tt
  arXiv:hep-ph/9710531 [hep-ph]}}].
%%CITATION = HEP-PH/9710531;%%

\bibitem{Kovchegov:1999kx}
Y.~V. Kovchegov and L.~D. McLerran, {\it {Diffractive structure function in a
  quasiclassical approximation}},
  \href{http://dx.doi.org/10.1103/PhysRevD.60.054025}{{\em Phys. Rev. D} {\bf
  60} (1999) 054025} [\href{http://arXiv.org/abs/hep-ph/9903246}{{\tt
  arXiv:hep-ph/9903246}}].
\newblock [Erratum: Phys.Rev.D 62, 019901 (2000)].

\bibitem{Gelis:2002ki}
F.~Gelis and J.~Jalilian-Marian, {\it {Photon production in high-energy proton
  nucleus collisions}},
  \href{http://dx.doi.org/10.1103/PhysRevD.66.014021}{{\em Phys. Rev. D} {\bf
  66} (2002) 014021} [\href{http://arXiv.org/abs/hep-ph/0205037}{{\tt
  arXiv:hep-ph/0205037}}].

\bibitem{JalilianMarian:2005jf}
J.~Jalilian-Marian and Y.~V. Kovchegov, {\it {Saturation physics and
  deuteron-Gold collisions at RHIC}},
  \href{http://dx.doi.org/10.1016/j.ppnp.2005.07.002}{{\em Prog. Part. Nucl.
  Phys.} {\bf 56} (2006) 104} [\href{http://arXiv.org/abs/hep-ph/0505052}{{\tt
  arXiv:hep-ph/0505052}}].

\bibitem{Cougoulic:2020tbc}
F.~Cougoulic and Y.~V. Kovchegov, {\it {Helicity-dependent extension of the
  McLerran\textendash{}Venugopalan model}},
  \href{http://dx.doi.org/10.1016/j.nuclphysa.2020.122051}{{\em Nucl. Phys. A}
  {\bf 1004} (2020) 122051} [\href{http://arXiv.org/abs/2005.14688}{{\tt
  arXiv:2005.14688 [hep-ph]}}].

\bibitem{Pauli:1995dt}
H.-C. Pauli, A.~C. Kalloniatis and S.~S. Pinsky, {\it {Towards solving QCD: The
  transverse zero modes in light cone quantization}},
  \href{http://dx.doi.org/10.1103/PhysRevD.52.1176}{{\em Phys. Rev.} {\bf D52}
  (1995) 1176} [\href{http://arXiv.org/abs/hep-th/9509020}{{\tt
  arXiv:hep-th/9509020 [hep-th]}}].
%%CITATION = HEP-TH/9509020;%%

\bibitem{Nielsen:1981hk}
H.~B. Nielsen and M.~Ninomiya, {\it {No Go Theorem for Regularizing Chiral
  Fermions}},  \href{http://dx.doi.org/10.1016/0370-2693(81)91026-1}{{\em Phys.
  Lett. B} {\bf 105} (1981) 219}.

\bibitem{Nielsen:1980rz}
H.~B. Nielsen and M.~Ninomiya, {\it {Absence of Neutrinos on a Lattice. 1.
  Proof by Homotopy Theory}},
  \href{http://dx.doi.org/10.1016/0550-3213(82)90011-6}{{\em Nucl. Phys. B}
  {\bf 185} (1981) 20}.
\newblock [Erratum: Nucl.Phys.B 195, 541 (1982)].

\bibitem{Faessler:2004yf}
S.~Chandrasekharan and U.~J. Wiese, {\it {An Introduction to chiral symmetry on
  the lattice}},  \href{http://dx.doi.org/10.1016/j.ppnp.2004.05.003}{{\em
  Prog. Part. Nucl. Phys.} {\bf 53} (2004) 373}
  [\href{http://arXiv.org/abs/hep-lat/0405024}{{\tt arXiv:hep-lat/0405024}}].

\bibitem{Wilson:1974sk}
K.~G. Wilson, {\it {Confinement of Quarks}},
  \href{http://dx.doi.org/10.1103/PhysRevD.10.2445}{{\em Phys. Rev. D} {\bf 10}
  (1974) 2445}.

\end{thebibliography}\endgroup
\end{document}